\documentclass{aa} % for a normal version
\usepackage{amsmath, amssymb,units}
\usepackage{graphicx}
\usepackage{array}
\usepackage{times}
\usepackage{calc}
\usepackage{epsfig}
\usepackage{natbib}
\bibpunct{(}{)}{;}{a}{}{,} % to follow the A&A style
\usepackage{txfonts}
\usepackage{txfonts}
\begin{document}
\title{Scaling properties of a complete X-ray selected galaxy group sample }
%\   \subtitle{I. Overviewing the $\kappa$-mechanism}
\author{L. Lovisari\inst{1} \and T.H. Reiprich\inst{1} \and G. Schellenberger\inst{1}
%\fnmsep\thanks{Just to show the usage  of the elements in the author field}
}

\institute{Argelander Institute for Astronomy (AIfA), University of Bonn,
           Auf dem H\"ugel 71, D-53121 Bonn \\
           \email{lorenzo@astro.uni-bonn.de}
%	   \and
%          University of Alexandria, Department of Geography, ...\\
%          \email{c.ptolemy@hipparch.uheaven.space}
%          \thanks{The university of heaven temporarily does not accept e-mails}
}

\date{Received ; accepted }

% \abstract{}{}{}{}{} 
% 5 {} token are mandatory
 
  \abstract
  % context heading (optional)
  % {} leave it empty if necessary  
  {Upcoming X-ray surveys like eROSITA require precise calibration between X-ray observables and mass down to the low-mass regime to set tight constraints on the
    fundamental cosmological parameters. Since an individual mass measurement is only possible for relatively few objects, it is crucial to have reliable and well understood scaling relations that relate the total mass to easily
    observable quantities.}
  % aims heading (mandatory)
  {The main goal of this work is to constrain the galaxy group scaling relations corrected for selection effects, and to quantify the influence of non-gravitational
physics at the low-mass regime.}
  % methods heading (mandatory)
  {We analyzed XMM-Newton observations for a complete sample of galaxy groups selected from the ROSAT All-Sky Survey and compared the derived scaling properties with a galaxy cluster sample. To investigate the role played by the different
    non-gravitational processes we then compared the observational
    data with the predictions of hydrodynamical simulations.}
  % results heading (mandatory)
  { After applying the correction for selection effects (e.g., Malmquist bias), the $L_\text{X}$-$M$ relation is steeper than the observed one. Its slope (1.66$\pm$0.22) is also steeper than the value obtained by using the more massive systems of the HIFLUGCS sample. This behavior can be explained by a gradual change of the true $L_\text{X}$-$M$ relation, which should be taken into account when converting the observational parameters into masses. The other observed scaling relations (not corrected for selection biases) do not show any break, although the comparison with the simulations suggests that feedback processes play an important role in the formation and evolution of galaxy groups. Thanks to our master sample of 82 objects spanning two order of magnitude in mass, we tightly constrain the dependence of the gas mass fraction on the total mass, finding a difference of almost a factor of two between groups and clusters. We also found that the use of different AtomDB versions in the calculation of the group properties (e.g., temperature and density) yields a gas fraction of up to 20$\%$ lower than an older version. 
}
  % conclusions heading (optional), leave it empty if necessary 
   {}

\keywords{Galaxies: clusters: general --
          Cosmology: observations --
          X-rays: galaxies: clusters
          }

\maketitle

%
%________________________________________________________________

\section{Introduction}
Groups and clusters of galaxies are the largest virialized structures
in the Universe. They form by continuous accretion of smaller mass
units such as field galaxies, groups, and small clusters. As a consequence
of this accretion, the mass function of virialized systems is quite
steep, and thereby, low-mass systems (often referred to as groups) are
more common than rich clusters.  Differently from massive clusters,
the impact of non-gravitational processes (cooling, AGN feedback, star
formation, and galactic winds) in galaxy groups begins to prevail over
the gravity because of their shallower potential well. These factors make
galaxy groups great laboratories through which to understand the complex baryonic
physics involved. Even though they play a central role in the process
of structure formation and evolution, they have been studied less frequently than massive clusters. Only recently, thanks to the high
sensitivity and resolution of X-ray observatories such as XMM-Newton,
Chandra, and Suzaku, the gas content of groups has been studied in more detail. Although several works indicate that groups are
consistent with being scaled-down clusters
(e.g.,  \citealt{2004MNRAS.350.1511O}; \citealt{2009ApJ...693.1142S}), many independent investigations conducted during the past decade have found that
 at low masses some scaling relations (e.g., $L_{\text{X}}$-$M$
and entropy-$T$) deviate from the relations of the most massive systems
(e.g., \citealt{2009ApJ...693.1142S};
\citealt{2011A&A...535A.105E}).\\
In the near future, eROSITA (e.g., \citealt{2010AIPC.1248..543P}, \citealt{2012arXiv1209.3114M}) is expected to detect $\sim$10$^5$ clusters, most of them in the low-mass regime \citep{2012MNRAS.422...44P}. Cosmological studies will at first require the knowledge of their redshift (e.g., from optical photometry or spectroscopy, or even X-ray spectroscopy, e.g., \citealt{2014A&A...567A..65B}) and total mass, which is not a direct observable. The hydrostatic cluster mass will be obtained only for a few objects because there will be too a few photons to determine the temperature and density profiles. This means that the mass will be inferred from other observable properties, and the cosmological studies will rely heavily on a detailed understanding of the scaling relations.\\
Different cluster properties have been suggested as mass proxies. Among them, the X-ray luminosity is the easiest to measure because it requires only a few tens of 
counts to be measured. Unfortunately,
the $L_{\text{X}}$-$M$ relation shows the largest scatter among the
scaling relations derived using a sample of galaxy clusters
(e.g., \citealt{2002ApJ...567..716R}), and the situation at the group
scale is even worse (e.g., \citealt{2011A&A...535A.105E}). A much lower
scatter can be obtained for this relation by removing the cluster cores
(e.g., \citealt{1998ApJ...504...27M}, \citealt{2007ApJ...668..772M},
\citealt{2009A&A...498..361P}). This implies that most of the scatter is
associated with the central part of the clusters, where cooling and
heating processes are stronger. For galaxy groups, the situation appears to be even more complex \citep{2014arXiv1402.0868B}. Furthermore, \citet{2009A&A...498..361P} showed that the intrinsic scatter is higher for disturbed than for relaxed systems. Since it is unlikely
that we will know the virialization state for all the objects detected
with eROSITA, it is essential to have mass proxies almost
independent of the dynamical state and morphology of the systems.\\
 \citet{2006ApJ...650..128K} showed that the thermal energy of the ICM,
 $Y_{\text{X}}$=$M_{\text{gas}} \times T$, is a reliable and low-scatter
 X-ray mass proxy independent of the dynamical state of the objects and is 
 insensitive to specific assumptions in modeling feedback processes in simulations (e.g., \citealt{2010ApJ...715.1508S}).
  The $M$-$T$ relation is also expected to have a small scatter once the
 central regions are removed from the analysis, because without non-gravitational heating and cooling, the temperature of a
 cluster is only determined by the depth of the potential well.  While for the  $M$-$Y_{\text{X}}$ relation the observed scatter ($\sim$20$\%$, e.g., \citealt{2010ApJ...721..875O}) is larger  than the scatter expected from simulations ($<$10$\%$), the $M$-$T$ relation indeed shows a small scatter (e.g., 13$\%$
, \citealt{2010MNRAS.406.1773M}). 
The scatter at the group scale has been found to be larger for both relations (e.g., \citealt{2011A&A...535A.105E}). On the other hand,
\citet{2009ApJ...693.1142S} showed that because of the large measurement errors
 of the group properties the intrinsic scatter is
consistent with zero in both relations. \\
Another mass proxy, that is often used in literature is $M_{\text{gas}}$. \citet{2010ApJ...721..875O} showed that it has a very small
scatter with total mass and is almost independent of the dynamical
state. A weak dependence on the redshift and lower gas mass in low
mass systems than expected from the self-similar scenario has been
found by \citet{2003ApJ...590...15V} and \citet{2008A&A...482..451Z},
respectively.\\
The observational determination of all the scaling relations can be significantly affected by selection biases (see the examples provided in Sect. 7 of \citealt{2013SSRv..177..247G}). The main goal of this paper is to correct for the selection bias effects the $L_\text{X}$-$M$ and the $L_\text{X}$-$T$ relations by studying a complete sample of X-ray galaxy groups for the first time. Furthermore, through a comparison with the HIFLUGCS results we test the breaking of the scaling relations and the relative importance
of gravitational and non-gravitational processes in the low-mass regime. To do this we excluded all the objects with a temperature lower than 3 keV from the HIFLUGCS sample so that only the most massive systems in the sample remain.\\
The paper is structured as follows: in Sect. \ref{sec_datared} we
describe the sample selection and data reduction. The
data analysis is presented in Sect. \ref{sec_dataanal}. The methodology for  quantifying the scaling relations is described in Sect. \ref{scallaw}. We present our
results in Sect. \ref{sec_results} and discuss them in Sect. \ref{sec_discussion}, including a comparison with other observational data and simulations. Sect. \ref{conclusion} contains the
summary and conclusions. Throughout this paper, we assume a flat
${\rm\Lambda}$CDM cosmology with $\Omega_m=0.27$ and $H_0=70 \ \rm{km
  \ s^{-1} \ Mpc^{-1}}$. Log is always base 10 here. Errorbars are at the 68$\%$ c.l..

%__________________________________________________________________

\section{X-ray observations and data reduction} \label{sec_datared}

\begin{table}[t]
  \caption{Galaxy groups in the sample sorted by redshift. $L_\text{X}$ is the luminosity in the 0.1-2.4 keV band.
 }
$$
\centering
\begin{array}{ccccc}
\hline\hline
\noalign{\smallskip}
{\rm Name} & z & {\rm N_H}  &  {\rm L_X} &   {\rm L_X \ error}  \\
     &   & {\rm 10^{20} \ cm^{-2} } & {\rm 10^{43} \ erg \ s^{-1} } & {\rm \% }          \\  
\hline
{\rm NGC4936} 	 	 & 0.012 & 6.79 & 0.22 & 13.6 \\ 
{\rm S0753}  		 & 0.013 & 5.31 & 0.31 & 14.4 \\ 
{\rm HCG62}  	  	 & 0.015 & 3.30 & 0.29 & 13.2 \\ 
{\rm S0805}  		 & 0.015 & 6.11 & 0.26 & 26.1 \\ 
{\rm NGC3402}		 & 0.016 & 3.87 & 0.55 & 6.8  \\ 
{\rm A3574E^{*}} 	 & 0.016 & 4.40 & 0.43 & 25.0 \\ 
{\rm A194 } 		 & 0.018 & 4.13 & 0.71 & 14.1 \\ 
{\rm RXCJ1840.6-7709}    & 0.019 & 8.45 & 0.84 & 10.4 \\ 
{\rm WBL154 	}	 & 0.023 & 5.10 & 0.68 & 6.2  \\ 
{\rm S0301  }		 & 0.023 & 2.39 & 0.84 & 8.6  \\ 
{\rm NGC1132 }	 	 & 0.024 & 5.46 & 0.73 & 16.2 \\ 
{\rm IC1633 }		 & 0.024 & 1.93 & 1.70 & 8.0  \\ 
{\rm NGC4325 }	 	 & 0.026 & 2.32 & 1.00 & 7.8  \\ 
{\rm RXCJ2315.7-0222 }   & 0.027 & 3.68 & 1.29 & 10.6 \\ 
{\rm NGC6338  }		 & 0.028 & 2.23 & 2.48 & 15.4 \\ 
{\rm IIIZw054	}	 & 0.029 & 15.0 & 3.86 & 7.5  \\ 
{\rm IC1262    }         & 0.031 & 1.78 & 2.39 & 5.1  \\ 
{\rm NGC6107^{*}  }	 & 0.032 & 1.50 & 1.72 & 8.0  \\ 
{\rm AWM4  	 }	 & 0.033 & 5.14 & 2.83 & 14.5 \\ 
{\rm A3390  	}	 & 0.033 & 7.03 & 1.41 & 11.3 \\ 
{\rm CID28 	}	 & 0.034 & 4.08 & 1.40 & 11.8 \\ 
{\rm AWM5^{*} }		 & 0.034 & 5.00 & 1.85 & 8.2  \\ 
{\rm UGC03957 	}	 & 0.034 & 4.27 & 4.71 & 7.5  \\ 
\hline
\end{array}
\label{tab:sample}
$$
\end{table}

\subsection{Sample selection}
A complete sample is required for any meaningful study of the scaling
relations, otherwise potentially important biases (e.g., Malmquist
bias) cannot be corrected for.\\
We constructed a complete sample of galaxy groups by applying a flux
limit and two redshift cuts to the NORAS and REFLEX catalogs
(\citealt{2000ApJS..129..435B} and \citealt{2004A&A...425..367B},
respectively). The lower $z$-cut ({\it z}$\ge$0.010) ensures that for most of the objects
$>$0.5$R_{\text{500}}$ fits into the XMM-Newton field of view. The upper $z$-cut ({\it z}$\le$0.035)
prevents galaxy clusters from being sampled because this limits the sampled volume, and massive systems are very rare. Alternatively, we could have applied an upper luminosity cut, but this would have enhanced the selection effects, particularly on the slope of the scaling relations, as shown by \citet{2011A&A...535A.105E}.  We continually decreased the
flux limit below the HIFLUGCS limit $f_{\text{lim}} (0.1-2.4 \ \text{keV}) =
2\times10^{-11} \ \text{erg/s/cm$^{2}$}$ until a sufficiently large number of low
$L_\text{X}$ groups was reached. This resulted in a flux limit of $f_{\text{lim}}
(0.1-2.4 \ \text{keV}) = 5\times10^{-12} \ \text{erg/s/cm$^{2}$}$ and 23 groups in the
sample (see Table \ref{tab:sample}). All 23 groups have been observed with XMM-Newton. Of these 23 objects, three 
(marked with a star in Table \ref{tab:sample}) had to be excluded for
technical reasons. NGC6107 and AWM5 have been observed in one
(Rev. 0672) and three pointings (Revs. 1039, 1040 and 1041),
 but all of them are completely flared. A3574E has been
observed in Revs. 210 and 670 in Small and Large window mode. With
these configurations most of the ICM extended emission is lost. Only
the pn detector during Rev. 210 was set to Full Frame mode, but the
observation suffer of pileup problems due to the presence of a strong
Seyfert galaxy at 3 arcmin from the galaxy group center (see
\citealt{2004A&A...422...65B} for more details). We analyzed
the data after excising the point spread function (PSF) to reduce the pileup effect, but the
data did not allow a good estimate of the cluster properties (e.g., k$T$), and we
 also excluded this object from the analysis. Thus, the results of this
work are based on the analysis of the remaining 20 objects. We decided
to include IIIZw54 in our sample, even though it is one of the HIFLUGCS
clusters, because its flux $f(0.1-2.4 \ \text{keV}) = 2\times10^{-11}
\ \text{erg/s/cm$^{2}$}$ lies exactly at the flux limit.\\
We did not lower the flux limit any further to avoid
compromising the quality of our sample by getting close to the flux limits of the input catalogs, where their incompleteness starts to
become significant. NORAS and REFLEX are estimated to be complete at
50$\%$ and $90\%$ at $f_{\text{lim}}
(0.1-2.4 \ \text{keV}) = 3\times10^{-12} \
\text{erg/s/cm$^{2}$}$, respectively (\citealt{2000ApJS..129..435B}, \citealt{2004A&A...425..367B}). The incompleteness increases at
high redshift and very low fluxes, which means that the effect is expected to be small for our sample. Indeed, it might be difficult to resolve
groups with a too compact X-ray emission, but also this effect should
be minor at the considered redshift range.

\subsection{Data reduction}
Observation data files (ODFs) were retrieved from the XMM-Newton archive and
reprocessed with the XMMSAS v11.0.0 software for data reduction. We
used the tasks {\it emchain} and {\it epchain} to generate calibrated event
files from raw data. We only considered event patterns 0-12 for MOS
and 0 for pn, and the data were cleaned using the standard procedures
for bright pixels and hot columns removal (by applying the expression
FLAG==0) and pn out-of-time correction.\\
The data were cleaned for periods of high background due to the soft
protons using a two-stage filtering process. We first accumulated in
100 s bins the light curve in the [10-12] keV band for MOS and [12-14]
keV for pn, where the particle background completely dominates because
there is little X-ray emission from clusters as a result of the small telescope
effective area at these energies. As in \citet{2002A&A...394..375P}, a Poisson distribution was fitted to a
histogram of this light curve, and a $\pm3\sigma$ thresholds
calculated. After filtering using the good time intervals (GTIs) from this
screening, the event list was then re-filtered in a second pass as a
safety check for possible flares with soft spectra
(e.g., \citealt{2004A&A...419..837D}; \citealt{2005ApJ...629..172N};
\citealt{2005A&A...443..721P}). In this case, light curves were made
with 10s bins in the full [0.3-10] keV band. Point sources were detected using the task {\it ewavelet} in the
energy band [0.3-10] keV and checked by eye on images generated for
each detector. We produced a list of selected point sources from all
available detectors, and the events in the corresponding regions were
removed from the event lists.\\ The background event
files were screened by the GTIs of the background
data, which were produced by applying the same PATTERN selection, flare
rejection, and point-source removal as for the corresponding target
observations.\\
For both images and spectra the vignetting was taken into account using
the weighting method described in \cite{2001A&A...365L..80A}, which
allowed us to use the on-axis response files.

\subsection{Background subtraction}
The background treatment is very important when fitting spectra
extracted from faint regions, where the emission flux of the object is
similar to that of the background. Correcting for all various background
components in XMM-Newton data is very challenging, in particular when
the source fills the entire field of view, as for the objects in our
sample.\\ After the flare rejection, the background consists mainly
of two components: the non-vignetted quiescent particle background
(QPB) and the cosmic X-ray background (CXB). The first is the sum of a
continuum component and fluorescence X-ray lines, with the continuum
dominating above 2 keV and lines dominating in the 1.3-1.9 keV band
(see \citealt{2008A&A...478..615S} for more details). Beyond $\sim$5 arcmin the background spectrum for the pn detector also show other strong fluorescence lines (e.g., Ni, Cu, and Zn lines); these are not important for this work because we excluded all the data above 7 keV. The CXB, showing
significant vignetting, consists of a thermal emission from the Local
Hot Bubble (LHB) and the Galactic halo and from an extragalactic component
representing the unresolved emission from AGNs. All these different
components exhibit different spectral and temporal characteristics
(e.g., \citealt{2002A&A...389...93L}; \citealt{2003A&A...409..395R};
\citealt{2008A&A...478..575K}; \citealt{2008A&A...478..615S}).\\
A good template for the instrumental background can be obtained using
the filter wheel closed (FWC) observations, as documented by
\citet{2008A&A...478..615S} for MOS and \cite{mekal...freyberg} for
pn. The intensity of this component can vary by typically $\pm10\%$
and must be accounted for by
renormalization. \citet{2004A&A...419..837D} pointed out that a simple
renormalization of the instrumental background using the high-energy
band count rate may lead to systematic errors in both the continuum
and the lines. \citet{2009ApJ...699.1178Z} found that the [3-10] keV
band is the best energy range to compute the renormalization
factors. We therefore re-normalized the FWC observations to
subtract the instrumental background using the count rate in the
[3-10] keV band. To determine the normalization, we computed the count
rate using events out of the field of view in the CCD 3 and
CCD 6 (when available) for MOS1, CCDs 3, 4 and 6 for MOS2, and CCDs 3,
6, 9 and 12 for pn. We excluded the other CCDs because
\citet{2004A&A...419..837D} and \citet{2008A&A...478..615S} found that
both X-ray photons and low-energy particles can reach the unexposed area of these CCDs.\\
Instead of subtracting the CXB from the spectra, we included the CXB
components during fitting following the method presented in
\citet{2008A&A...478..615S}. We used the spectra extracted from the
region just beyond the virial radius from RASS data to model the CXB
using the available tool\footnote
{http://heasarc.gsfc.nasa.gov/cgi-bin/Tools/xraybg/xraybg.pl} at
the HEASARC webpage. These data were simultaneously fit with the XMM-Newton
spectra after proper corrections for the observed solid angle. The
best-fits of the spectra show that the CXB can typically be well described using
a model that includes: 1) an absorbed $\sim$0.2 keV thermal component
representing the Galactic halo emission; 2) an unabsorbed $\sim$0.1
keV representing the LHB; and 3) an absorbed power-law model with its slope
set to 1.41 representing the unresolved point sources
(\citealt{2004A&A...419..837D}). In a few cases, a second absorbed
thermal emission component was needed to properly fit the CXB
background.  To derive the normalization (metallicity and redshift of the different background components were frozen to 1 and 0) of the CXB and to measure the
cluster emission, we made a joint fit of the RASS spectrum and
all the spectra for each observation with the fore- and background
components left free to vary and linked across all regions and
detectors. The temperatures, metallicities
(\citealt{2009ARA&A..47..481A}), and normalizations of the cluster
emission were left free and linked across the three EPIC detectors.\\
To check the consistency between the XMM-Newton and RASS background, we also obtained the background properties by using only the XMM-Newton data for all the objects with a source-free area beyond 9 arcmin. In general, the fits are quite unstable, and in some cases, the final results seem to depend on the initial fit values. However, when freezing the background temperatures to the best-fit values obtained by \cite{2000ApJ...543..195K}, the normalizations of the background components agree well with the results we obtained by fitting the XMM-Newton data together with the ROSAT data.

\section{Data analysis} \label{sec_dataanal}

\subsection{Emission peak and emission-weighted center}
The sample contains clusters with a wide variety of dynamical states
that can have a strong influence on the derived global properties.
Objects with a relatively high projected separation between the X-ray
emission peak (EP) and the centroid of the emission (e.g., emission-weighted center, EWC) are considered irregular or unrelaxed
(e.g., \citealt{1993ApJ...413..492M}).  These offsets are probably
caused by substructure, as a result of the infall of smaller groups of
galaxies. The EP position is probably associated with the minimum of
the potential well of the objects, but when studying the global
properties of groups and clusters, we are more interested in the
behavior of the slope of the surface brightness (SB) and temperature profiles at large
radii.  The EP and the EWC for each group were determined from
adaptively smoothed, background-subtracted, vignetting-corrected
images taking into account CCD gaps and point-source exclusion. We
used a 10 arcmin circle to derive the EWC by iterating the process
untill the coordinates of the flux-weighted center ceased to vary. Fewer than ten iterations were typically needed to reach the
goal. In most cases, the EP and EWC coincide, although in groups with
irregular morphology they can be separated by several arcmin. One
group (NGC6338) is centered on the FOV border so that the EWC cannot
be determined, and we only used the EP.

\subsection{Surface brightness}
We computed background-subtracted, vignetting-corrected, radial
SB profiles centered on the EWC in the [0.7-2] keV energy band, which provides an optimal ratio of the source and background flux in XMM-Newton data. This also 
ensures an almost temperature-independent X-ray emission coefficient
over the expected temperature range, although 
this is not strictly true at very low temperatures. The surface brightness profiles were 
convolved with the XMM-Newton PSF.  Both the single
($N$=1) and double ($N$=2) $\beta$-models were used in fitting:
\begin{eqnarray}
S_X &=& \displaystyle\sum_{i=1}^N S_i
\left[1+\left(\frac{r}{r_{c,i}}\right)^2\right]^{-3\beta_i+0.5},
\end{eqnarray}
and using the F-test functionality of Sherpa, we determined whether the
addition of extra model components was justified given the degrees of
freedom and $\chi^2$ values of each fit. If the significance was lower
than 0.05, the extra components were justified and the double
$\beta$-model was used. Apart from IIIZw054 and NGC4325, all the groups are better
fitted by a double $\beta$-component, therefore we only show the
analysis performed with the two $\beta$ parameters.\\
When fitting the SB profiles of irregular groups extracted from the
EWC center, the fit statistics may be poor because of the excess associated
with the EP. Whem these spikes are excluded the reduced $\chi^2$ improves significantly 
without strongly modifying the final fit values, therefor we decided to retain them. The derived SB profiles and the best-fit parameters are shown in Appendix \ref{SBprofiles}.

\subsection{Spectral analysis}
All the extracted spectra were re-binned to ensure a signal-to-noise
ratio of at least 3 and at least 20 counts per bin which is necessary for the
validity of the $\chi^2$ minimization method.  Spectra were fit
in the 0.5-7 keV energy range, excluding the 1.4-1.6 keV band because of
the strong Al line in all three detectors.  We fitted the particle
background-subtracted spectrum with an absorbed APEC thermal plasma
(\citealt{2001ApJ...556L..91S}). The EPIC spectra were fitted
simultaneously, with temperatures and metallicities tied and enforcing
the same normalization value for MOS, and allowing the pn normalization
to vary. In the outer bins the abundances are sometimes unconstrained, 
therefore we linked the metallicity of those bin to that of the next inner
annulus, as in \citet{2008A&A...478..615S}.

\subsection{Temperature profiles}
For the radial temperature profiles, successive annular regions were created around the EP and EWC. The annuli for spectral analysis were determined by requiring that the width of each annulus is larger than 0.5$^{\prime}$ and a S/N$>$50. The first requirement ensures that the
redistribution fraction of the flux is at most about 20$\%$
(\citealt{2009ApJ...699.1178Z}), the second that the uncertainty in
the spectrally resolved temperature (and consequently in the fitted temperature
profiles) is $<$10$\%$. The outermost annuli were selected to ensure that we obtained
the largest extraction radius with a S/N$>$50.\\
The temperature profiles were modeled using the parametrization
proposed by \citet{2005A&A...432..809D}:
\begin{eqnarray}
\bullet \ \ \ T(r)=T_0+2T_0\frac{(R/r_t)^{1/2}}{1+(R/r_t)^2},
\end{eqnarray} 
and by \citet{2007ApJ...669..158G}: 

\begin{eqnarray}
\bullet \ \ \ \lefteqn{ T(r)=\frac{1}{\left[\left(\frac{1}{t_1(r)}\right)^s + \left( \frac{1}{t_2(r)}\right)^s \right]^{\frac{1}{s}}} \nonumber
} \\
\lefteqn{ t_i(r)=T_{i,100}\left(\frac{r}{100 \ {\rm kpc}}\right)^{p_i} \ \ \ i=1,2 }
\end{eqnarray}
\begin{eqnarray}
\bullet \ \ \ \lefteqn{ T(r)=T_0+t_1(r)e^{-\left(\frac{r}{rp}\right)^{\gamma}}+t_2(r)\left(1-e^{-\left(\frac{r}{rp}\right)^{\gamma}}\right) \nonumber
} \\
\lefteqn{ t_i(r)=T_i\left(\frac{r}{r_0}\right)^{p_i} \ \ \ i=1,2. }
\end{eqnarray} 
These different parameterizations have enough flexibility to describe
the temperature profiles of all the systems in our sample. For every object we used the parameterization that returns the best $\chi^2$. In some cases, the profiles are quite well fitted by at least two functionals, but the change in the temperature at R$_{500}$ is only around 5$\%$.\\
When comparing our temperature profiles (Appendix \ref{kTprofiles}) with those in literature we
sometimes see a higher temperature than reported in previous results. This is
mainly because most of the previous works are based on
AtomDB\footnote{http://www.atomdb.org/index.php} 1.3.1 or older
versions. Version 2.0.1\footnote{Since February 2012 
version 2.0.2 is available, which includes some new corrections that only have
 a weak effect at X-ray wavelengths.} (\citealt{2012ApJ...756..128F}),
released in 2011 and used in this work, includes significant changes
in the iron L-shell complex, which is particularly important when
studying galaxy groups. The new temperatures are higher by up to 20$\%$
and the normalization lower by $\sim$10$\%$ (see the discussion in Section
\ref{fgasdiscuss} and Appendix \ref{app_atom}). In general, the low-mass objects are more affected than the more massive systems.

\begin{table*}[htp]
\caption{Derived properties for the galaxy groups.}
$$
\centering
\setlength\extrarowheight{2pt}
\begin{array}{ccccccccccc}
\hline\hline
\noalign{\smallskip}
{\rm Group \ name}  & kT & R_{spec} & R_{500} & M_{500} & M_{gas,500} & R_{2500}  & M_{2500} & M_{gas,2500} & t_{cool} & L_{X,xmm}\\ 
				   & {\rm kev} & {\rm h^{-1}_{70} \ kpc} & {\rm h^{-1}_{70} \ kpc}  & {\rm 10^{13} \ h^{-1}_{70} \ M_{\sun} } & {\rm 10^{12} \ h^{-5/2}_{70} \ M_{\sun}} & {\rm h^{-1}_{70} \ kpc} & {\rm 10^{13} \ h^{-1}_{70} \ M_{\odot} } & {\rm 10^{12} \ h^{-5/2}_{70}} & {\rm Gyr} & {\rm 10^{43} \ erg \ s^{-1}} \\ 
\noalign{\smallskip}
\hline
{\rm NGC4936} & 0.85\pm0.03 & 167 & 417\pm21 & 2.07\pm0.32 & 1.69\pm0.24 & 187\pm9 & 0.93\pm0.14 & 0.49\pm0.02 & 0.69\pm0.15 & 0.42\pm0.04 \\ 
{\rm S0753}   & 1.51\pm0.03 & 219 & 547\pm36 & 4.67\pm0.94 & 3.53\pm0.19 & 211\pm5 & 1.33\pm0.11 & 0.62\pm0.04 & 1.30\pm0.26 & 0.69\pm0.08  \\ 
{\rm HCG62}   & 1.05\pm0.01 & 245 & 437\pm6 & 2.39\pm0.54 & 2.01\pm0.22 & 202\pm1 & 1.17\pm0.24 & 0.49\pm0.08 & 0.10\pm0.14 & 0.62\pm0.06 \\ 
{\rm S0805}   & 1.01\pm0.01 & 201 & 427\pm58 & 2.22\pm0.97 & 1.71\pm0.42 & 155\pm13 & 0.53\pm0.13 & 0.17\pm0.04 & 1.03\pm0.07 & 0.62\pm0.05 \\ 
{\rm NGC3402} & 0.96\pm0.02 & 127 & 470\pm9 & 2.95\pm0.18 & 1.35\pm0.16 & 210\pm2 & 1.32\pm0.29 & 0.49\pm0.02 & 0.17\pm0.01 & 0.66\pm0.03 \\ 
{\rm A194 }   & 1.37\pm0.04 & 278 & 463\pm23 & 2.83\pm0.43 & 2.72\pm0.26 & 197\pm11 & 1.09\pm0.13 & 0.47\pm0.06 & 21.6\pm4.41 & 0.71\pm0.07 \\ 
{\rm RXCJ1840.6-7709} & 1.17\pm0.04 & 112 & 488\pm12 & 3.31\pm0.24 & 1.71\pm0.17 & 219\pm5 & 1.49\pm0.10 & 0.53\pm0.04 & 0.05\pm0.01 & 1.01\pm0.06 \\ 
{\rm WBL154^{*}} & 1.19\pm0.02 & 254 & 508\pm8 & 3.53\pm0.18 & 2.71\pm0.16 & 233\pm4 & 1.71\pm0.07 & 0.60\pm0.04 & - & 0.68\pm0.04 \\ 
{\rm S0301  } & 1.35\pm0.03 & 348 & 497\pm47 & 3.49\pm0.86 & 3.06\pm0.44 & 230\pm8 & 1.73\pm0.32 & 0.91\pm0.05 & 0.48\pm0.03 & 0.95\pm0.08 \\ 
{\rm NGC1132} & 1.08\pm0.01 & 206 & 490\pm9 & 3.35\pm0.19 & 2.47\pm0.06 & 215\pm2 & 1.41\pm0.04 & 0.69\pm0.01 & 0.63\pm0.03 & 0.84\pm0.04 \\ 
{\rm IC1633 }  & 2.80\pm0.06 & 263 & 797\pm80 & 14.4\pm2.62 & 11.5\pm1.65 & 289\pm35 & 3.43\pm1.25 & 1.59\pm0.43 & 5.26\pm1.21 & 2.07\pm0.64 \\ 
{\rm NGC4325} & 1.00\pm0.01 & 252 & 435\pm3 & 2.34\pm0.05 & 1.69\pm0.03 & 205\pm1 & 1.22\pm0.01 & 0.62\pm0.01 & 0.14\pm0.01 & 1.29\pm0.07 \\ 
{\rm RXCJ2315.7-0222} & 1.39\pm0.02 & 425 & 552\pm25 & 4.78\pm0.68 & 3.79\pm0.15 & 253\pm10 & 2.30\pm0.30 & 1.14\pm0.09 & 0.92\pm0.44 & 1.31\pm0.10 \\ 
{\rm NGC6338} & 1.97\pm0.09 & 436 & 671\pm39 & 8.59\pm1.50 & 6.34\pm0.33 & 295\pm15 & 3.66\pm0.13 & 1.96\pm0.17 & 0.69\pm0.11 & 2.58\pm0.40 \\ 
{\rm IIIZw054}& 2.17\pm0.02 & 312 & 694\pm14 & 9.51\pm0.57 & 9.06\pm0.37 & 307\pm5 & 4.11\pm0.19 & 2.42\pm0.10 & 4.95\pm0.97 & 3.90\pm0.24 \\ 
{\rm IC1262 } & 1.87\pm0.02 & 480 & 625\pm11 & 6.96\pm0.37 & 5.34\pm0.29 & 208\pm13 & 1.28\pm0.23 & 0.62\pm0.11 & 0.88\pm0.09 & 3.28\pm0.58 \\ 
{\rm AWM4  	} & 2.45\pm0.03 & 333 & 724\pm38 & 10.8\pm1.72 & 6.50\pm0.35 & 309\pm12 & 4.19\pm0.49 & 2.16\pm0.15 & 4.41\pm0.25 & 2.87\pm0.25 \\ 
{\rm A3390^{*}} & 1.58\pm0.06 & 272 & 543\pm21 & 4.37\pm0.51 & 1.94\pm0.19 & 253\pm14 & 2.22\pm0.37 & 0.51\pm0.07 & - & 1.41\pm0.11 \\ 
{\rm CID28}   & 2.01\pm0.02 & 360 & 655\pm11 & 8.00\pm0.42 & 6.27\pm0.16 & 295\pm3 & 3.66\pm0.13 & 1.77\pm0.04 & 2.50\pm0.22 & 1.82\pm0.09 \\ 
{\rm UGC03957}& 2.66\pm0.03 & 496 & 751\pm20 & 12.1\pm0.95 & 8.17\pm0.46 & 357\pm5 & 6.47\pm0.28 & 3.05\pm0.13 & 0.81\pm0.03 & 5.27\pm0.35 \\ 
\hline
\end{array}
\label{tab:propgroups}
$$
\tablefoot{The total masses for WBL154 and A3390 (marked with a star) which show a double peak have been calculated by summing up the total mass of each of their two components. The $R_{500}$ and $R_{2500}$ have been then estimated from their total $M_{500}$ and $M_{2500}$, respectively. $R_{spec}$ is the radius within which the global temperature of the object has been determined. $L_{X,xmm}$ is the luminosity in the 0.1-2.4 keV band.}
\end{table*}

\subsection{Global temperature}
When galaxy groups and clusters are in hydrostatic equilibrium, the gas
temperature is a direct measure of the potential depth of the
system. For some groups it generally declines in the central regions,
so that this central drop can bias the estimation of the virial
temperature. To reduce this bias, the cooler central region
has to be removed. Since we have evidence that the central drop does
not scale uniformly with $R_{\rm vir}$ (\citealt{2010A&A...513A..37H}),
the sizes of the central regions were estimated by investigating of
the temperature profiles centered on the EP, following the method
presented in \citet{2011A&A...535A.105E}. In practice, we fitted the
profiles with a two-component function, consisting of a power-law for
the central region and a constant beyond the cutoff radius that was
varied iteratively, starting from the center, until the best-fit
statistics was obtained. The global temperatures were
then determined fitting the whole observed area
beyond the cutoff radius to a single spectrum. When the temperature profiles consisted of
five or fewer bins, the whole region was used to determine the
global temperature. The overall determined temperatures are summarized in Table \ref{tab:propgroups}.

\subsection{Mass modeling}
In addition to the temperature profile, the mass determination requires the
knowledge of the gas density profile. Under the assumption of
spherical symmetry, the gas density profile for a double $\beta$-model
is described by
\begin{equation}\label{gasdens}
n_e(r)=\left\{n_{e,1}^2\left[1+\left(\frac{r}{r_{c,1}}\right)^2\right]^{-3\beta_1}+n_{e,2}^2\left[1+\left(\frac{r}{r_{c,2}}\right)^2\right]^{-3\beta_2}\right\}^{\frac{1}{2}},
\end{equation}
where $n_{e,1}$ and $n_{e,2}$ are the central densities of the two components derived by using the EWC. \\
The central electron densities are then obtained using the spectral
information. In particular, when fitting spectra with an APEC model,
one of the free parameters is the normalization K, defined as
\begin{equation}\label{normapec}
K=\frac{10^{-14}}{4\pi D_A^2 (1+z)^2}\int{n_en_HdV},
\end{equation}
where $D_A$ is the angular distance of the source, and $n_e$ and $n_H$ are
the electron and proton densities in units of cm$^{-3}$. By combining Eq. \ref{gasdens} and \ref{normapec}, and
given that in an ionized plasma {\it n$_H$}$\approx$0.82${\it n_e}$, we
can recover the gas density by integrating the equation over a volume
within a radius $R_{\rm extr}$ and matching it with the observed
emission measure in the same region. The largest radius $R_{\rm extr}$
was chosen for each group to
optimize the S/N ratio in the [0.5-7] keV band.\\
When the gas density and temperature profiles are known, the total mass
within a radius $r$ can be estimated by solving the equation of
hydrostatic equilibrium. Assuming spherical symmetry, one can write
\begin{equation}
M(<r)=-\frac{k_BTr}{G\mu m_p}\left\{ \frac{r}{\rho} \frac{d\rho}{dr} + \frac{r}{T} \frac{dT}{dr} \right\},
\end{equation}
where $k_{\rm B}$ is the Boltzmann constant, $\mu\approx$ 0.6 is the mean
particle weight in units of the proton mass, $m_{\text{p}}$, and G is the
gravitational constant. To evaluate the relative errors on the mass estimation we randomly varied 
the observational data points of the SB and k$T$ profiles 1,000 times to determine a new best fit. The randomization was
driven from Gaussian distribution with mean and standard deviation in
accordance with the observed data points and the associated uncertainties. To estimate the error on the gas mass we additionally varied the normalization $K$ 1,000 times. The masses were estimated
for two characteristic radii: $r_{\text{2500}}$ and $r_{\text{500}}$, which correspond to the radii within which the overdensity of the galaxy groups is 2,500 and 500 times the critical density of the Universe.  In Table \ref{tab:propgroups} we list the estimated total masses for all the objects in our sample.

\subsection{Cooling time}
\cite{2010A&A...513A..37H} used the cooling time to classify galaxy
clusters in the HIFLUGCS sample as strong, weak, and non-cool-core
objects.  To
compare the fraction of cool cores in the two samples (i.e., at low and
high masses) we determined the cooling time for the objects in our
sample. For a direct comparison with the results of the HIFLUGCS
sample we used the same equation as used by \cite{2010A&A...513A..37H},
\begin{equation}
t_{\rm {cool}}=\frac{3}{2}\frac{(n_{i0}+n_{e0})kT_{cc}}{n_e\Lambda(T_{cc})},
\end{equation}
where $n_{\text{i0}}$ and $n_{\text{e0}}$ are the central ion and electron
densities, while $T_{\text{cc}}$ and $\Lambda({\rm {T_{cc}}})$ are the average
temperature and the cooling function at $r=0.4\%R_{\text{500}}$. Note that
for a nearly fully ionized plasma with typical cluster abundance {\it n$_i$}$\approx$1.1{\it n$_H$}. The properties used in
this calculation were derived using the EP as the center of the
galaxy group. In Table \ref{tab:propgroups} we list the derived cooling
times for the objects in our sample.

\section{Scaling relations} \label{scallaw}
We investigated the following relations: $L_\text{X}$-$T$,  $L_\text{X}$-$M$, $M$-$T$, $f_{\text{gas}}$-$T$, $f_{\text{gas}}$-$M$, $M$-$Y_{\text{X}}$, $L_\text{X}$-$M_{\text{gas}}$, $M_{\text{gas}}$-$M$, and
$L_\text{X}$-$Y_\text{X}$. For each set of parameters (Y,X) we fitted a power-law
relation in the form of
\begin{equation}
\log(Y/C1)=a\cdot\log(X/C2)+b,
\end{equation} 
with $C1$ and $C2$ listed in Table \ref{tab:constrel} chosen to have
approximately uncorrelated results of the slope, normalization, and
their errors for the galaxy groups sample. The fits were performed
using linear regressions in $\log$-$\log$ space using the $Y|X$
statistics of the code BCES$\_$REGRESS 
(\citealt{1996ApJ...470..706A}). The choice of the $Y|X$ statistic was driven by the new method presented here for correcting the selection bias effects (see Sect. \ref{malmsubsect} for more details). For an easier comparison with  literature
results, we also list the bisector and
orthogonal best fits for some of the observed relations when the differences between the different estimators
are significant. The total logarithmic scatter on $Y$ is measured as
${\rm \sigma_{tot}^Y=\sqrt{<(\log Y - b - a\log X)^2>} }$, while along
the X-axis it is ${\rm \sigma_{tot}^X=\sigma_{tot}^Y/a }$. The intrinsic
logarithmic scatter along the Y and X axis is ${\rm
  \sigma_{intr}^Y=\sqrt{\rm (\sigma_{tot}^Y)^2 - (\sigma_{stat}^Y)^2 -
    a^2\cdot (\sigma_{stat}^X)^2} }$ and ${\rm
  \sigma_{intr}^X=\sqrt{\rm (\sigma_{tot}^X)^2 - (\sigma_{stat}^X)^2 -
    a^{-2}\cdot (\sigma_{stat}^Y)^2} }$, where the statistical errors
are ${\rm \sigma_{stat}^Y=<\log(e) \cdot \Delta Y/Y> }$ and ${\rm
  \sigma_{stat}^X=<\log(e) \cdot \Delta X/X> }$, with ${\rm \Delta X}$ and ${\rm \Delta Y}$ the symmetrical errors along the two axes. This procedure is identical to the procedure presented by \citet{2011A&A...535A.105E}, but note their typos on the intrinsic and total scatter equations. The equations they used for the fitting were the same as presented here (private communication).

\begin{table}[!tp]
\caption{Normalization values used in the scaling relations.}
$$
\centering
\begin{array}{lll}
\hline\hline
\noalign{\smallskip}
{\rm Relation} &  {\rm C1} & {\rm C2}   \\
\hline
L_{\text{X}}-T & {\rm 10^{43} \ h_{70}^{-2} \ erg \ s^{-1}} & {\rm 2 \ keV} \\ 
L_{\text{X}}-M & {\rm 10^{43} \ h_{70}^{-2} \ erg \ s^{-1}} & {\rm 5\cdot 10^{13} \ h_{70}^{-1} \ M_{\odot}} \\ 
M-T  	 & {\rm 5\cdot 10^{13}  \ h_{70}^{-1} \ M_{\odot}} & {\rm 2 \ keV} \\ 
M-Y_{\text{X}} & {\rm 5\cdot 10^{13}  \ h_{70}^{-1} \ M_{\odot}} & {\rm 5\cdot 10^{12}  \ h_{70}^{-2} \ M_{\odot} \ keV} \\ 
M_{\text{gas}}-M & {\rm 5\cdot 10^{12} \ h_{70}^{-5/2} \ M_{\odot}} &  {\rm 5\cdot 10^{13}  \ h_{70}^{-1} \ M_{\odot}}  \\ 
L_{\text{X}}-Y_{\text{X}} & {\rm 10^{43} \ h_{70}^{-2} \ erg \ s^{-1}} & {\rm 5\cdot 10^{12}  \ h_{70}^{-2} \ M_{\odot} \ keV}   \\
L_{\text{X}}-M_{\text{gas}} & {\rm 10^{43} \ h_{70}^{-2} \ erg \ s^{-1}} & {\rm 5\cdot 10^{12} \ h_{70}^{-5/2} \ M_{\odot}} \\  
f_{\text{gas}}-T & {\rm 0.1  \ h_{70}^{-3/2} } & {\rm 2 \ keV} \\ 
f_{\text{gas}}-M & {\rm 0.1  \ h_{70}^{-3/2} } & {\rm 5\cdot 10^{13} \ h_{70}^{-1} \ M_{\odot}} \\ 
\hline
\end{array}
\label{tab:constrel}
$$
\end{table}

\section{Results}  \label{sec_results}

The derived properties for individual groups are listed in Table
\ref{tab:propgroups}. Combining the 20 groups in our sample with
those from HIFLUGCS\footnote{In this first version the sample included 63
  clusters. Another cluster was included later after we recalculated 
  the flux (see Sect. C.43 of \citealt{2010A&A...513A..37H}).}  (\citealt{2002ApJ...567..716R}) creates a master sample
of 82\footnote{For the common object IIIZw54, we used the properties
  derived in this paper, which agree within the error bars with
  the properties derived by \cite{2002ApJ...567..716R}.} groups and clusters
ranging over more than three orders of magnitude in luminosity. The
luminosity values for the groups listed in Table \ref{tab:sample} are
all taken from the input catalogs. Except for the temperature values
taken from \citet{2010A&A...513A..37H}, all the
other HIFLUGCS parameters were taken from \citet{2002ApJ...567..716R}. \\
We obtained the temperature profile for 17 out of 20 groups, while for
the other 3 groups (IC1262, NGC6338, and RXCJ1840) we calculated the
overall cluster temperatures, and the masses were obtained assuming
isothermality. The data allowed us to derive the temperature profile
out to or beyond $R_{2500}$ for 12 objects.

\begin{table*}[ht]
\caption{Fit results for the scaling relations. BC indicates the relations corrected for selection bias. In the last two columns we list the slopes and normalizations derived using all the groups and HIFLUGCS objects.}
\vspace{-10pt}
$$
\centering
\setlength\extrarowheight{2pt}
\begin{array}{lcc@{\hspace{1.2em}}c@{\hspace{1.2em}}c@{\hspace{1.2em}}c@{\hspace{1.2em}}c@{\hspace{1.2em}}c@{\hspace{1.2em}}}
\hline\hline
\noalign{\smallskip}
{\rm Relation \ (Y-X)} & {\rm BCES \ estimator} & \multicolumn{2}{c}{ {\rm groups}} & \multicolumn{2}{c}{ {\rm HIFLUGCS \ (kT>3 \ keV)}} & \multicolumn{2}{c}{ {\rm all}} \\
\noalign{\smallskip}
\hline \\ [-2.0ex]
\multicolumn{2}{c}{} & {\rm a} & {\rm b} & {\rm a} & {\rm b} & {\rm a} & {\rm b} \\
\hline \\ [-1.0ex]
{\rm L_{\text{X}}-M_{\text{500}} \ BC} & {\rm Y|X} & 1.66\pm0.22 & -0.03\pm0.04 & 1.08\pm0.21 & 0.18\pm0.18 & 1.39\pm0.05 & -0.12\pm0.04 \\[0.1cm]   
{\rm L_{\text{X}}-M_{\text{500}}} & {\rm Y|X} & 1.32\pm0.24 &  0.04\pm0.05 & 1.22\pm0.21 &  0.26\pm0.20 & 1.40\pm0.06 &  0.09\pm0.05  \\[0.1cm]  
{\rm L_{\text{X}}-M_{\text{500}}} & {\rm bisector} & 1.49\pm0.20 &  0.05\pm0.06 & 1.49\pm0.17 &  0.01\pm0.16 & 1.46\pm0.06 &  0.06\pm0.04  \\[0.1cm] 			
{\rm L_{\text{X}}-M_{\text{500}}} & {\rm orthogonal} & 1.57\pm0.24 &  0.04\pm0.06 & 1.61\pm0.19 & -0.10\pm0.18 & 1.49\pm0.06 &  0.05\pm0.05  \\[0.1cm] 
\hline \\ [-1.0ex]
{\rm L_{\text{X}}-T \ BC} & {\rm Y|X} & 2.86\pm0.29 &  0.37\pm0.06 & 2.55\pm0.27 &  0.35\pm0.10 & 2.67\pm0.11 &  0.34\pm0.04 \\[0.1cm]  
{\rm L_{\text{X}}-T }     & {\rm Y|X} & 2.05\pm0.32 &  0.27\pm0.07 & 1.91\pm0.27 &  0.58\pm0.11 & 2.36\pm0.10 &  0.38\pm0.03 \\[0.1cm] 
{\rm L_{\text{X}}-T }     & {\rm bisector} & 2.41\pm0.30 &  0.32\pm0.06 & 2.34\pm0.16 &  0.40\pm0.10 & 2.49\pm0.23 &  0.40\pm0.09 \\[0.1cm]   
{\rm L_{\text{X}}-T }     & {\rm orthogonal} & 2.76\pm0.43 &  0.36\pm0.08 & 2.78\pm0.27 &  0.22\pm0.11 & 2.60\pm0.10 &  0.33\pm0.03 \\[0.1cm]   
\hline \\ [-1.0ex]
{\rm M_{\text{500}}-T}    & {\rm Y|X} & 1.65\pm0.07 & 0.19\pm0.02 & 1.62\pm0.08 & 0.24\pm0.04 & 1.71\pm0.04 & 0.20\pm0.02 \\[0.1cm]
\hline \\ [-1.0ex]
{\rm M_{\text{gas,500}}-M_{\text{500}}} & {\rm Y|X} & 1.09\pm0.08 & -0.14\pm0.02 & 1.27\pm0.14 & -0.20\pm0.13 & 1.22\pm0.04 & -0.16\pm0.03 \\[0.1cm]   
{\rm M_{\text{gas,2500}}-M_{\text{2500}}} & {\rm Y|X} & 1.19\pm0.07 & -0.27\pm0.04 & - & - & - & - \\[0.1cm]
\hline \\ [-1.0ex]
{\rm M_{\text{500}}-Y_{\text{X}}} & {\rm Y|X} & 0.60\pm0.03 & -0.03\pm0.02 & 0.59\pm0.03 & -0.08\pm0.05 & 0.57\pm0.01 & -0.03\pm0.02 \\[0.1cm] 
\hline \\ [-1.0ex]
{\rm L_{\text{X}}-Y_{\text{X}}}   & {\rm Y|X} & 0.72\pm0.14 & -0.01\pm0.05 & 0.75\pm0.06 &  0.12\pm0.06 & 0.79\pm0.03 &  0.06\pm0.04  \\[0.1cm]
\hline \\ [-1.0ex]
{\rm f_{\text{gas,500}}-T} & {\rm Y|X} & 0.08\pm0.12 & -0.13\pm0.02 & 0.15\pm0.13 & -0.02\pm0.05 & 0.32\pm0.06 & -0.10\pm0.02  \\[0.1cm]   
{\rm f_{\text{gas,2500}}-T} & {\rm Y|X} & 0.21\pm0.11 & -0.32\pm0.02 & - & - & - & - \\[0.1cm]
\hline \\ [-1.0ex]
{\rm f_{\text{gas,500}}-M} & {\rm Y|X} & 0.01\pm0.07 & -0.13\pm0.02 & 0.03\pm0.08 &  0.07\pm0.08 & 0.16\pm0.04 & -0.122\pm0.03  \\[0.1cm]   
\hline \\ [-1.0ex]
{\rm L_{\text{X}}-M_{\text{gas,500}}} & {\rm Y|X} & 1.02\pm0.24 & 0.16\pm0.07 & 1.18\pm0.07 & 0.24\pm0.07 & 1.18\pm0.04 & 0.26\pm0.03  \\[0.1cm]   
\hline
\end{array}
\label{tab:scalawmalm}
$$
\end{table*}

\subsection{Cool cores}
In the HIFLUGCS sample 44$\%$ of the objects were classified as strong
cool cores (SCC), 28$\%$ as weak cool cores (WCC), and 28$\%$ as non-cool cores (NCC) (\citealt{2010A&A...513A..37H}). In
our sample of galaxy groups the fraction of SCC is
higher (55$\%$), while the fraction of NCC is only
$15\%$, half of the fraction found in galaxy clusters. This agrees with the standard
scenario of the structure formation, where galaxy groups tend to be older
than their massive counterparts and therefore are more
relaxed in general. \citet{2009A&A...501..835M} showed that 75$\%$ of the galaxy
clusters host a central radio source (CRS), with a higher probability
to host an AGN for the CC objects and the lowest probability for the NCC. A
similar trend is also observed for our galaxy group sample, but
with a slightly higher fraction (85$\%$) of objects hosting a CRS.

%-----------------------------Figure Start------------------------------
\begin{figure*}[!ht]
\begin{center}
\epsfig{figure=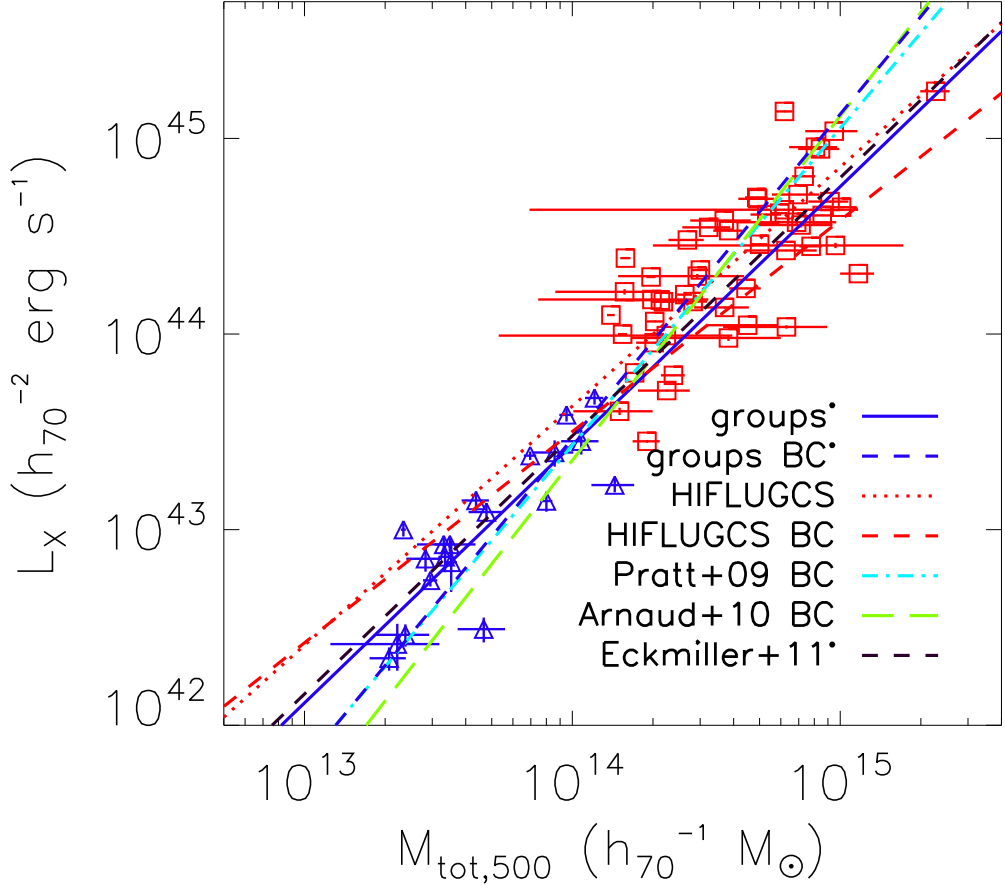,width=0.85\textwidth,angle=0}
\end{center}
\caption{
$L_{\text{X}}$-$M$ relation. Blue triangles are groups and red boxes are HIFLUGCS clusters with a temperature higher than 3 keV. The blue lines represent the best-fit values obtained in this work. They are compared with the best-fit results obtained with different samples. BC indicates the relation corrected for the selection bias effects.  The stars indicate the works that studied galaxy groups.
}
\label{fig:LM}
\end{figure*}
%-----------------------------Figure End--------------------------------

\subsection{Selection bias effect}\label{malmsubsect}
\subsubsection{$L_{\text{X}}$-$M$ relation}
One of the most important X-ray scaling laws for cosmology with galaxy groups and clusters is the
$L_{\text{X}}$-$M$ relation, because it can be used to directly
convert the
easiest to derive observable to the total mass. \\
Complete samples are required to constrain the cosmological
mass function because the cluster number density must be
calculated. If on one hand a flux-limited sample matches this
requirement, on the other hand it suffers from the well-known Malmquist bias: brighter objects can be seen out to farther
distances. From the statistical point of view, this implies that the
intrinsically brighter sources will be detected more often than they
ought, which distorts the sample composition. This effect has
previously been taken into account for scaling relations, for example by \citet{2002A&A...383..773I}, \citet{2006ApJ...648..956S},
\citet{2007MNRAS.382.1289P}, \citet{2009ApJ...692.1033V},
\citet{2009A&A...498..361P}, and \citet{2011A&A...532A.133M}. It is important to note that proper corrections cannot be calculated for incomplete samples. \\
To estimate the effect of applying the sample selection ($5\times10^{-12} \ \text{erg/s/cm$^{2}$} \le f_{\text{lim}} (0.1-2.4 \ \text{keV}) \le 2\times10^{-11} \ \text{erg/s/cm$^{2}$}$ and 0.01$\le${\it z}$\le$0.035) we applied the same flux and redshift thresholds to a set of simulated
samples. By using the halo mass function derived by
\cite{2008ApJ...688..709T} with the transfer-function from
\cite{1998ApJ...496..605E}, the density fluctuation amplitude at 8
Mpc/$h$ $\sigma_8$=0.811 and a spectral index of the primordial power
spectrum $n_s$=0.967 (\citealt{2011ApJS..192...18K}), we obtained the mass and
redshift for all the simulated objects. We applied a lower mass
threshold of {\it M}$>$5$\times10^{12}${\it M$_{\sun}$} to ensure that we
selected groups and not galaxies, and an upper threshold {\it
  M}$<$5$\times10^{15}${\it M$_{\sun}$} (above this mass there are
only a few clusters that are not important for this
work). 
We then assigned a luminosity through the $L_{\text{X}}$-$M$ relation to every object and also introduced the total scatter we
derived in Sect. \ref{secscatter}. We note, that the scatter should only be introduced in the $L_{\text{X}}$ direction because otherwise the value of the total masses that we derived directly from the mass function would be changed as well. Since for all the BCES estimator except $Y|X$ the minimization is not purely performed in the Y (i.e., $L_{\text{X}}$) direction, they were not used for the selection bias correction.\\
We assigned the same error (i.e., the mean relative error derived in our
analysis) to every simulated object because we did not see any particular trend in the distribution of the statistical measurements errors as function of mass or luminosity. The slope and normalization of the
input $L_{\text{X}}$-$M$ relation were varied in the range [1.20:2.20]\footnote{
  We first ran a set of low-resolution simulations to identify the
  interval of values with the lowest $\chi^2$ of Eq. \ref{eq:malm}. These intervals of values refer to the group sample only.}
(with steps of 0.01) and [-0.15:0.05] (with steps of 0.01),
respectively. For each grid point (i.e., every combination of slopes and
normalizations) 300 artificial flux-limited group samples were simulated. The input slope ($a_{{\rm sim}}$) and
normalization ($b_{{\rm sim}}$) that after applying the flux and redshifts cuts (to
reproduce the same selection effects of our sample) yields an
$L_{\text{X}}$-$M$ relation that matched the observed relation are the
values corrected for the selection bias.
We searched for the best combination of values by minimizing the following equation:\\
\begin{equation}\label{eq:malm}
\chi^2_{{\rm tot}}= \frac{(\tilde{b}_{{\rm sim}}-b_{{\rm obs}})^2}{\Delta b_{{\rm obs}}^2}+\frac{(\tilde{a}_{{\rm sim}}-a_{{\rm obs}})^2}{\Delta a_{{\rm obs}}^2},
\end{equation}
where $\tilde{b}_{sim}$ and $\tilde{a}_{sim}$ are the median values for the normalization and slope of the 300 output relations of each grid point.
The total number of objects obtained by using the halo mass function was scaled such that the distribution of the simulated samples peaked at about 20 objects as the real sample. The scatter of the best-fit output relation after the flux and redshift cuts agrees with the observed scatter. We also verified that the luminosity and mass distribution of the simulated objects after the flux and redshift cuts matched the
observed one. The correction was then also applied to the HIFLUGCS (kT$>$3 keV) and full sample (i.e., groups plus all the HIFLUGCS objects). \\
The $L_{\text{X}}$-$M$ relations corrected for selection bias  are shown in
Fig. \ref{fig:LM} and are compared with the observed relations.  The corrected
relation for galaxy groups is steeper (slope of 1.66$\pm$0.22) than the observed relation
($a$=1.32$\pm$0.24). In contrast the corrected relation for massive systems was found to be slightly shallower than what is observed. Interestingly, the slope of the corrected relation remains unchanged when including all the objects in the sample (groups and HIFLUGCS). The errors of the corrected slopes were obtained from the distribution of the $\chi^2_{{\rm tot}}$ in the grid. For each parameter (i.e., slope and normalization) the error was derived by keeping the other interesting parameter frozen and by searching for the range of values with a $\chi^2_{tot}<\chi^2_{min}+1$.

%-----------------------------Figure Start------------------------------
\begin{figure*}[!t]
\begin{center}
\hbox{
\epsfig{figure=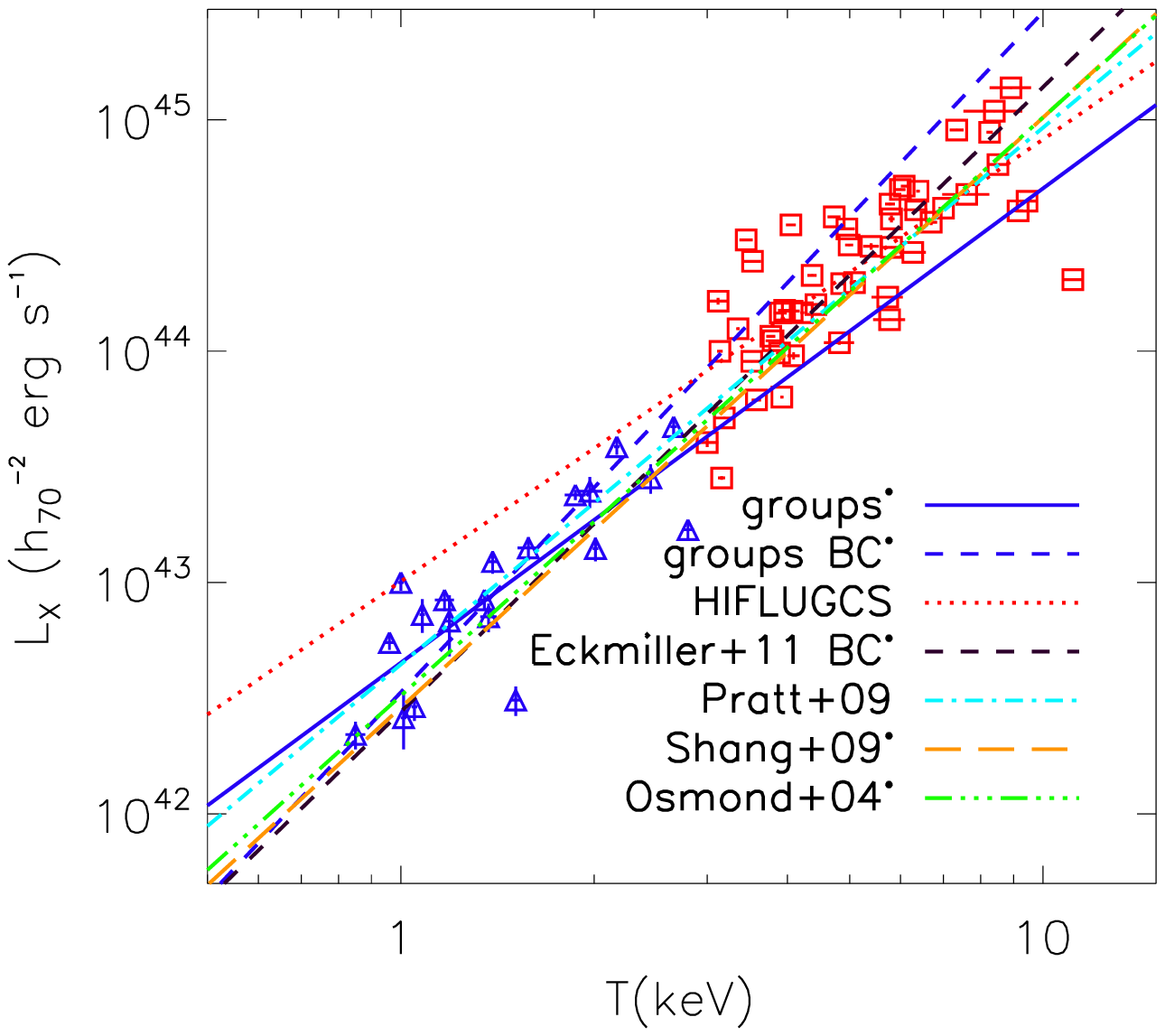,height=10cm,width=0.5\textwidth,angle=0}
\epsfig{figure=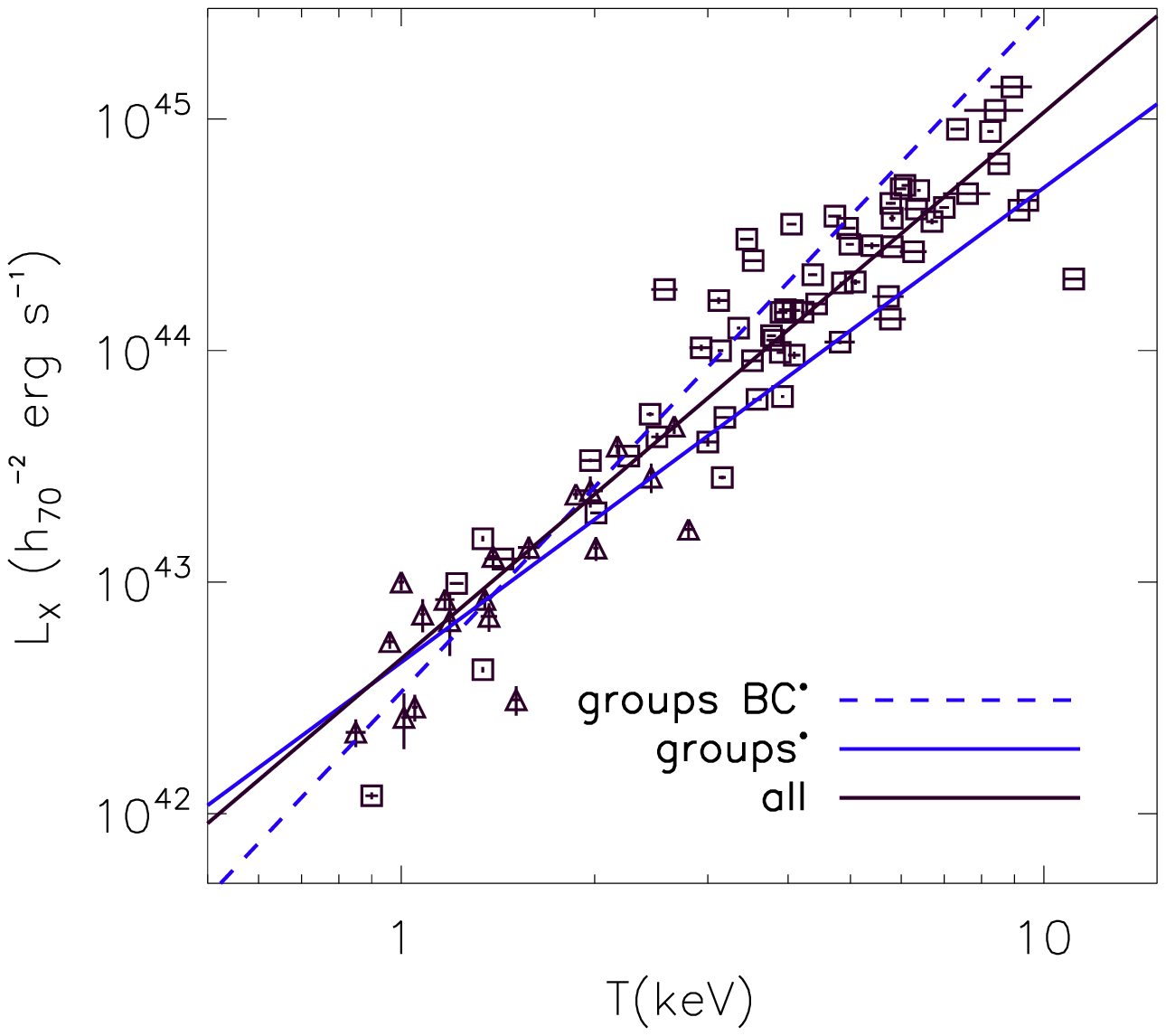,height=10cm,width=0.5\textwidth,angle=0}
}
\end{center}
\caption{{\it Left:} $L_{\text{X}}-T$ relation. Blue triangles are groups, red boxes are HIFLUGCS clusters with a temperature higher than 3 keV. The stars indicate the works that studied galaxy groups. {\it Right:} Same points as in the left panel plus the HIFLUGCS clusters with a temperature lower than 3 keV. BC indicates the relation corrected for the selection bias effects. }
\label{fig:LT}
\end{figure*}
%-----------------------------Figure End--------------------------------

%-----------------------------Figure Start------------------------------
\begin{figure*}[!t]
\begin{center}
\hbox{
% un-comment the following line to include your fig1a.ps postscript file
\epsfig{figure=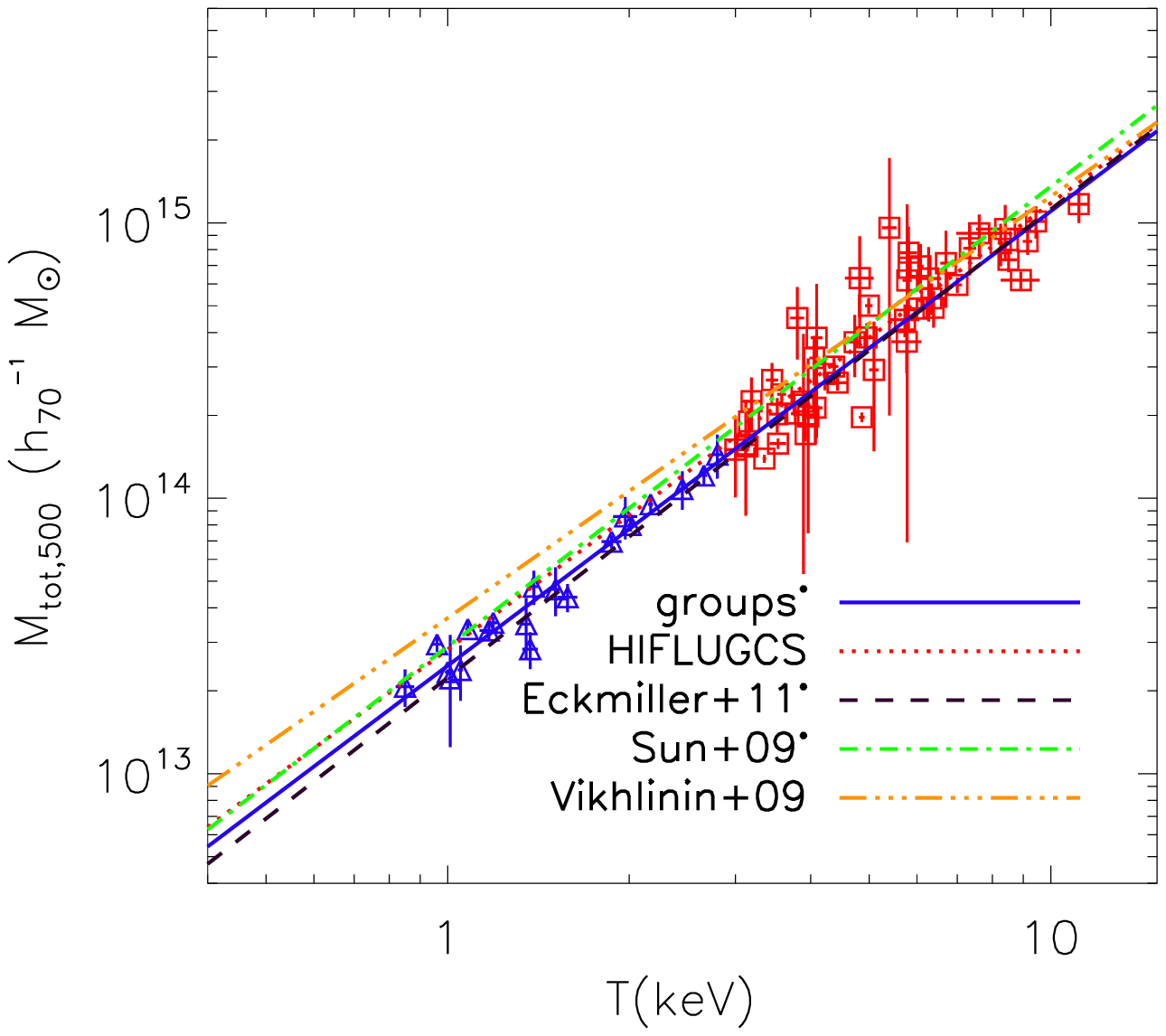,height=10cm,width=0.5\textwidth,angle=0}
% un-comment the following line to include your fig1b.ps postscript file
\epsfig{figure=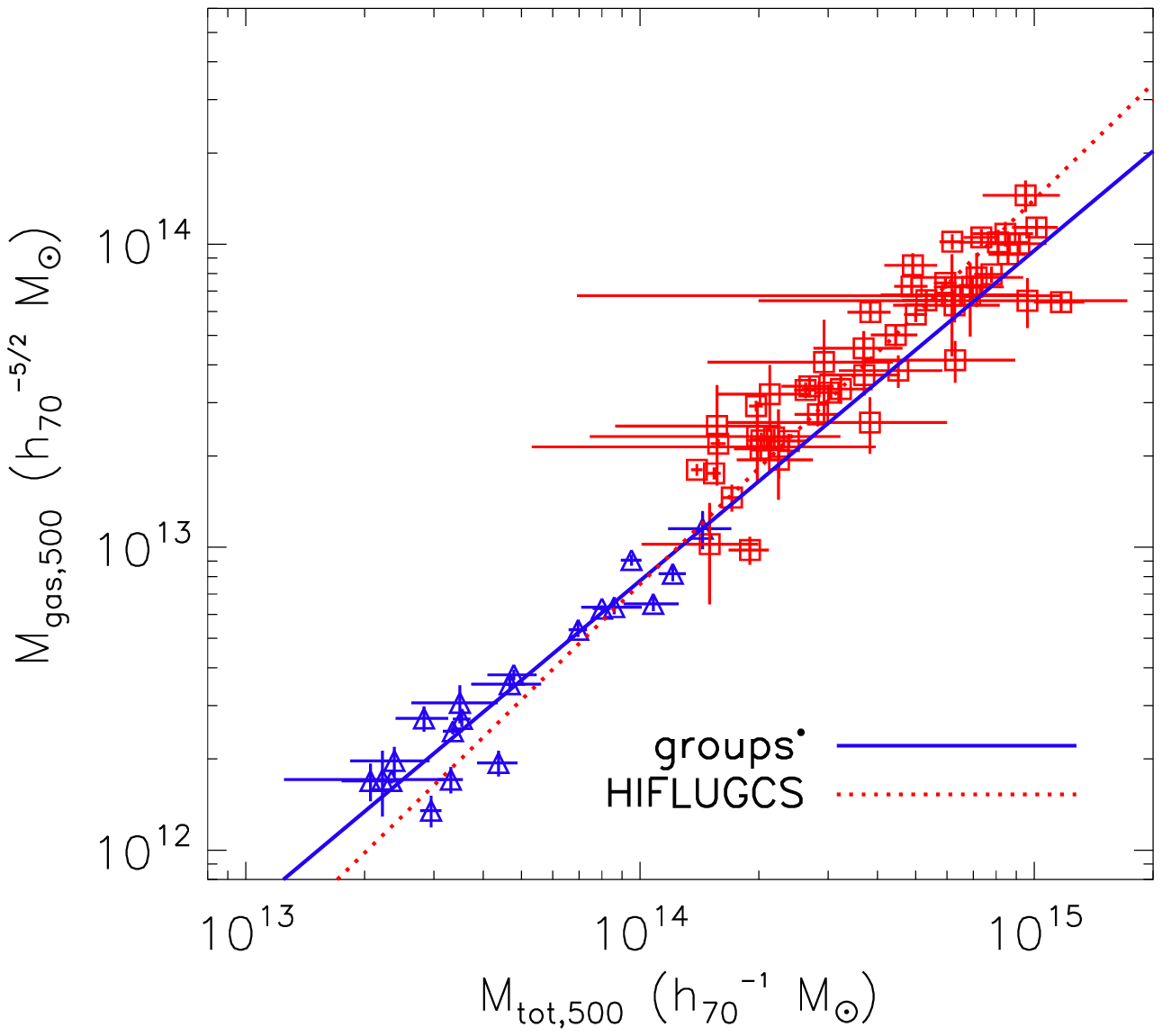,height=10cm,width=0.5\textwidth,angle=0}
}
\end{center}
\caption{
{\it Left:} $M$-$T$ relation. Blue triangles are groups, red boxes are HIFLUGCS clusters with a temperature higher than 3 keV.  
{\it Right:} same as in the left panel, but for the $M_{\text{gas}}$-$M$ relation. 
}
\label{fig:MgasMT}
\end{figure*}
%-----------------------------Figure End--------------------------------

%-----------------------------Figure Start------------------------------
\begin{figure*}[!t]
\begin{center}
\hbox{
% un-comment the following line to include your fig1a.ps postscript file
\epsfig{figure=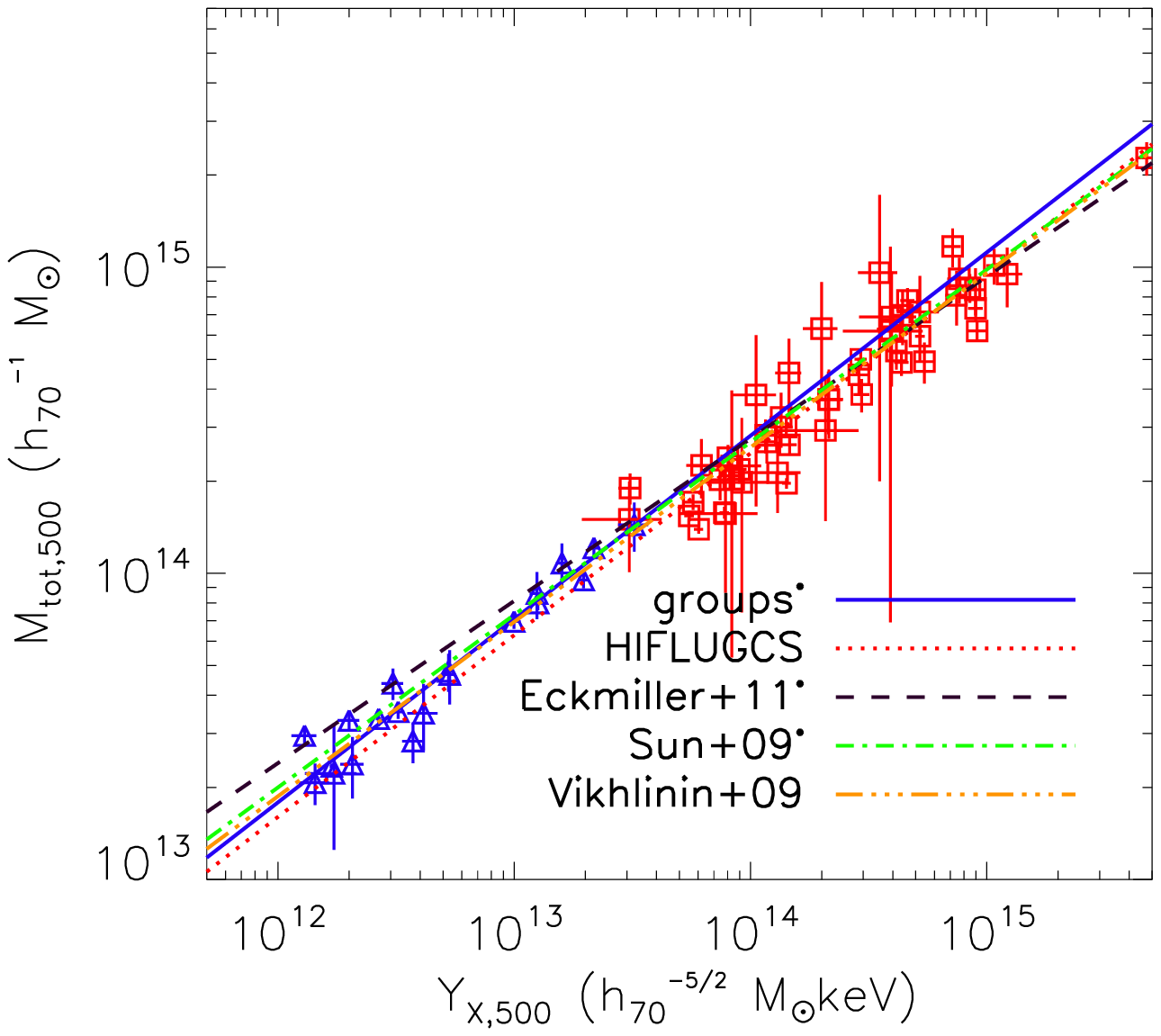,height=10cm,width=0.5\textwidth,angle=0}
\epsfig{figure=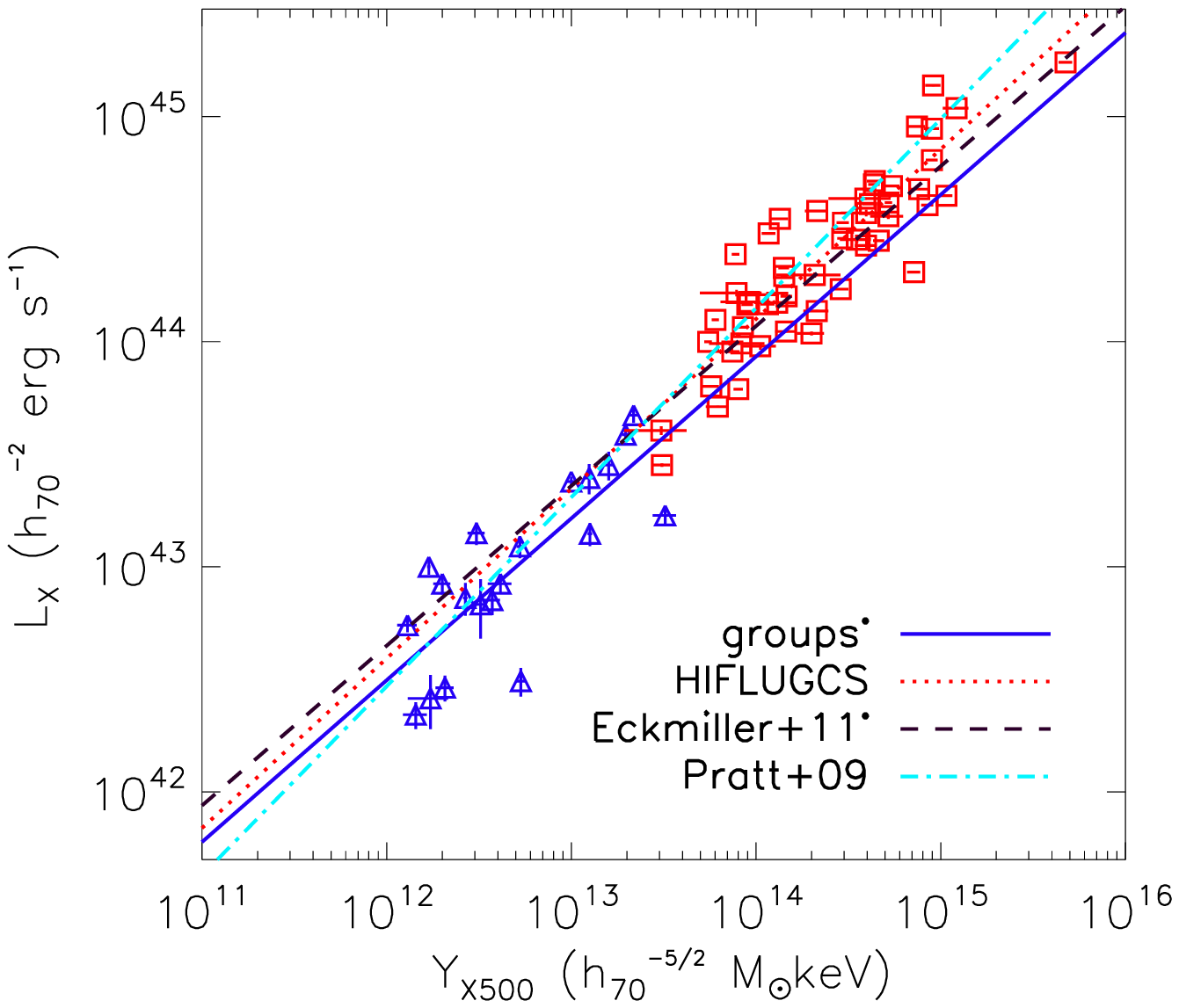,height=10cm,width=0.5\textwidth,angle=0}
% un-comment the following line to include your fig1b.ps postscript file
}
\end{center}
\caption{{\it Left:} $M$-$Y_{\text{X}}$ relation. Blue triangles are groups, red boxes are HIFLUGCS clusters.   
 {\it Right:} same as in the left panel, but for the $L$-$Y_{\text{X}}$.}
\label{fig:LMY}
\end{figure*}
%-----------------------------Figure End--------------------------------

\subsubsection{$L_{\text{X}}$-$T$ relation}
When the ``true'' $L_{\text{X}}$-$M$ relation is recovered, the result can be
used to derive the corrected $L_{\text{X}}$-$T$ relation. Following
the procedure presented in the previous section, we assigned a luminosity to all the
objects using the $L_{\text{X}}$-$M$ relation corrected for selection biases 
 by also introducing the total scatter
along the Y($L_{\text{X}}$) direction. We then assigned the temperatures through an
input $L_{\text{X}}$-$T$ relation by creating a grid with a slope and normalization in
the range [2.5:3.5] and [0.10:0.55]. After applying the flux and
redshift cuts in the same way as for our group sample, we compared the simulated
and observed $M$-$T$ relations under the assumption that this relation
is unbiased. The best-fit values are the ones that minimize the
$\chi^2_{tot}$ of Eq. \ref{eq:malm}.\\
In Fig. \ref{fig:LT} ({\it left panel}) we compare the observed $L_{\text{X}}$-$T$ relation for galaxy groups with the one determined using the HIFLUGCS sample. Again, as for the $L_{\text{X}}$-$M$ relation, we do not see any steepening at the group scale. In Fig. \ref{fig:LT} ({\it right panel}) we compare the corrected
luminosity-temperature relation derived for the group sample with the
observed relation. While the observed slope of our group sample, given the large errors (see Table \ref{tab:scalawmalm}), is consistent with the slope derived with massive systems, the
corrected slope shows a steepening.\\
By fitting all the objects of our group sample and the HIFLUGCS objects, the relation becomes steeper (see {\it right panel} of Fig. \ref{fig:LT}), probably because of the different normalizations of the two samples. This effect is not significant given the uncertainties.

\subsection{$M$-$T$, $M_{\text{gas}}$-$M$, and  $L_\text{X}$-$M_{\text{gas}}$ relations}
The $M$-$T$ relation is expected to follow the same behavior for
galaxy groups and galaxy clusters because it is less affected by heating and
cooling processes, which are thought to be responsible for the steeper relations observed in other analysis at the group scale (e.g., \citealt{2011A&A...535A.105E}). 
Indeed, the slopes we found for the group and cluster
samples are very similar (see Table
\ref{tab:scalawmalm}). Even when fitting the HIFLUGCS together with the group sample we
do not see any steepening. Both groups and clusters show a slope slightly steeper than the one predicted by the self-similar scenario. \citealt{2013ApJ...778...74K} suggested that X-ray masses are biased down due to the assumption of hydrostatic equilibrium with a larger bias for low mass systems that cause the steepening.  However, a stronger bias for groups appears to be in tension with the finding of \citealt{2014A&A...564A.129I} who find the opposite trend. \\
\cite{2005A&A...441..893A} analyzed a sample of massive clusters and
showed that the slope of the $M$-$T$ relation is stable at all the
overdensities. We verified whether or not this is also true at the
group scale by fitting the relation at $R_{2500}$ and
$R_{1000}$ as well. We found that the slope is quite
stable: 1.61$\pm$0.07 at $R_{2500}$, 1.71$\pm$0.13 at $R_{1000}$,
and 1.65$\pm$0.11 at $R_{500}$.\\
In Fig. \ref{fig:MgasMT} ({\it right panel}) we show the fit to the
$M_{\text{gas}}$-$M$ relation. Galaxy groups have a shallower slope than clusters, 
but the slopes agree well within the errors. 
The slope of the galaxy group sample also agrees well with the slope from clusters moving
from $R_{2500}$ to $R_{500}$.
If the gas fraction of galaxy clusters
is universal. we would expect that the gas mass is linearly
related to the total mass. A slope greater than one of this relation
implies a trend to lower gas fraction for objects with lower
temperature.\\
In Table \ref{tab:scalawmalm} we also summarize the best-fit results for the $L_\text{X}$-$M_{\text{gas}}$ relation. Although the relation is slightly shallower at the group scale, the result still agrees within the error bars with the value obtained for the more massive systems. 

\subsection{$M$-$Y_{\text{X}}$ and $L_{\text{X}}$-$Y_{\text{X}}$  relations}\label{MYrel}
The $Y_{\text{X}}$ parameter defined by \citet{2006ApJ...650..128K} is
considered one of the less scattered mass proxies, although this is
still a matter of debate (see \citealt{2010ApJ...715.1508S}). In
Fig. \ref{fig:LMY} ({\it left panel}) we show the
$M$-$Y_{\text{X}}$ relation obtained for the groups and the
HIFLUGCS samples. Our best fit for the slope (0.60$\pm$0.03) is aligned
well with the slope of the massive systems (0.59$\pm$0.03). This means that for this relation we do not observe any hint of steepening at low masses either.  
The slopes are also very close to the value predicted by the self-similar scenario. Even when fitting galaxy groups and HIFLUGCS together, the best-fit is close to the self-similar prediction.\\
An indirect way of using the $Y_{\text{X}}$ parameter to constrain the mass is to use the
$L_{\text{X}}$-$Y_{\text{X}}$ relation to reduce the scatter in the
$L_{\text{X}}$-$M$ relation (\citealt{2007ApJ...668..772M}). The
result is shown in Fig. \ref{fig:LMY} ({\it right panel}). We do not observe a steepening at low masses for this relation either (see Table \ref{tab:scalawmalm}).

%-----------------------------Figure Start------------------------------
\begin{figure*}[!t]
\begin{center}
\hbox{
% un-comment the following line to include your fig1a.ps postscript file
\epsfig{figure=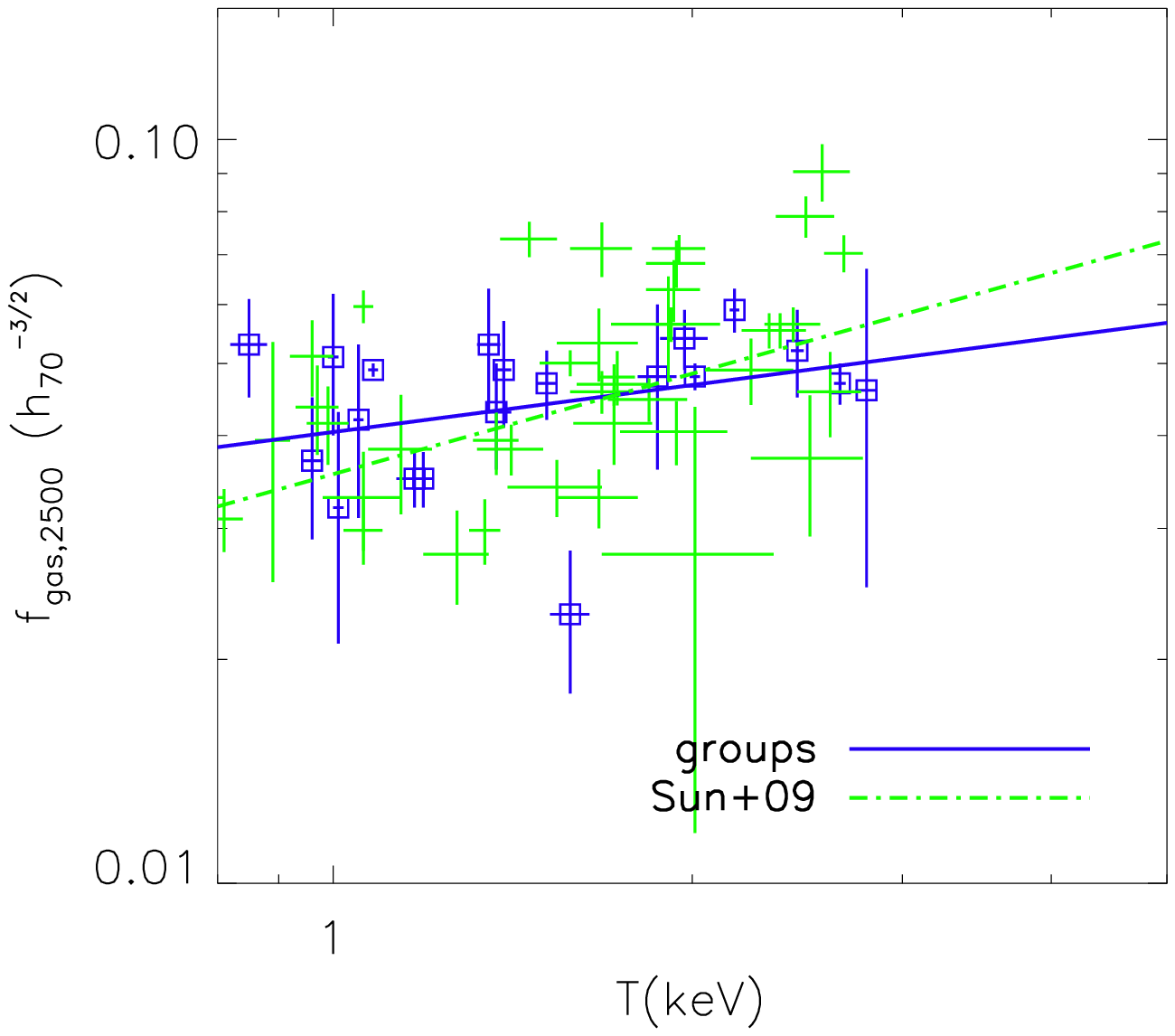,width=0.5\textwidth,angle=0}
% un-comment the following line to include your fig1b.ps postscript file
\epsfig{figure=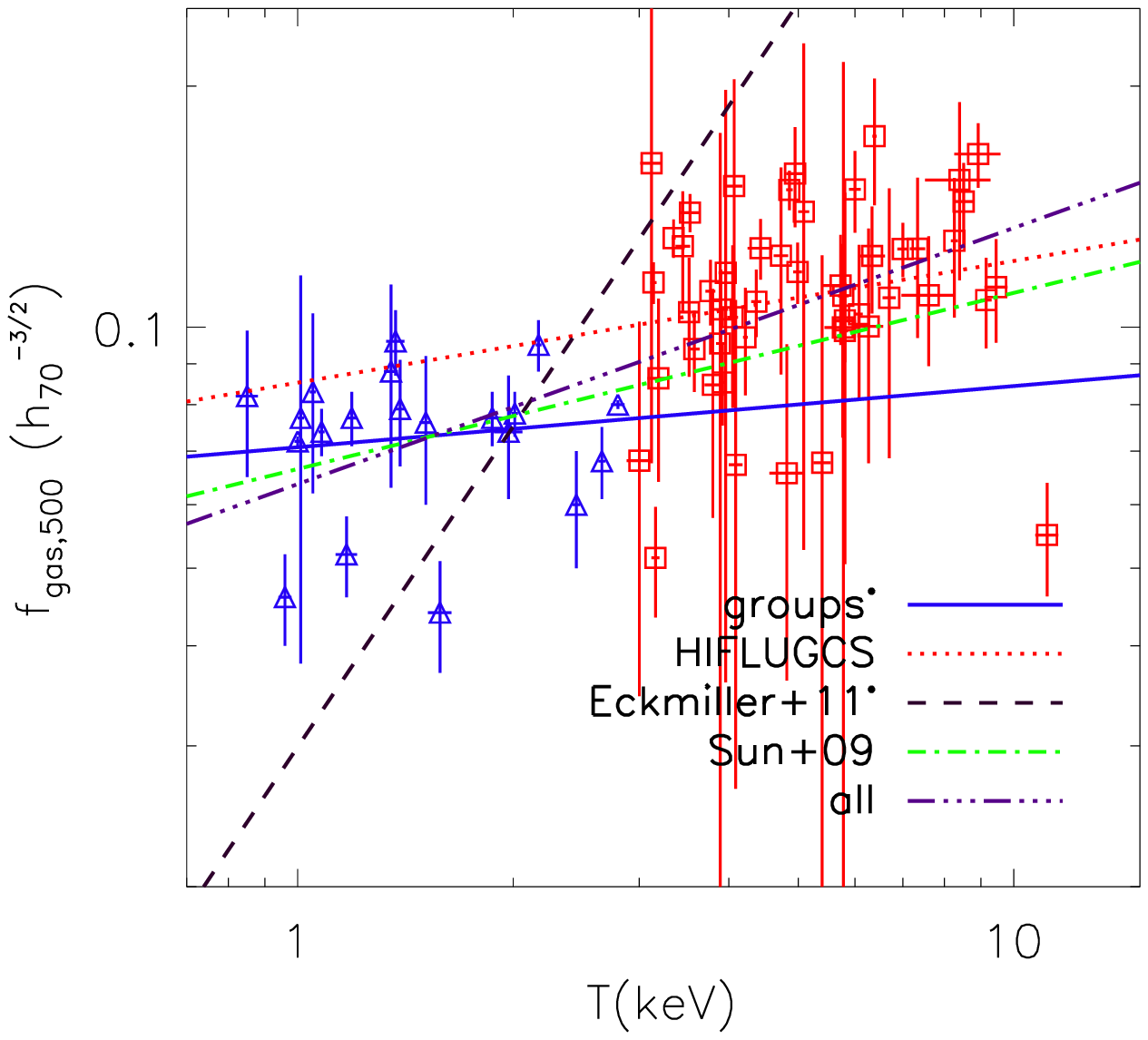,width=0.5\textwidth,angle=0}
}
\end{center}
\caption{{\it Left:} $f_{\text{gas}}$-$T$ relation at $R_{2500}$ compared with the result by \cite{2009ApJ...693.1142S}. {\it Right:} $f_{\text{gas}}$-$T$ relation at $R_{500}$.} 
\label{fig:fgas}
\end{figure*}
%-----------------------------Figure End--------------------------------

\subsection{Gas fraction}
In Fig. \ref{fig:fgas} we show the $f_{\text{gas}}$-$T$ relation at $R_{2500}$ ({\it left panel})
and $R_{500}$ ({\it right panel}). The best-fit relation is in good agreement with that
obtained by \citet{2009ApJ...693.1142S} analyzing 43 galaxy
groups. The weighted mean gas fraction within $R_{2500}$ obtained for our
sample is 0.049$\pm$0.001, which is slightly higher than the weighted mean of 0.047$\pm$0.001 found by \citet{2009ApJ...693.1142S}. This can also be seen from the marginally higher
normalization at 1 keV found in this work compared with the result by
\citet{2009ApJ...693.1142S} in Fig. \ref{fig:fgas} ({\it left
  panel}). However, because the gas fraction is temperature dependent, the weighted mean gas fraction depends on the temperature distribution of the objects in the samples.  \\ The $f_{\text{gas}}$ at $R_{2500}$ that we determined is on
average 37$\%$ lower than the $f_{\text{gas}}$ at $R_{500}$. Both groups and high mass systems have a slope that agrees within the errors, although with a lower normalization for the low-mass objects. This indicates a higher gas fraction for the most massive objects. This trend of higher gas fraction for increasing masses can also be seen in the steepening of the relation when all the objects were fitted together (see Fig. \ref{fig:fgas}, {\it right panel}).

\subsection{Scatter} \label{secscatter} 
Due to their shallower
potential well, galaxy groups are expected to be much more affected by non-gravitational processes than galaxy clusters. Therefore it is a
common thought that galaxy groups show a larger intrinsic scatter in the X-ray scaling relations,
although to our knowledge only \cite{2011A&A...535A.105E} extensively
quantified this for galaxy groups for several scaling relations and directly compared this with the scatter obtained with HIFLUGCS sample. Indeed, they found that
galaxy groups in general show a larger dispersion from the best fit of
the scaling relations. \\
In Table \ref{tab:scatter} we summarize the results for the analysis
of the scatter obtained in this work. In general, the sample of galaxy groups seems to be less scattered than the sample of galaxy clusters, although the values are similar  and might be consistent within the errors. \\
The statistical scatter in the $M$-$T$, $M_{\text{gas}}$-$M$, and $M$-$Y_{\text{X}}$ relations is lower for low-mass
systems, which might be due to the better determination of the low temperatures because of the larger effective area of the current
instruments at low energies and the stronger line emission. This then translates into  lower mass
uncertainty  and a lower total scatter.  Anyway, the statistical scatter for these relations dominates 
and the intrinsic scatter is consistent with zero.

\section{Discussion} \label{sec_discussion} 
The galaxy group sample we
studied together with the HIFLUGCS data have allowed us
to investigate the effect of the selection bias and to study
systematic differences between the scaling properties of low- and high-mass systems. We did not perform any
morphological study because of the poor statistics in our sample, in
particular for the unrelaxed objects. By using the central
cooling time to classify the objects, we found that only three groups in our
sample can be considered disturbed.  
In the next sections we discuss the results of our analysis in more detail.\\

\begin{table}[!t]
\caption{Scatter results using the values from the $Y|X$ fits. The scatter for the $M$-$T$, $M_{\text{gas}}$-$M$ and $M$-$Y_{\text{X}}$ relations is not listed because the intrinsic scatter is consistent with zero. }
$$
\centering
\setlength\extrarowheight{2pt}
\begin{array}{cccccc}
\hline\hline
\noalign{\smallskip}
{\rm Relation} & {\rm Sample} & {\rm \sigma_{tot}(X)} & {\rm \sigma_{int}(X)} & {\rm \sigma_{tot}(Y)}  & {\rm \sigma_{int}(Y)} \\ 
\noalign{\smallskip}
\hline
								 & {\rm groups} 					& 0.157 & 0.139 & 0.207 & 0.184 \\
{\rm L_{\text{X}}-M} 			 & {\rm HIFLUGCS \ (kT>3 \ keV)}	& 0.210 & 0.185 & 0.265 & 0.234 \\
	 		 	  				 & {\rm all} 						& 0.197 & 0.175 & 0.275 & 0.245 \\
\hline
					 			 & {\rm groups}  					& 0.103 & 0.099 & 0.211 & 0.204 \\
{\rm L_{\text{X}}-T}			 & {\rm HIFLUGCS \ (kT>3 \ keV)}	& 0.119 & 0.118 & 0.228 & 0.227 \\
	 		 					 & {\rm all}						& 0.104 & 0.103 & 0.245 & 0.243 \\ 
\hline
					 			 & {\rm groups}  					& 0.317 & 0.305 & 0.228 & 0.219 \\
{\rm L_{\text{X}}-Y_{\text{X}}}  & {\rm HIFLUGCS \ (kT>3 \ keV)}    & 0.231 & 0.224 & 0.174 & 0.168 \\
	 		 	  				 & {\rm all}					 	& 0.284 & 0.278 & 0.225 & 0.220 \\
\hline 
\end{array}
\label{tab:scatter}
$$
\end{table}

\subsection{Selection bias}
Any survey of galaxy systems provides catalogs that are somewhat
biased because of the chosen survey strategy or simply because of the
finite sensitivity of the instruments (see
\citealt{2013SSRv..177..247G}).  A complete sample is required to be
able to calculate and correct for these biases. Thanks to our simple
selection, we were able to construct simulated samples, which were required
to apply the corrections. The $L_{\text{X}}$-$M$ relation for the
groups we analyzed, corrected for selection effects, is adequately described by a power
law with a slope and normalization equal to 1.66$\pm$0.22 and
-0.03$\pm$0.04. The corrected slope is steeper than the
observed slope (i.e.,  1.32$\pm$0.24) and has a lower normalization. This change in slope is larger than any other systematic effect we studied (e.g., cluster sample, fitting method). This result highlights the importance of correcting the observed scaling relations for selection effects. In fact, X-ray surveys such as eROSITA require proper and precise scaling relations to determine the proper cluster number density and so constrain the correct cosmological mass function.\\ 
As a result of  the
relatively small sample we analyzed, the uncertainty on the slope is unfortunately still quite
large and a larger complete sample is required to place a stronger
constraint on the necessary correction. Unlike the galaxy groups, the more massive systems (i.e., HIFLUGCS clusters with a $kT>3$ keV) return a shallower corrected slope. One possible explanation is that the true $L_{\text{X}}$-$M$ relation is gradually steepening when moving toward the low-mass objects. In this case, higher temperature cuts would result in shallower relations.  
To verify this, we tested  by applying different cuts whether the true relation that we obtained after the bias correction is temperature dependent. Indeed, the corrected $L_{\text{X}}$-$M$ relation steepens when lowering the temperature cut. For comparison with the value quoted in Table \ref{tab:scalawmalm}, the corrected slope is 1.25$\pm$0.22 and 1.13$\pm$0.21 when applying a cut to the HIFLUGCS sample at 1 and 2 keV. 
This result would suggest a break in the $L_{\text{X}}$-$M$ relation after correcting for the selection bias effects. This would be very important for future X-ray surveys because it would imply that a simple power law cannot be used to convert the measured luminosities (or temperatures) into masses. For a quick reference we provide here the corrected  $L_{\text{X}}$-$M$ relation:
\begin{equation}
L_{\text{X}} =
\bigg \{
\begin{array}{rl}
1.70\cdot10^{20}M^{1.66}; & M \lesssim  10^{14} M_{\odot} \\
2.43\cdot10^{28}M^{1.08}; & M \gtrsim 	10^{14} M_{\odot}. \\
\end{array}
\end{equation}
Moreover, we note that our corrected slope for the massive systems is much shallower than the slope obtained for example, by \cite{2009A&A...498..361P} and \cite{2010A&A...517A..92A}. \\
The uncorrected observed $L_{\text{X}}$-$T$ relation behaves quite similarly to
that of the uncorrected $L_{\text{X}}$-$M$. The observed slope
of our group sample is consistent within the errors with that of the
clusters and in general with the results from other papers
investigating galaxy groups (e.g., \citealt{2004MNRAS.350.1511O};
\citealt{2009ApJ...690..879S}; \citealt{2011A&A...535A.105E}). Because the emissivity at low temperatures scales with $T^{-0.6}$ (\citealt{1977ApJ...215..213M}), the relation predicted by the self-similar scenario for galaxy groups is $L_{\text{X}}\propto T^{1.1}$, which is flatter than the observed relation. Thanks to the
corrected $L_{\text{X}}$-$M$ relation, we were also able to correct the $L_{\text{X}}$-$T$
relation for
the selection bias effects. Similarly to the $L_{\text{X}}$-$M$ relation, the
$L_{\text{X}}$-$T$ is steeper that the observed relation when the selection bias effects are
taken into account with a steepening in the low-mass regime. Our result agrees qualitatively quite well with the findings of \cite{2011A&A...535A.105E} (but note that their group sample is incomplete) and \cite{2011A&A...532A.133M}, who corrected the $L_{\text{X}}$-$T$
relation for the HIFLUGCS clusters. Since they used the bolometric
luminosities, we cannot directly compare their results with our
relations.

\subsubsection{Group luminosities and completeness of the sample}
The corrected relations we obtained are based on the assumption that we deteced all the objects above a certain flux. Because of the shallow observation of the RASS data  some of the faint objects might be missing from the REFLEX and NORAS input catalogs, or that their estimated flux fell below our limit. To check this, we estimated the luminosities using the SB and k$T$ profiles derived in this work and compared them with the values in the input catalogs. The new X-ray luminosities were estimated by integrating the X-ray surface brightness up to the $R_{500}$. Basically, for all the annuli used to derive the temperature profiles we estimated the total count rate from the SB profile and converted this to the X-ray luminosity using the best-fit temperature and abundance values obtained during the spectral analysis. Since our data only cover a fraction of $R_{500}$, the temperatures to convert the count rate to the luminosity beyond the detection radius were set to an average value given by the extrapolated temperature profiles with  an abundance frozen to 0.3 solar. The results are summarized in the Table \ref{tab:propgroups}. While for most of the objects the luminosity estimated using the XMM-Newton data agree quite well within the errors with the ROSAT luminosities, for some very low mass objects our estimated luminosity is much higher. If on one hand the large extrapolation of the profiles makes these measurements quite uncertain, it might be that ROSAT was only able to detect a small fraction of the $R_{500}$ because of their faint emission in the outer region . As a consequence, is also possible that  some of the objects with the lowest flux are missing and that the input catalogs are more incomplete than previously thought.

\subsection{Mass proxies}
Among all the mass proxies, the $Y_{\text{X}}$ parameter seems the most
promising one to be used with the next X-ray surveys. In contrast to
$M_{\text{gas}}$, $L_{\text{X}}$, and $T$  the agreement between the slope and normalization of different works 
(e.g., \citealt{2007A&A...474L..37A}; \citealt{2007ApJ...668..772M};
\citealt{2009ApJ...692.1033V}; \citealt{2009ApJ...693.1142S}) is very
good and is very close to the self-similar scenario independently of the fitting method. This is probably because the $Y_{\text{X}}$ parameter is related to the total thermal energy of the ICM, which is mainly associated with the gravitational processes and so is less sensitive to any feedback. It is also
useful to note that despite the different fraction of unrelaxed
systems in the samples, the slope remains stable, which confirms that indeed
the $M$-$Y_{\text{X}}$ relation is quite insensitive to the dynamical
state of the objects. A direct implication is that if the eROSITA data will allow us to measure the temperature and SB profiles for many galaxy groups and clusters we will be able to
estimate the total mass from the $M$-$Y_{\text{X}}$ relation. While \citet{2014A&A...567A..65B}  found that at least an overall temperature can be obtained with  good accuracy (errors lower than 10$\%$) for $\sim2,000$ objects, determining the surface brightness profiles might be more complicated because of the eROSITA PSF. \\
Indeed, given the expectation for the temperature determination with
eROSITA, it would be much more straightforward to use the $M$-$T$
relation to estimate the mass. Unfortunately, simulations
(e.g., \citealt{1996ApJ...469..494E}, \citealt{2007ApJ...655...98N})
show that masses are underestimated by up to 20$\%$ for merging
systems because the assumptions of hydrostatic equilibrium and
spherical symmetry are invalid. Thus, a different fraction of merging
systems in the analyzed sample would result in a different slope and
normalization.  Although our slopes agree well with the slopes in
literature, in particular the slopes obtained using samples of galaxy
groups, the normalization of the $M$-$T$ relation is $\sim$15$\%$ lower
than the normalization obtained by \citet{2009ApJ...693.1142S} and $\sim$5$\%$
higher than that obtained by \citet{2011A&A...535A.105E}. \\
The $M_{\text{gas}}$-$M$ relation has been suggested as the lowest
scattered mass proxy. Although the slopes for galaxy groups and clusters are quite similar, there is some indication of steepening for high-mass systems. This then translates into an even higher gas mass and so higher gas fractions for a given total mass, than is actually
observed. \\

%-----------------------------Figure Start------------------------------
\begin{figure}[!t]
\begin{center}
\epsfig{figure=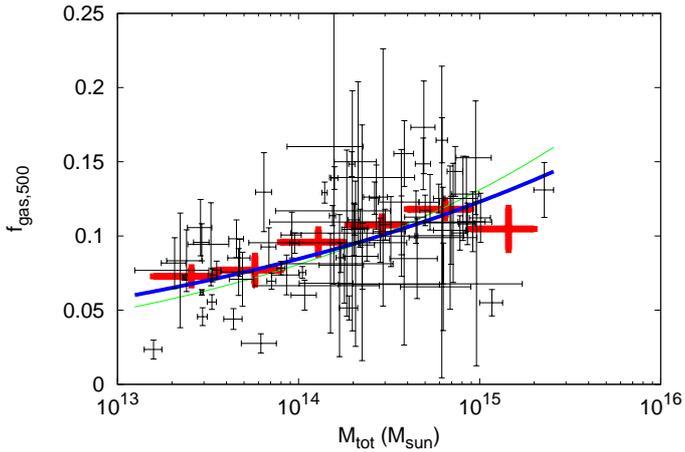,width=0.35\textwidth,angle=270}
\end{center}
\caption{Gas fraction as a function of the total mass. The black points represent the single objects (groups and HIFLUGCS), the red points are the mean values. The error bars are the standard errors. The solid blue line represents the best fit to the unbinned data. The parameters are listed in Table \ref{tab:scalawmalm}. The green line represents the best fit obtained by \citet{2009A&A...498..361P}.}
\label{fig:fgasbinned}
\end{figure}
%-----------------------------Figure End--------------------------------

%-----------------------------Figure Start------------------------------
\begin{figure*}[!t]
\begin{center}
\epsfig{figure=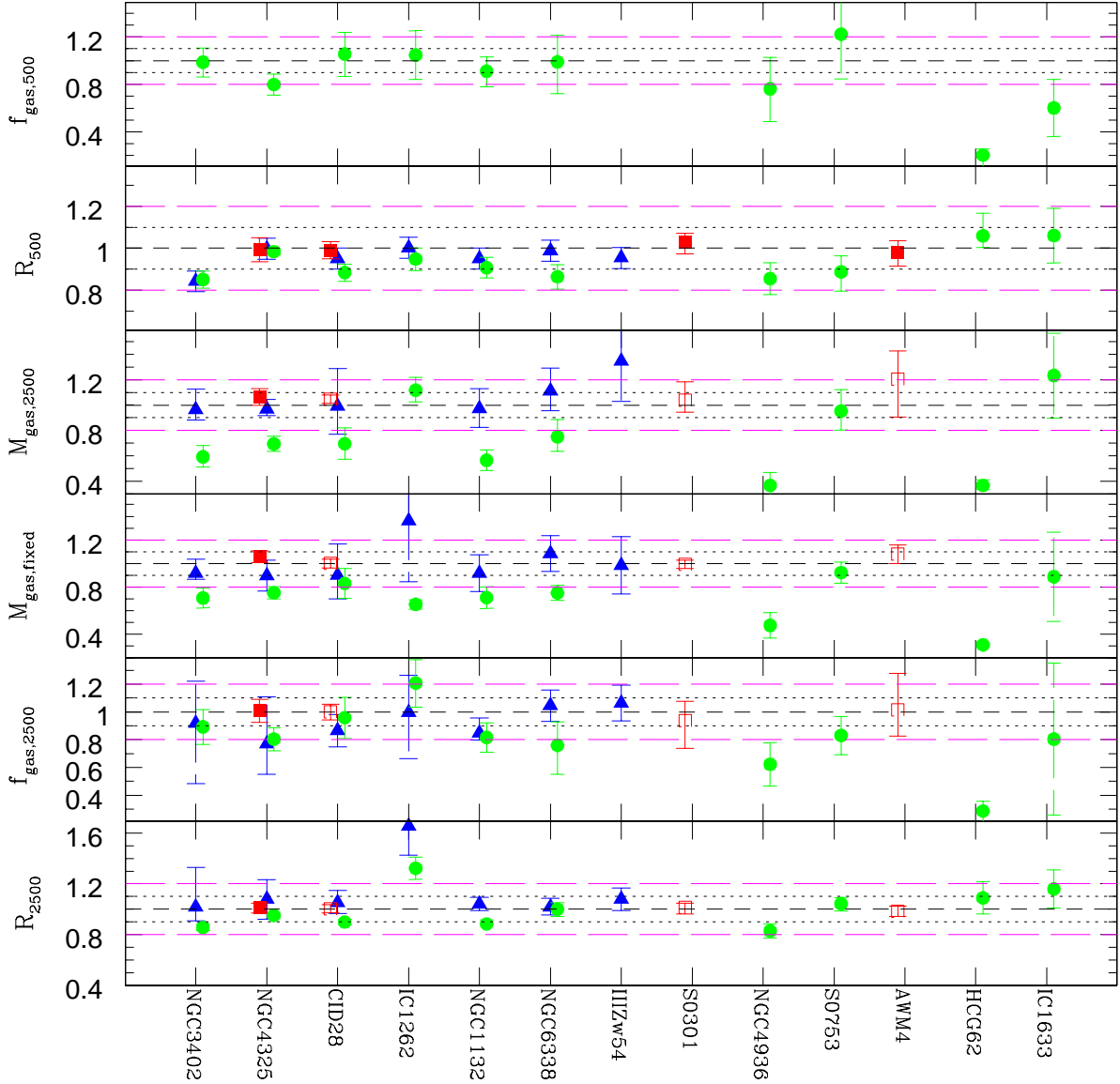,width=0.9\textwidth}
\end{center}
\caption{Comparison between the ratio of the results obtained in this work with the results reported by \citealt{2007ApJ...669..158G} ({\it $R_{2500}$: filled red squares; $R_{1250}$: empty red squares}), \citealt{2009ApJ...693.1142S} ({\it blue triangles}), and \citet{2011A&A...535A.105E} ({\it green circles}). 
From bottom to top we compare the ratio of $R_{\text{2500}}$; $f_{\text{gas,2500}}$; $M_{\text{gas}}$ at the fixed $R_{\text{2500}}$ by \citet{2007ApJ...669..158G}, \citet{2009ApJ...693.1142S}, and \citet{2011A&A...535A.105E} with this work; $f_{\text{gas,2500}}$; $R_{\text{500}}$; and $f_{\text{gas,500}}$, respectively.   }
\label{fig:fgascomp}
\end{figure*}
%-----------------------------Figure End--------------------------------

\subsection{Gas fraction} \label{fgasdiscuss} Several independent
analyses have found that the fraction of gas in galaxy clusters
decreases when moving toward lower mass systems
(e.g., \citealt{2003MNRAS.340..989S}; \citealt{2006ApJ...640..691V};
\citealt{2007ApJ...669..158G}; \citealt{2009A&A...498..361P}). \cite{2009A&A...498..361P} showed the tendency of the gas mass
fraction versus total mass by using data from different
works. While the massive system domain was well represented, the
sample they built had only a few low-mass systems (e.g., only five objects
with $M<5\times10^{13}M_{\odot}$). Our master sample has almost four times more objects at low masses and is more than two times larger in
total. In Fig. \ref{fig:fgasbinned} we show the results for our
sample. The behavior is similar to the one found by \citet{2009A&A...498..361P} with an increase of the gas fraction within $R_{500}$ with mass and a hint of flattening at the lowest ($<$$10^{14}M_{\odot}$) and highest masses ($>$$10^{15}M_{\odot}$). \\
In the review by \citet{2012NJPh...14d5004S} about galaxy groups, the
author compared the gas fractions obtained in different papers (i.e., \citealt{2006ApJ...640..691V}; 
 \citealt{2007ApJ...669..158G}; \citealt{2009ApJ...693.1142S};
 \citealt{2011A&A...535A.105E}) and found that while the first three agree relatively well, the gas fraction estimated by \citet{2011A&A...535A.105E} has a mean nonsystematic deviation of $\sim20\%$ for most groups and a much larger offset toward the low side for four groups (which are not in our sample). \citet{2012NJPh...14d5004S} suggested that the difference may come from the limited coverage of the group emission by the Chandra data, the background treatment performed by \citet{2011A&A...535A.105E}, and the simpler models adopted by
\citet{2011A&A...535A.105E} to describe the temperature and surface brightness profiles. 
Apart from the parametrization of the
temperature profiles, our approach is quite similar to the approach 
presented by \citet{2011A&A...535A.105E}, therefore we update the plot by
\citet{2012NJPh...14d5004S} with our results (see Fig. \ref{fig:fgascomp}) to check whether they are consistent with previous works. 
Our estimated radii agrees generally well with the other works. Nevertheless, by comparing the  temperatures of the objects in common between the samples we found that our global temperature is 13$\%$ higher than the temperatures obtained by \citet{2009ApJ...693.1142S} (in agreement with the different AtomDB results, see Appendix \ref{app_atom}), while the temperatures by \citet{2011A&A...535A.105E} agree  well within $5\%$ (except for three objects) with the temperatures in this work. Thus, although our global temperatures are in general higher than the ones derived by \citet{2009ApJ...693.1142S}, the estimated total masses are slightly lower. A possible explanation is that our density profiles are flatter in the outer region, causing slighter lower total masses. \\
At $R_{2500}$ the gas fraction of all the objects in common with \citet{2009ApJ...693.1142S}, except for IC1262,  agree to at maximum $20\%$
with a mean nonsystematic deviation of $\sim$10$\%$. Four out of ten objects in common with \citet{2011A&A...535A.105E} have a gas fraction systematically lower than our finding with a deviation larger than 25$\%$, and the mean deviation for the others 6 is
$\sim$15$\%$ (only for IC1262 our gas fraction is lower than the gas fraction found by \citealt{2011A&A...535A.105E}). We have only four groups in common with \citet{2007ApJ...669..158G}, and for three of them they provided only
the gas fraction at an overdensity of 1250. Therefore we computed the gas fraction for these three groups at the same overdensity and obtained a very good agreement. For a more direct comparison we then recomputed our gas masses at the radii ($R_{\Delta}$) derived by \citet{2007ApJ...669..158G}, \citet{2009ApJ...693.1142S}, and  \citet{2011A&A...535A.105E}.  The result is shown in Fig. \ref{fig:fgascomp} (third panel). The gas masses from \citet{2009ApJ...693.1142S}, \citet{2007ApJ...669..158G}, and this work agree to $\sim$10$\%$, while the gas masses from \citet{2011A&A...535A.105E} are very low (seven out of ten objects have a gas mass lower by 25-75$\%$ than our finding). To investigate the cause of the difference with \citet{2011A&A...535A.105E}, we calculated the gas masses at the $R_{2500}$ quoted in their paper using the best fit values of their surface brightness and central electron densities  (private communication) with our code. For most of the objects the gas masses we obtained are different from the masses calculated by \citet{2011A&A...535A.105E} 
but agree with the ones obtained in this work. There are still a few objects (HCG62, IC1633, NGC3402, and S0753) for which the differences are still significantly high. Thus, apart from a possible inaccuracy in the code, there might be other effects (e.g., the effects suggested by \citealt{2012NJPh...14d5004S}) that cause the observed discrepancies between our results and those of \citet{2011A&A...535A.105E}. However, the result shows that the strong
difference for the $f_{\text{gas}}$ values found by
\citet{2011A&A...535A.105E} does not mainly arise from a simpler analysis, as our agreement with \citet{2009ApJ...693.1142S} and \citet{2007ApJ...669..158G} suggests, but probably  by an incorrect calculation of the gas masses, which may also affect their scatter estimates (see Sect. \ref{secscatter}).\\
We then investigated which kind of groups contributed more to the scatter of the scaling relations. Interestingly, we found that the groups with higher gas fraction within $R_{2500}$ deviate more from the best fit of the scaling relations. For example, the mean deviation from the best fit of the L-T relation for the ten objects with lower gas fraction is 0.31, while for the objects with higher gas fraction within $R_{2500}$ is 0.82 (0.58 if we do not consider IC1633 which deviates more).

\subsection{Scatter}
Low-mass systems show a similar and sometimes even smaller
intrinsic scatter than their massive counterparts. If, on one hand, this
result contradicts the common thought of groups having a larger
scatter, on the other hand it is expected from the result of the
previous section. In fact, since the groups contributing more to the
scatter are the ones with the higher gas fraction, it is expected
that galaxy clusters that generally have an even higher gas fraction
show a larger dispersion in the scaling
relations. \citet{2011A&A...532A.133M} analyzed the HIFLUGCS sample
and found that objects with a CRS have a smaller scatter than objects
without. Since the fraction of objects in the group sample with a CRS
is larger that the fraction in the massive systems, our result 
agree with the finding of \citet{2011A&A...532A.133M}.\\
However, the properties of the massive clusters
were obtained using an isothermal model and a single $beta$-model
(\citealt{2002ApJ...567..716R}) and not a temperature profile like for
the groups in this work. Together with the larger fraction of
disturbed systems in the HIFLUGCS sample and the fact that in this
analysis a strong extrapolation is required to estimate the group properties, this might partially explain the lower scatter observed at the group scale.

%-----------------------------Figure Start------------------------------
\begin{figure*}[!t]
\begin{center}
\hbox{
% un-comment the following line to include your fig1a.ps postscript file
\epsfig{figure=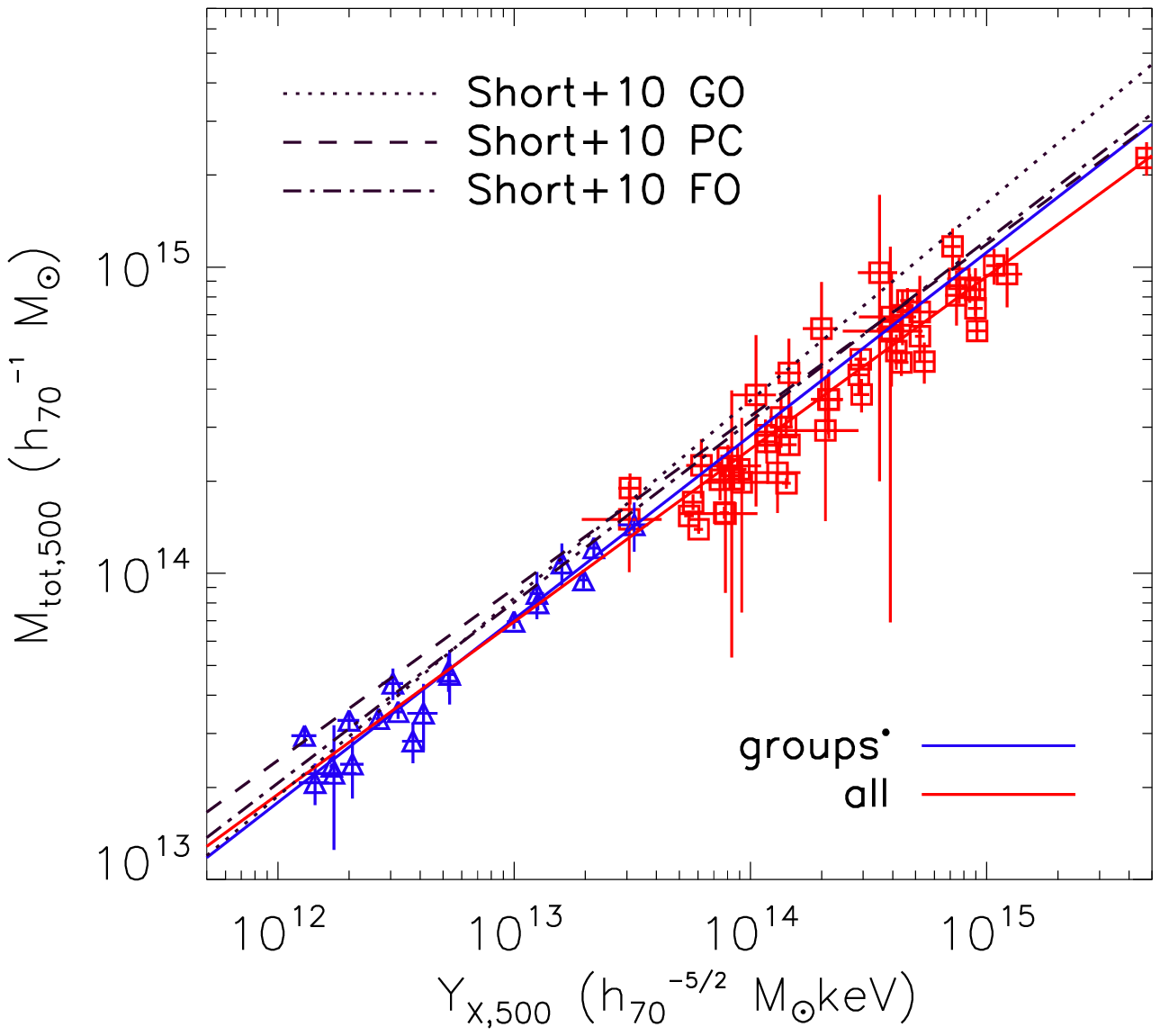,height=8cm,width=0.33\textwidth,angle=0}
\epsfig{figure=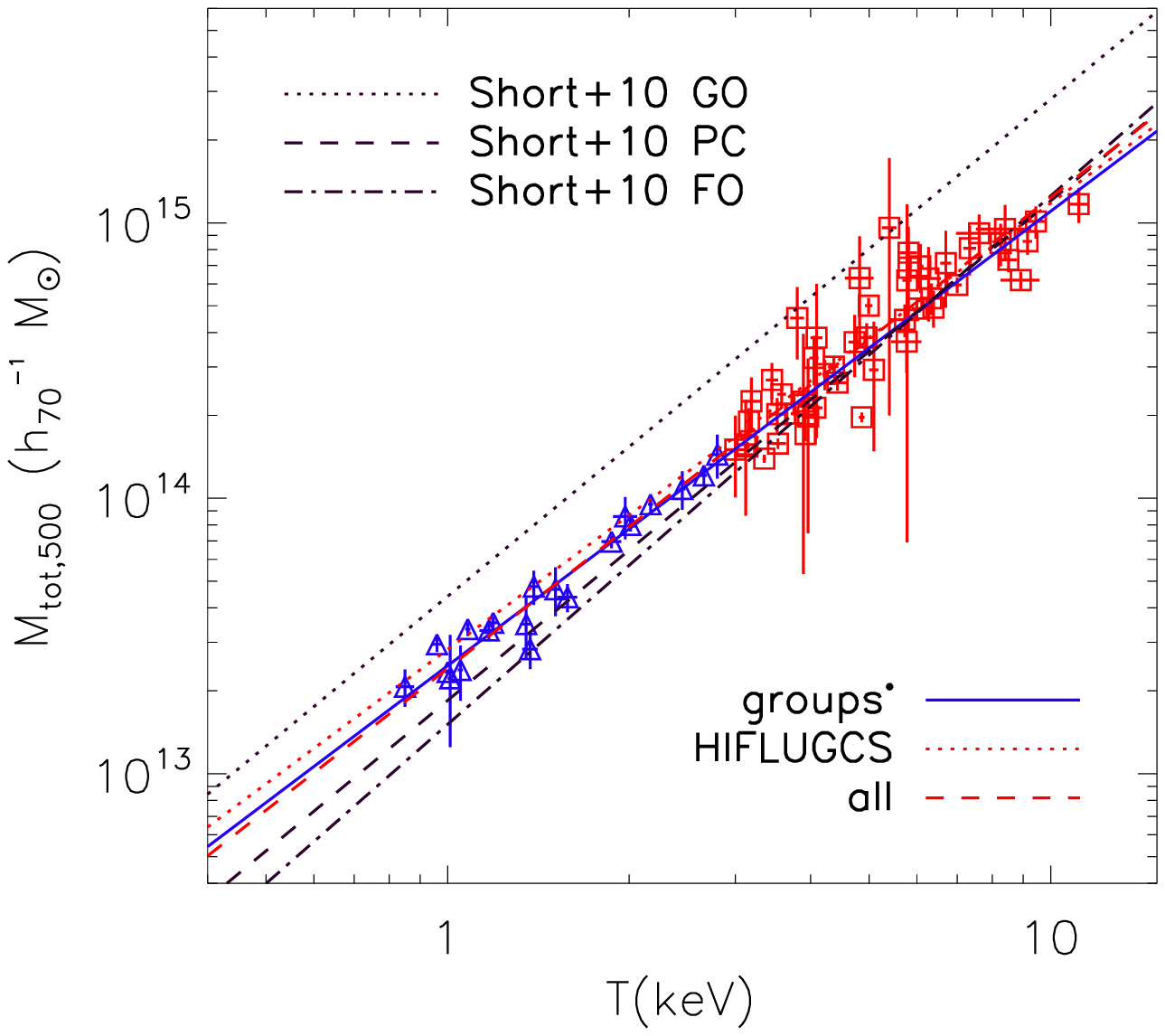,height=8cm,width=0.33\textwidth,angle=0}
\epsfig{figure=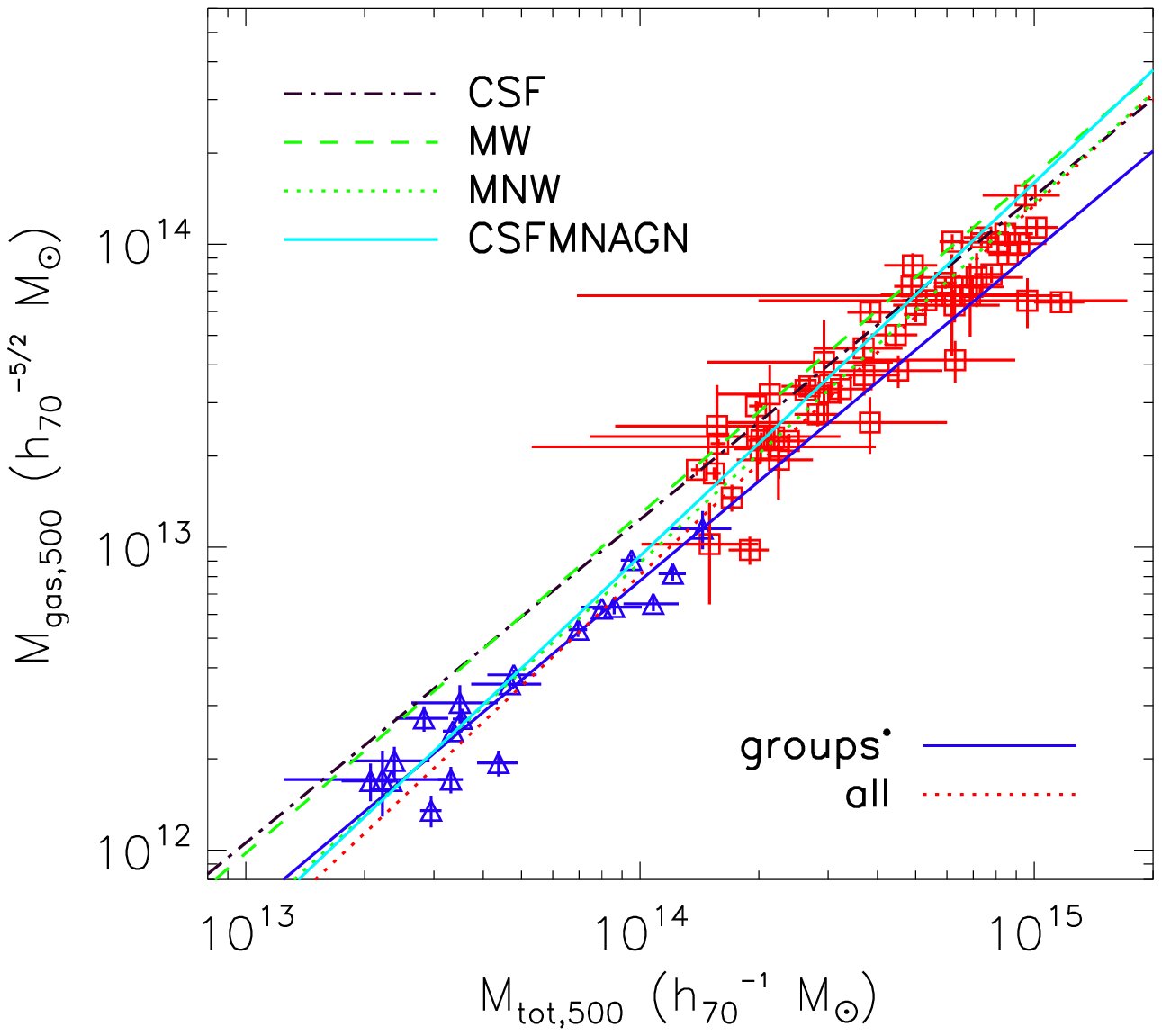,height=8cm,width=0.33\textwidth,angle=0}
% un-comment the following line to include your fig1b.ps postscript file
}
\end{center}
\caption{{\it Left:} The best-fit $M$-$Y_{\text{X}}$ relations determined in this work are compared with the results from \citet{2010MNRAS.408.2213S}. {\it Center:} As in the left panel, but for the $M$-$T$ relation.  {\it Right:} The best-fit $M_{\text{tot}}$-$M_{\text{gas}}$ relations determined in this work are compared with the results from \citet{2010MNRAS.401.1670F}.}
\label{fig:sim}
\end{figure*}
%-----------------------------Figure End--------------------------------

\subsection{Comparison with simulations}
A direct comparison between the observed scaling relations and the
results from hydrodynamical simulations can give us information about
the processes that play a role in the ICM at different
masses. We investigate the mass-proxies relations by comparing the
data with the simulation of \citet{2010MNRAS.408.2213S} for the
$M$-$T$ and $M$-$Y_{\text{X}}$ relations, and
\cite{2011MNRAS.416..801F} for the $M_{\text{gas}}$-$M$. The results for
the $M$-$T$ and $M$-$Y_{\text{X}}$ relations by
\cite{2011MNRAS.416..801F} were not used because of the known
difference between the
mass-weighted temperature used in their paper and the spectroscopic temperature (\citealt{2004MNRAS.354...10M}). \\
In the $M$-$Y_{\text{X}}$ relation, the gravitational
heating alone (black dotted line in the left panel of Fig. \ref{fig:sim}) is not enough to match the observational data, and
additional processes have to be included. The slope obtained by
\citet{2010MNRAS.408.2213S} when including only the
gravitational processes in the simulations is steeper than the slope predicted by the
self-similar scenario, which was quite well reproduced when the
non-gravitational heating was included in the simulations. Thus, on
one hand the observed relation seems weakly affected by the non
gravitational processes (confirmed also by the fact that the relation
shows the same slope at all the masses), on the other hand, simulations
need to include feedback to match the observed slope. Although our slope agrees well  with the prediction from
\citet{2010MNRAS.408.2213S} when the energy input from supernovae and AGNs
(FO run) were included, the normalization is $\sim$14$\%$ lower than
the
prediction from their simulation.\\
The observational results for the $M$-$T$ relation compared with the
predictions by \citet{2010MNRAS.408.2213S} are shown in the center panel
of Fig. \ref{fig:sim}. Although at high masses there is a good
agreement (difference $<$10$\%$ at 10 keV) in the low-mass regime, there is a strong difference ($\sim25\%$ at 1 keV) between
our observational results and the simulations. On the other hand, they neglect cooling processes
so that systems with a cool-core are not formed, whiche prevents a comparison with our group sample that contains a large fraction of such
systems. Thus, the mismatch
between observation and simulations requires further investigations.\\
The $M_{\text{gas}}$-$M$ relation (right panel of Fig. \ref{fig:sim}) of our group sample agrees to better than 10$\%$ at 5$\times$10$^{13}M_{\rm sun}$ with
the finding of \citet{2011MNRAS.416..801F} when they include AGNs
feedback. In contrast, when only galactic winds are included in
the simulations, there is a strong disagreement with the
observations (e.g., $>$50$\%$ at 5$\times$10$^{13}M_{\rm sun}$). Together with the lower gas fraction in galaxy groups, these
results suggest that indeed AGN activity is at work and that this is 
probably responsible
for transporting the gas away from the galaxy group center.\\
In general, simulations including feedback processes are able to
reproduce the observed slope and normalization better than the simulations
including only gravitational effects. On the other hand, their effect
does not strongly affects the slope (i.e., we do not observe any break in the scaling relations as also found  for the L-T relation by \citealt{2014ApJ...783L..10G} when the self-regulated  kinetic feedback model is adopted in the simulations) in the low-mass regime, although it is
possible that their contribution decreases gradually with the mass of
the systems and is then hidden by the scatter.

\section{Conclusions} \label{conclusion} 
The complete sample of galaxy
groups studied in this paper allowed us to correct the
$L_{\text{X}}$-$M$ and $L_{\text{X}}$-$T$ relation for selection bias
effects. While selection biases have been taken into account in
several papers analyzing complete samples of galaxy clusters
(including some groups), this is the first time that this was done for a
complete sample of local X-ray selected groups. We summarize our results as follows:\\
\begin{enumerate}
\item[-] The slope (1.66$\pm$0.22) of the $L_{\text{X}}$-$M$ relation corrected for the selection
  bias effects, derived at the group scale, is steeper than the corrected slope (1.08$\pm$0.21) obtained with massive systems.
  If confirmed with larger samples, this would imply that for future X-ray surveys such as eROSITA a relation with more freedom than a single power law to convert the   luminosities into the total masses is required to constrain the
  cosmological parameters.
\item[-] For the other mass proxies we found that the
  $M$-$Y_{\text{X}}$ relation seems less sensitive to the dynamical state of the objects and consequently, to the sample
  properties.
\item[-] In general, we did not observe any steepening of the observed uncorrected scaling  relations in the low-mass regime.
\item[-] Groups have an intrinsic scatter similar or even smaller
  than the scatter derived for galaxy clusters.
\item[-] The observed scaling relations for galaxy groups agree well with the simulations including AGNs, although it depends
  strongly on the physical processes included in the simulations. This
  indicates that non-gravitational processes play an important
  role in the evolution of galaxy groups.
\item[-] The gas mass fraction in galaxy groups is almost a factor of two lower than the gas fraction in galaxy clusters.  
\item[-] The new improved AtomDB version yields a gas fraction up to 20$\%$ lower than an older version.
\end{enumerate}

\begin{acknowledgements}
  We acknowledge useful discussions with V. Bharadwaj, E. Torresi, and Y.-Y. Zhang. We thank H. Eckmiller, D. Fabjan, and M. Sun for providing details of their published results and the anonymous referee for the suggestions that improved the quality of the paper. LL acknowledges support by the DFG through grant RE 1462/6 and
  LO 2009/1-1, by the German Aerospace Agency (DLR) with funds from the
  Ministry of Economy and Technology (BMWi) through grant 50 OR
  1102. THR acknowledges support from the DFG through Heisenberg grant
  RE 1462/5 and grant RE 1462/6. GS acknowledges support from the DFG through grant
  RE 1462/6. The program for calculating the
  $t_{cool}$ was kindly provided by P. Nulsen; it is based on
  spline interpolation on a table of values for the APEC model
  assuming an optically thin plasma by R.~K.~Smith.
\end{acknowledgements}

\bibliographystyle{aa} \bibliography{Lovisari_groups}

%\begin{thebibliography}{}

% \bibitem[1966]{baker} Baker, N. 1966,
%      in Stellar Evolution,
%      ed.\ R. F. Stein,\& A. G. W. Cameron
%      (Plenum, New York) 333

%\end{thebibliography}

\begin{appendix}
\section{AtomDB} \label{app_atom}

%-----------------------------Figure Start------------------------------
\begin{figure*}[!t]
\begin{center}
\hbox{
\epsfig{figure=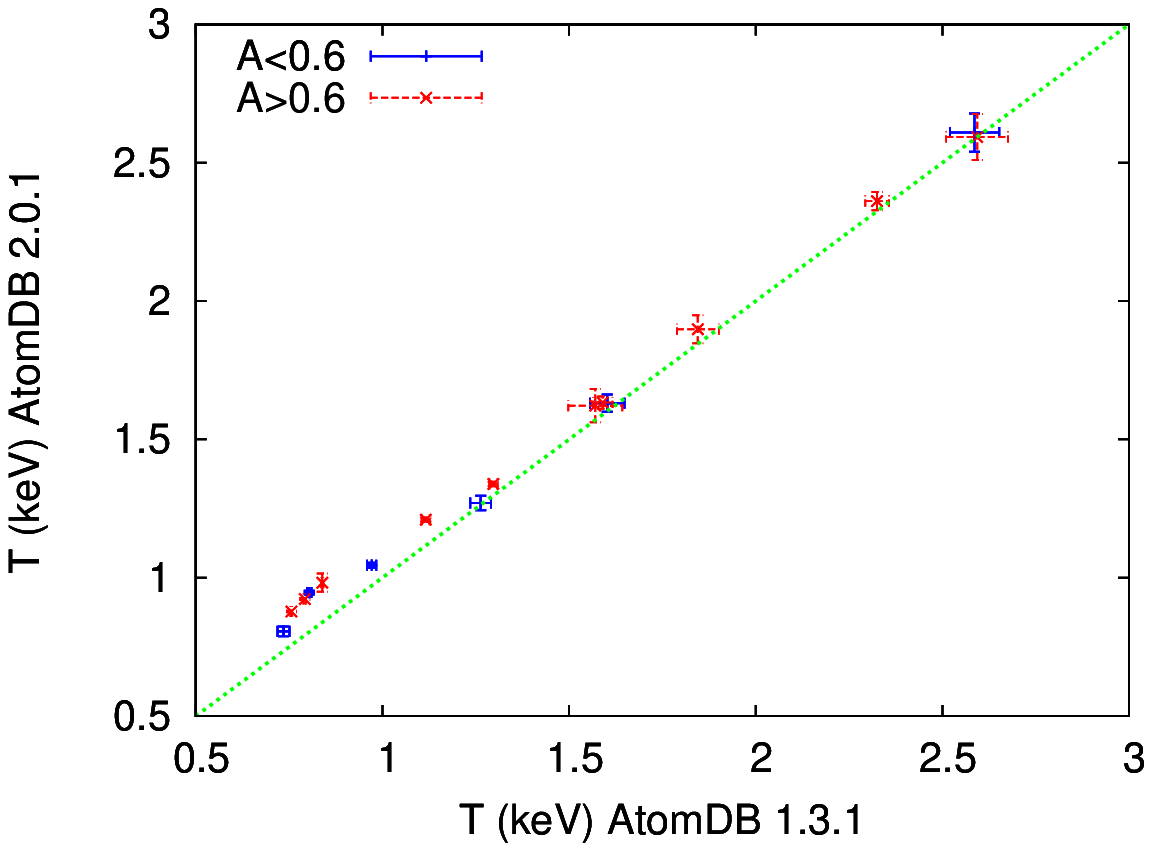,width=0.5\textwidth,angle=0}
\epsfig{figure=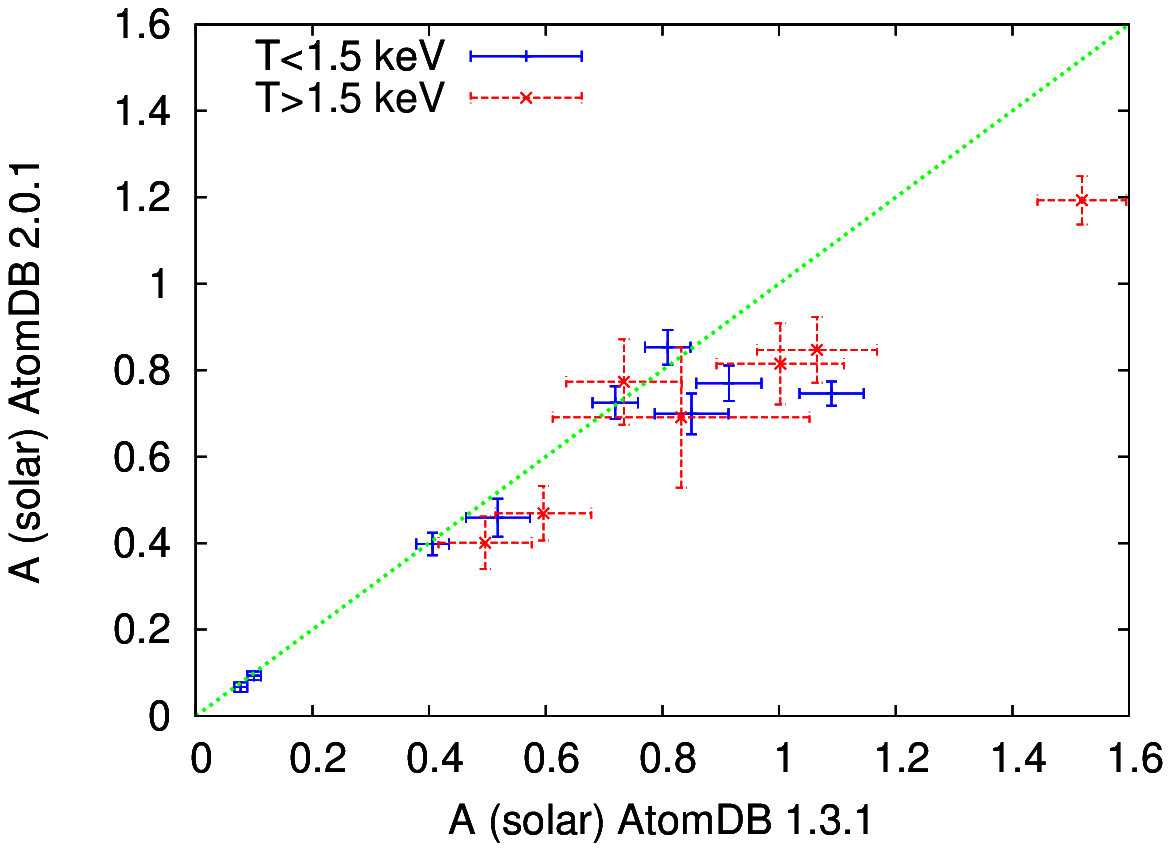,width=0.5\textwidth,angle=0}
}
\end{center}
\caption{\footnotesize
{{\it Left:} comparison between the temperatures determined using the old version 1.3.1 and the new version 2.0.1 of AtomDB. Different colors represent groups with an abundance higher ({\it red}) and lower ({\it blue}) than 0.6 solar. Here, we plotted only the innermost temperature bin, and excluded the groups for which we determined only a global temperature: IC1262, NGC6338, RXCJ1840. {\it Right:} the same as in the left panel, but comparing the abundances instead of the temperatures. Different colors represent groups with a temperature higher ({\it red}) and lower ({\it blue}) than 1.5 keV.}}
\label{fig:atomDB}
\end{figure*}
%-----------------------------Figure End--------------------------------

%-----------------------------Figure Start------------------------------
\begin{figure*}[!t]
\begin{center}
\hbox{
\epsfig{figure=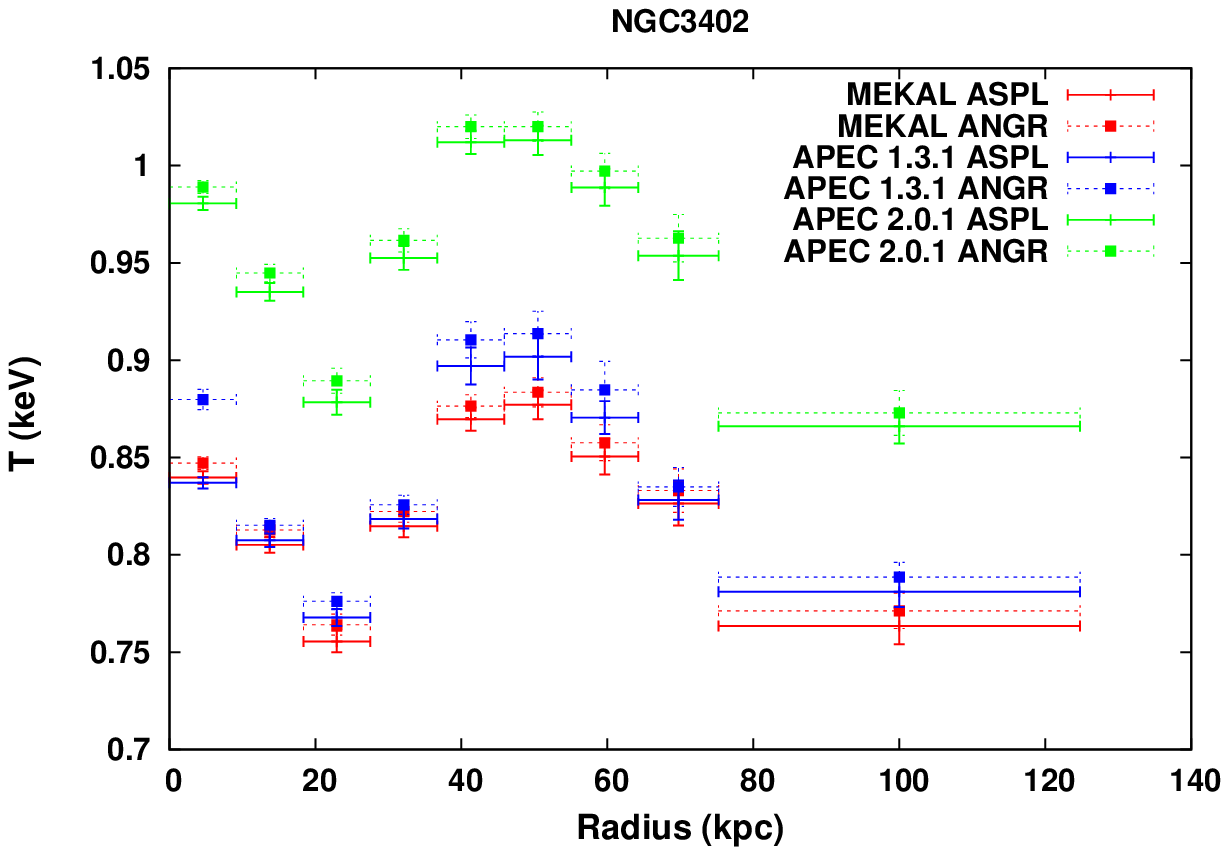,width=0.5\textwidth,angle=0}
\epsfig{figure=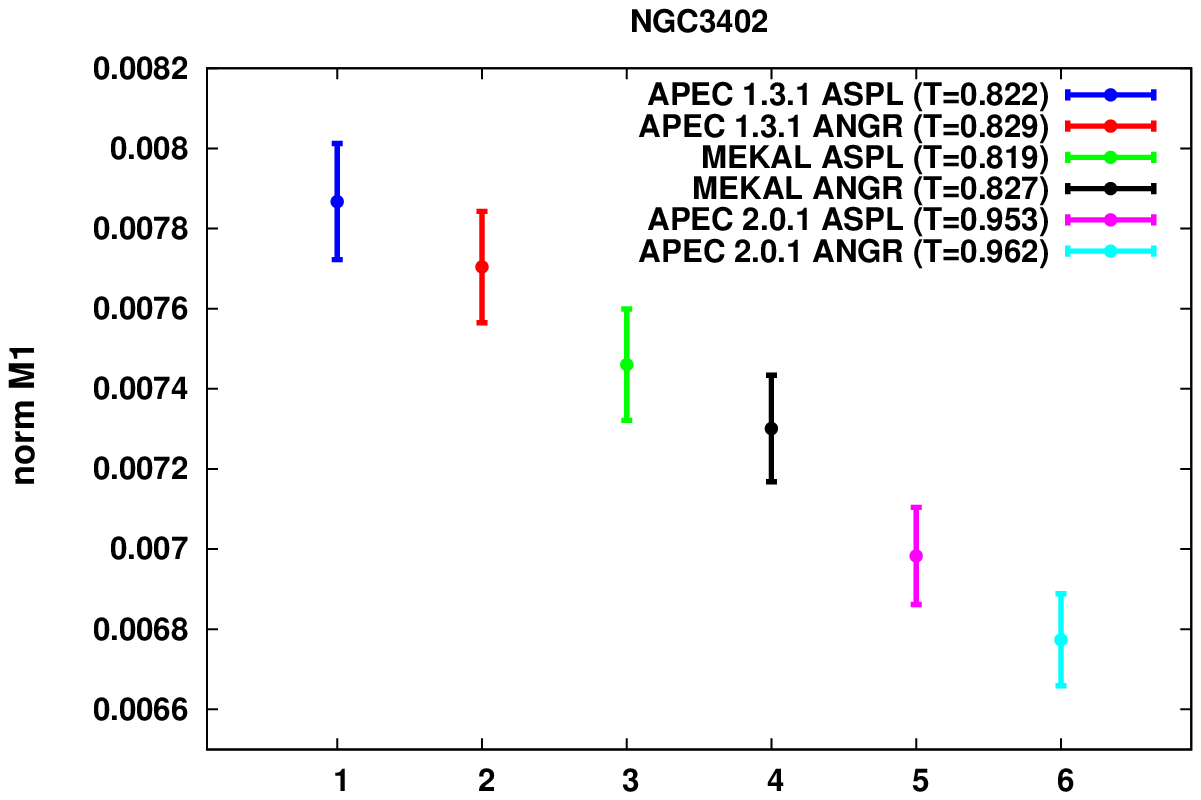,width=0.5\textwidth,angle=0}
}
\end{center}
\caption{\footnotesize
{{\it Left:} temperature profile of NGC3402 derived by fitting the spectra with different plasma models and abundance tables. {\it Right:} normalization values obtained by fitting a spectrum with different plasma models and abundance tables.}}
\label{fig:ngc3402}
\end{figure*}
%-----------------------------Figure End--------------------------------

Version 2.0 of AtomDB is available since 2011. With respect to the
older version 1.3, it includes significant improvements on the iron
L-shell data. As we show below, this change strongly affects the
temperature and abundance determination for low-mass systems. We used
 the temperature and the APEC normalization from the spectral
fits to determine the gas and total masses. Thus, the
use of different AtomDB versions has to be taken into account when
comparing our results with the ones in literature. Here, we analyze
the main effects.
\subsection{Temperature and total mass}
To show how the temperature and abundance determination change, we
fitted the innermost bin of the galaxy groups in our sample using the
two AtomDB versions. The results are shown in
Fig. \ref{fig:atomDB}. While at temperatures higher than 1.5 keV the
temperature difference is quite small, at very low temperatures
(i.e., $kT$$<$1 keV) the temperatures obtained using the version 2.0
are up to 18$\%$ higher than the ones obtained using version 1.3. At
the same time, the obtained abundance is 20-30$\%$ lower with a trend
of a larger deviation for higher temperatures. More in detail, we note
that when the group abundance is relatively low
($A$$<$0.6$A$$_{\sun}$), the temperature deviation arises only for
$kT$$<$ 1 keV. In contrast, when the abundance is relatively high
($A$$>$0.6$A$$_{\sun}$), small differences can be observed
already at a temperature of $\sim$2 keV.\\
To investigate how much this influences the total mass estimate 
we used NGC3402 as a test case because of its low temperature and good
quality of data. Since many authors use a MEKAL model
instead of the APEC model, we also included this thermal plasma model
in our analysis. We determined the temperature profile for the
different models and for different abundance tables (i.e., from
\citealt{1989GeCoA..53..197A} and \citealt{2009ARA&A..47..481A}). As
shown in Fig. \ref{fig:ngc3402} ({\it left panel}), while the profile
from MEKAL and the old AtomDB version agree quite well, the new AtomDB
has a higher temperature at all radii. This translates into a 
total mass higher by $\sim$10$\%$. Although the effect is weak, a higher temperature is also
obtained when the old abundance table from \citet{1989GeCoA..53..197A}
instead of the most recent table from \citet{2009ARA&A..47..481A} is
used. 

\subsection{Gas density and gas mass}	
In Fig. \ref{fig:fgascomp} we compared the gas mass at a given radius
for different works and found that the difference for most of
the objects is of $\sim$10$\%$. Since we used the APEC normalization
to estimate the central electron densities to better understand whether
the new AtomDB can explain part of the difference, we compared the
normalization of the spectrum extracted from an annulus of $\sim$7
arcmin (to maximize the S/N) and fitted with the different models. As
shown in the {\it right panel} of Fig. \ref{fig:ngc3402} (for display
purposes we only show the MOS1 normalization, but the trend is similar
 for MOS2 and pn, although with different values), the normalization
with the new AtomDB is $\sim$10$\%$ lower than the older one with the
MEKAL one lying in between. Depending on the combination of abundance
tables and plasma models used in a particular paper, the difference can
be up to $\sim$15-20$\%$. Since the central electron density scales
with the square root of the normalization from XSPEC, using the new
AtomDB can give a lower central density (and so a lower integrated gas
mass) of up to 7-10$\%$. The difference in gas mass for NGC3402 is
$\sim$6$\%$ at $R_{2500}$ and $\sim$10$\%$ at $R_{500}$, so the use of
different AtomDB versions alone can explain the different gas mass
shown in Fig. \ref{fig:fgascomp}.

\subsection{Gas fraction}
The use of the new AtomDB version results in a total mass higher by 10$\%$ than the mass derived using an older version. At the same time, the gas mass will be up to $10\%$ lower than the value obtained using the old AtomDB versions. Given these results, the gas fraction for the less massive galaxy groups obtained with the most recent version of the AtomDB can be up to 20$\%$ lower than the mass derived with the old AtomDB version. Of course, this is an upper limit because we used NGC3402 for the calculation, one of the coolest groups in our sample, which implies a larger difference between the different AtomDB versions, and not all the low temperature objects show such a large difference. Furthermore, in general, the temperature profiles obtained with the new AtomDB version cannot be simply scaled up because, as we showed, the difference in temperature depends both on the real temperature and on the associated metallicity. For example, in the outer regions where the temperature is lower, the metallicity is lower as well, which mitigates the real difference in the temperature estimation (see, e.g., \ref{fig:atomDB}). This result highlights the
importance of taking this problem into account for comparisons between
different papers.

\section{A few details on some galaxy groups}
\subsection*{A194}
At first glance, A194 can be confused with a merging system because it
shows three X-ray peaks: the main one in the NE, a second one in the
SW, and a third in the center.  \cite{2005ApJ...622..187M} argued that
the SW source is a background cluster of galaxies at
$z$=0.15. \citet{2008MNRAS.384...87S} confirmed that although it might
be possible that the SW source is a background cluster, it is not
possible with the XMM-Newton data to exclude that the source
is at the same redshift as A194. We used an extraction area with a radius of 1 arcmin 
centered on the source and found that it is better
fitted by a thermal plasma at redshift 0.15 than by a model with a
redshift fixed at 0.018, in agreement with the finding of
\cite{2005ApJ...622..187M}. In particular, we obtained a temperature of
$1.82^{+0.10}_{-0.13}$ and metallicity of $0.36^{+0.05}_{-0.05}$ with
$\chi^2/dof=137/114$ when $z$=0.15 and a temperature of
$1.28^{+0.04}_{-0.04}$ and metallicity $0.10^{+0.02}_{-0.02}$ with
$\chi^2/dof=173/114$ if $z$=0.018. Thus, since the peak is probably
the BCG of a background cluster, we decided to exclude a region
corresponding to $R_{500}$ from the analysis of A194 to minimize the
effect that it would have on the derived properties. By using the
$M$-$T$ relation derived only using the other objects in the
sample, we then estimated for A194 a mass of {\it
  M}$\sim$6$\times10^{13}$ and {\it R$_{500}$}$\approx$500 kpc which
corresponds to about 3.5 arcmin at the A194 redshift. To be on the
safe side, we excluded 4 arcmin around the SW peak. The flux in the
0.1-2.4 keV band from this 4 arcmin region is $\sim$10$^{-12}$
erg/s/cm$^2$, so
even excluding this region, the net flux of A194 is $\sim$8.7$\times$10$^{-12}$
erg/s/cm$^2$, well above the flux limit threshold. \\
By extracting a spectrum from a region with a radius of
15$^{\prime\prime}$ around the NW source, we found that it is
consistent with that of an AGN type 2 (an intrinsic absorption
component was needed to fit the spectrum). The redshift of the source
is 0.0182 consistent with the redshift of the cluster and a luminosity of
3$\times10^{41}$ erg/s, suggesting that it is
accreting inefficiently. The estimated flux is 1.3$\times10^{-13}$ ergs/s/cm$^2$. \\
Both regions were excluded from the analysis of the group
properties.

\subsection*{A3390}
A3390 shows two X-ray peaks that are centered on two bright
galaxies at the cluster redshift. We extracted a spectrum from a
region of 4$^{\prime}$ around the two X-ray peaks to estimate the
redshift of the two clumps with the X-ray data alone. We did not find
any evidence that the two clumps have a different redshift. We estimated the temperature and surface brightness profiles of each component independently by excluding a region of 10 arcmin around the second subcluster.

\subsection*{IC1633}
IC1633 appears as a relaxed group with no usual signs of a
merger, such as a radio halo or a mismatch between the X-ray peak and
the cD galaxy. Instead, from the exposure-corrected,
background- and point-source-subtracted image we note that there is
a strong elongation to the north of the emission peak (i.e., higher
surface brightness). This feature together with the separation of
more than 30 kpc between the EP and EWC suggests we are probably observing an unrelaxed system. 

\subsection*{A3574E}
A3574 has two components separated by $\sim$0.6 Mpc that are accepted as
independent clumps (\citealt{2004A&A...425..367B}).  The main one is the
eastern clump (A3574E), whose central galaxy is a Seyfert galaxy: IC 4329A. This
galaxy carries 75$\%$ of the total flux of the clump
(\citealt{2004A&A...425..367B}), but the net flux of $\sim$7.3$\times$10$^{-12}$ ergs/s/cm$^2$ is still above the flux
threshold of this paper. 

\subsection*{WBL154}
This system is clearly in the process of merging, with a bright
subclump just to the south of the main X-ray peak.The subclump
corresponds to a small group of galaxies apparently falling into the
gravitational potential of the main group. \\

\subsection*{NGC4936}
NGC4936 is the lowest redshift group analyzed in this sample.
Its surface brightness profile has an unusual outer $\beta$ value of 0.32$\pm$0.01. The X-ray image 
shows that the X-ray peak, which is centered on the cD galaxy, is surrounded by a very faint extended emission. 

\subsection*{NGC3402}
Despite its overall regular and spherical X-ray emission, this galaxy group shows an anomalous temperature distribution with a central temperature peak surrounded by a relatively cool shell.
This remarkable feature has also been observed by different authors (e.g., \citealt{2006ApJ...640..691V}; \citealt{2007ApJ...658..299O}; \citealt{2009ApJ...693.1142S} \citealt{2011A&A...535A.105E}) and different instruments (e.g., XMM-Newton and Chandra). Combining XMM-Newton, Chandra, and VLA observations  \citet{2007ApJ...658..299O} concluded that the most likely explanation for this feature is the interplay between the cool-core region and a previous period of AGN activity.

\onecolumn 

\section{Surface brightness profiles} \label{SBprofiles} 
%-----------------------------Figure Start------------------------------
\begin{figure*}[ht]
\begin{center}
\hbox{
\epsfig{figure=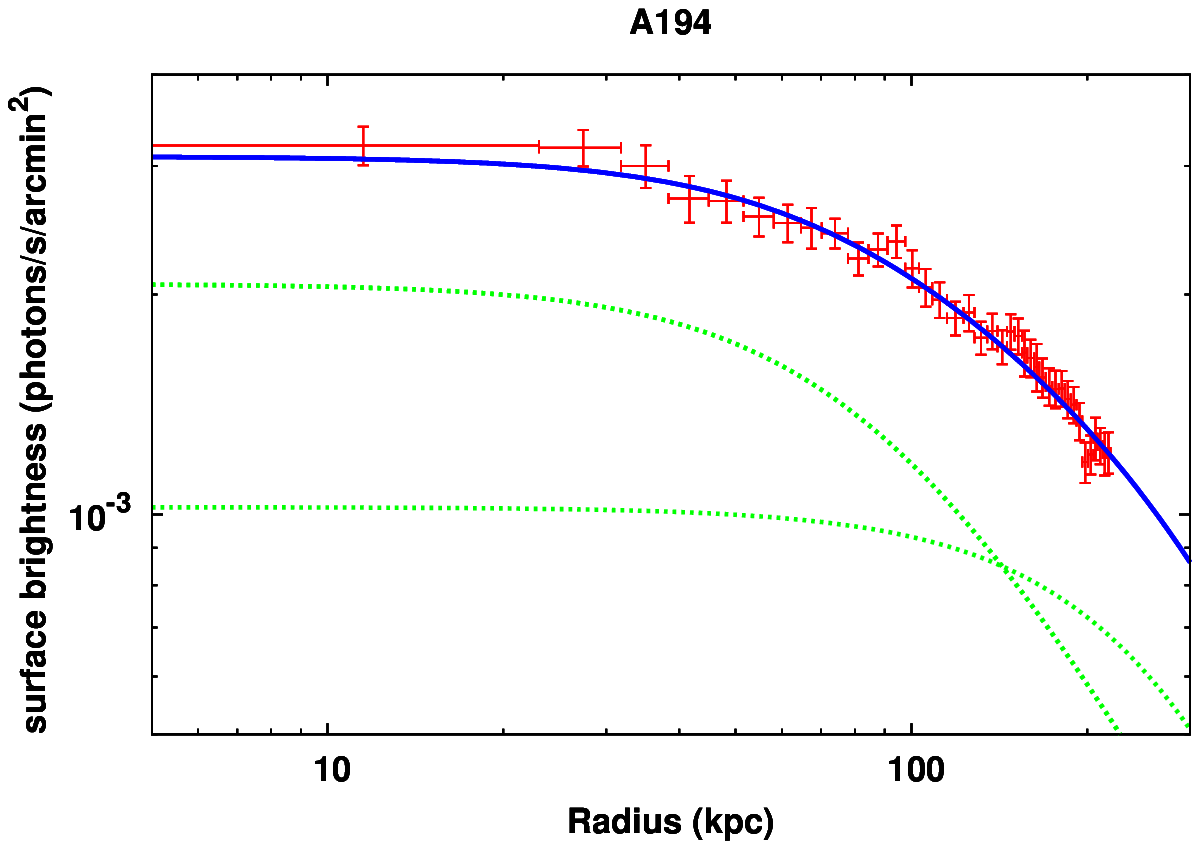,height=5cm,width=0.3\textwidth,angle=0}
\epsfig{figure=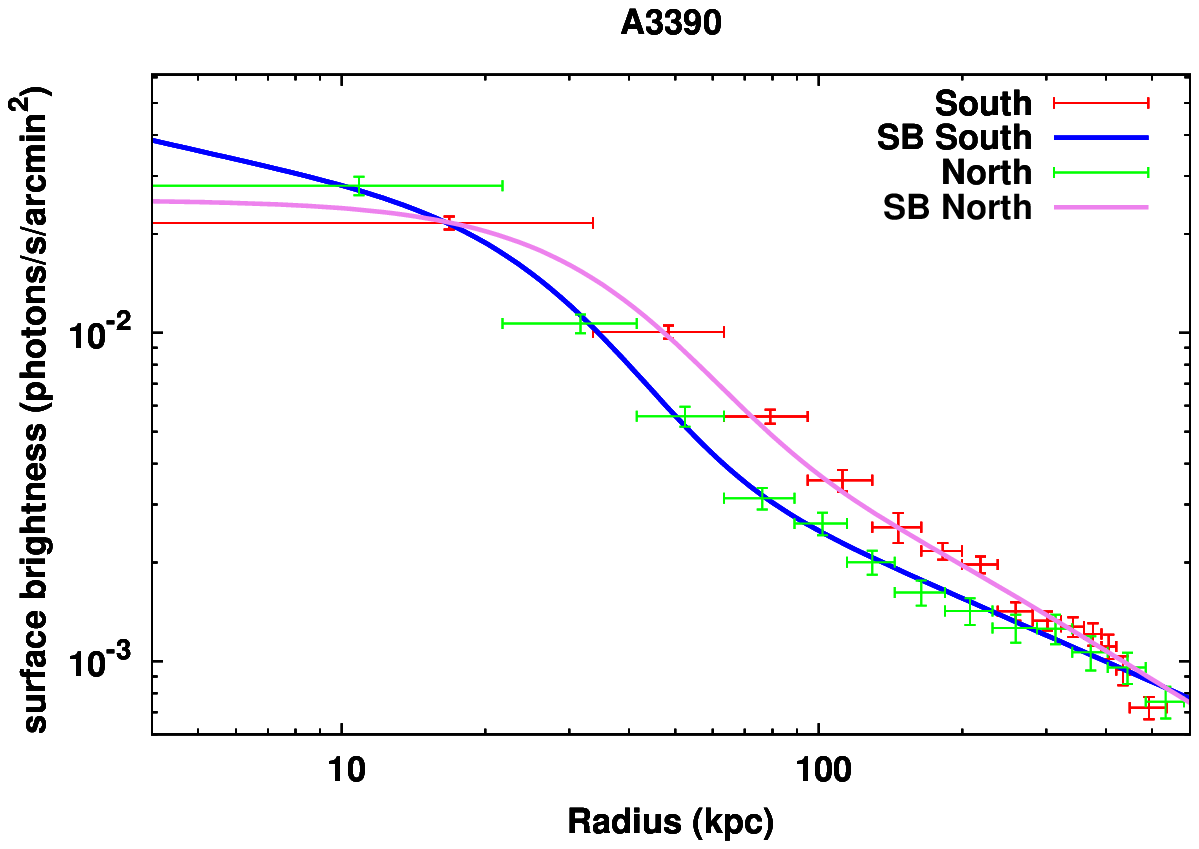,height=5cm,width=0.3\textwidth,angle=0}
\epsfig{figure=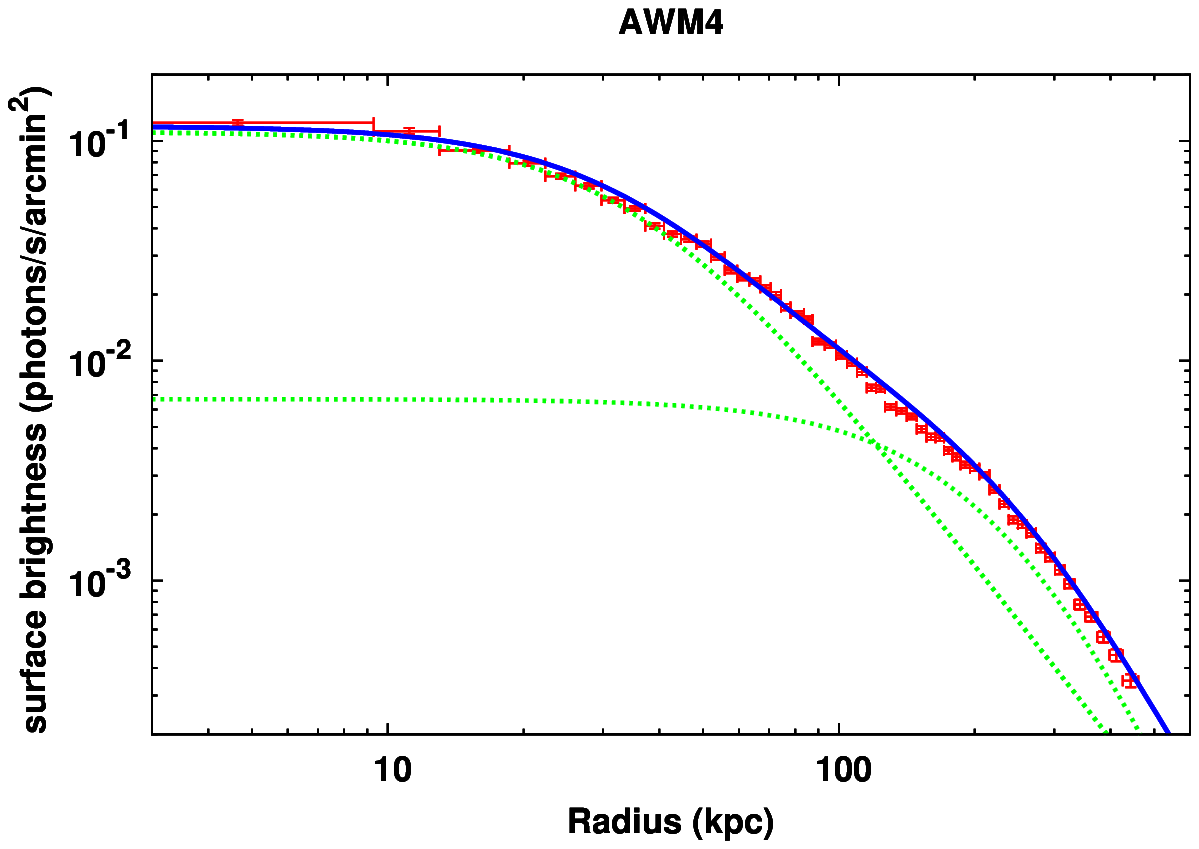,height=5cm,width=0.3\textwidth,angle=0}
}
\end{center}
\vspace{-20pt}
\caption{\footnotesize{\it Surface brightness profiles for A194, A3390, and AWM4.}}
\end{figure*}

%-----------------------------Figure End--------------------------------

%-----------------------------Figure Start------------------------------
\begin{figure*}[ht]
\begin{center}
\hbox{
\epsfig{figure=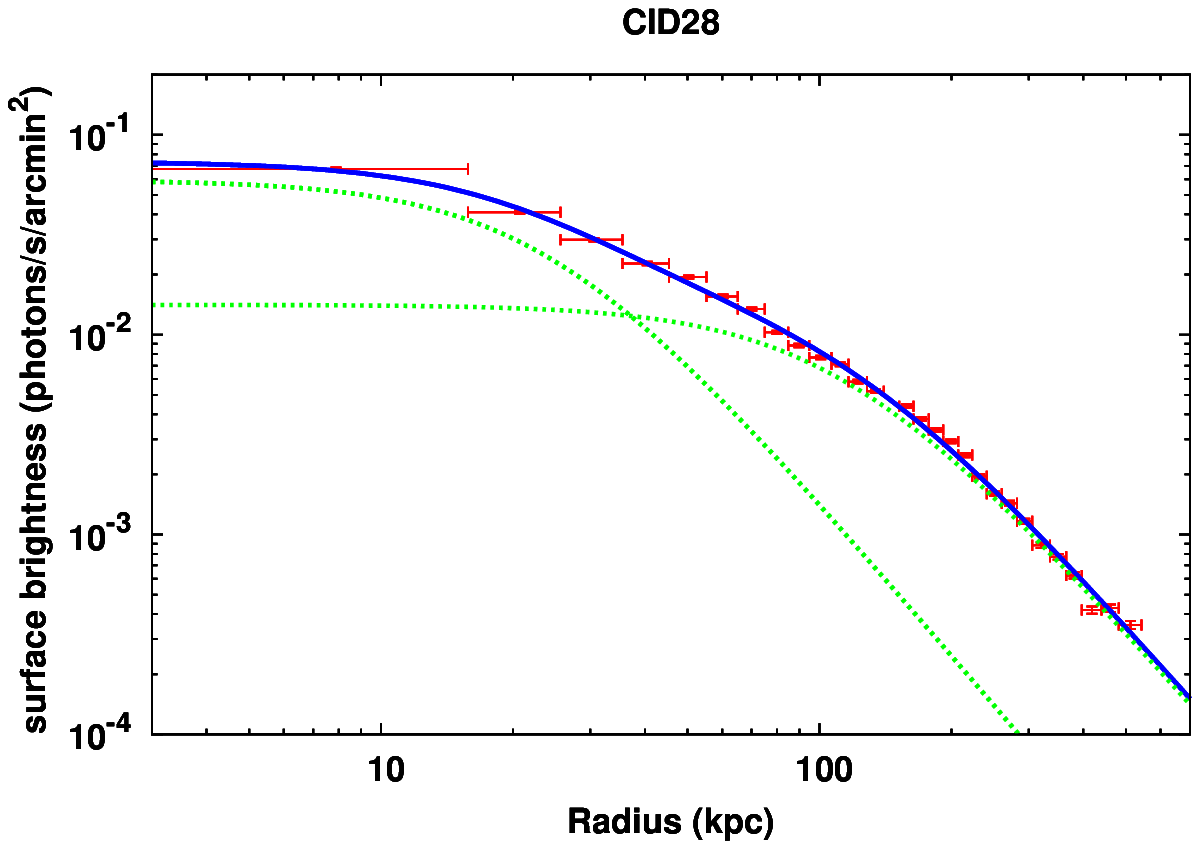,height=5cm,width=0.3\textwidth,angle=0}
\epsfig{figure=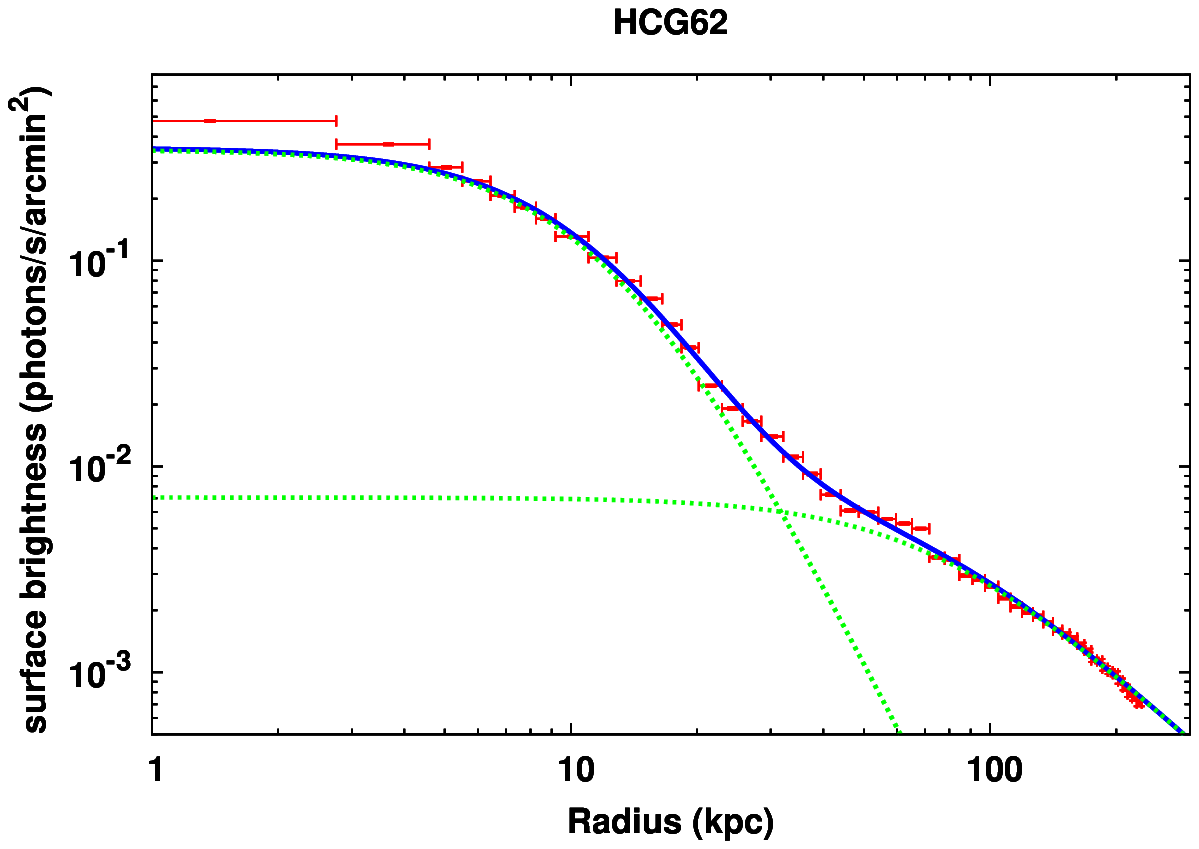,height=5cm,width=0.3\textwidth,angle=0}
\epsfig{figure=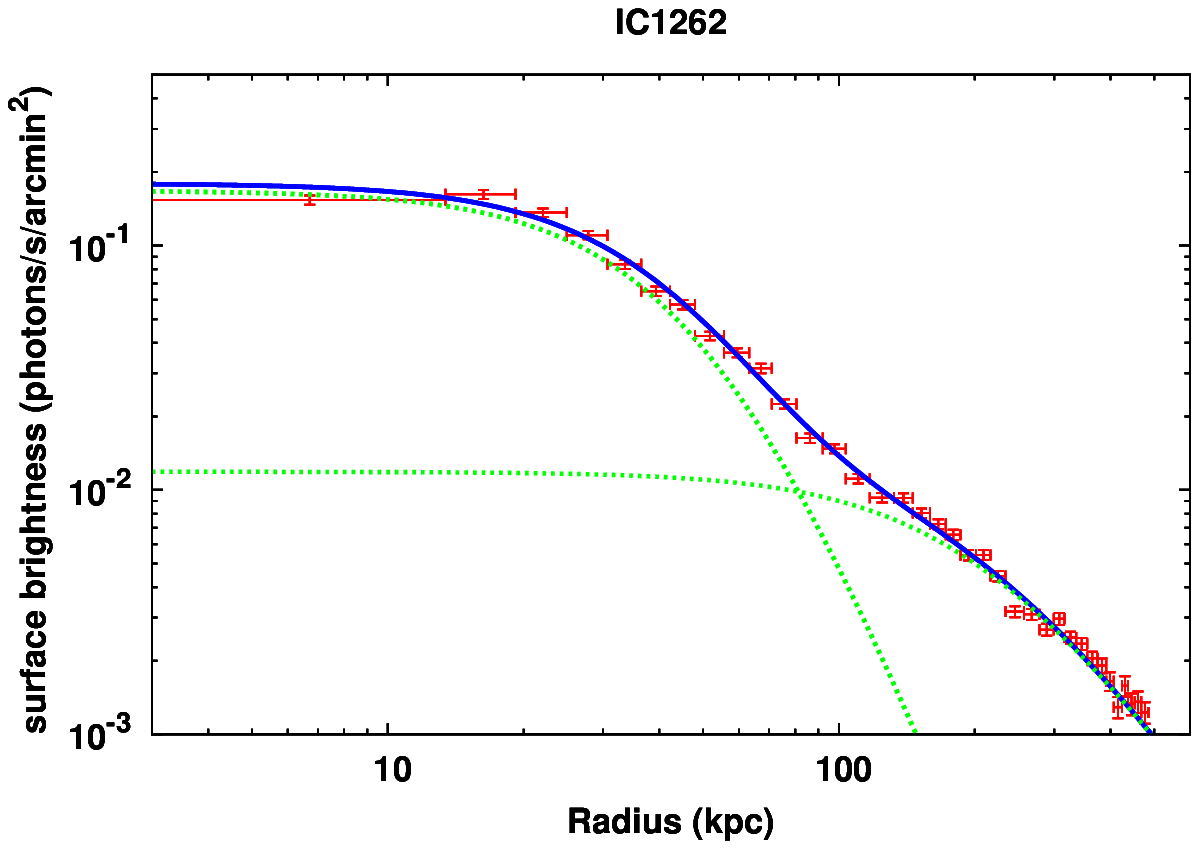,height=5cm,width=0.3\textwidth,angle=0}
}
\end{center}
\vspace{-20pt}
\caption{\footnotesize{\it Surface brightness profiles for CID28, HCG62, and IC1262.}}
\end{figure*}

%-----------------------------Figure End--------------------------------

%-----------------------------Figure Start------------------------------
\begin{figure*}[ht]
\begin{center}
\hbox{
\epsfig{figure=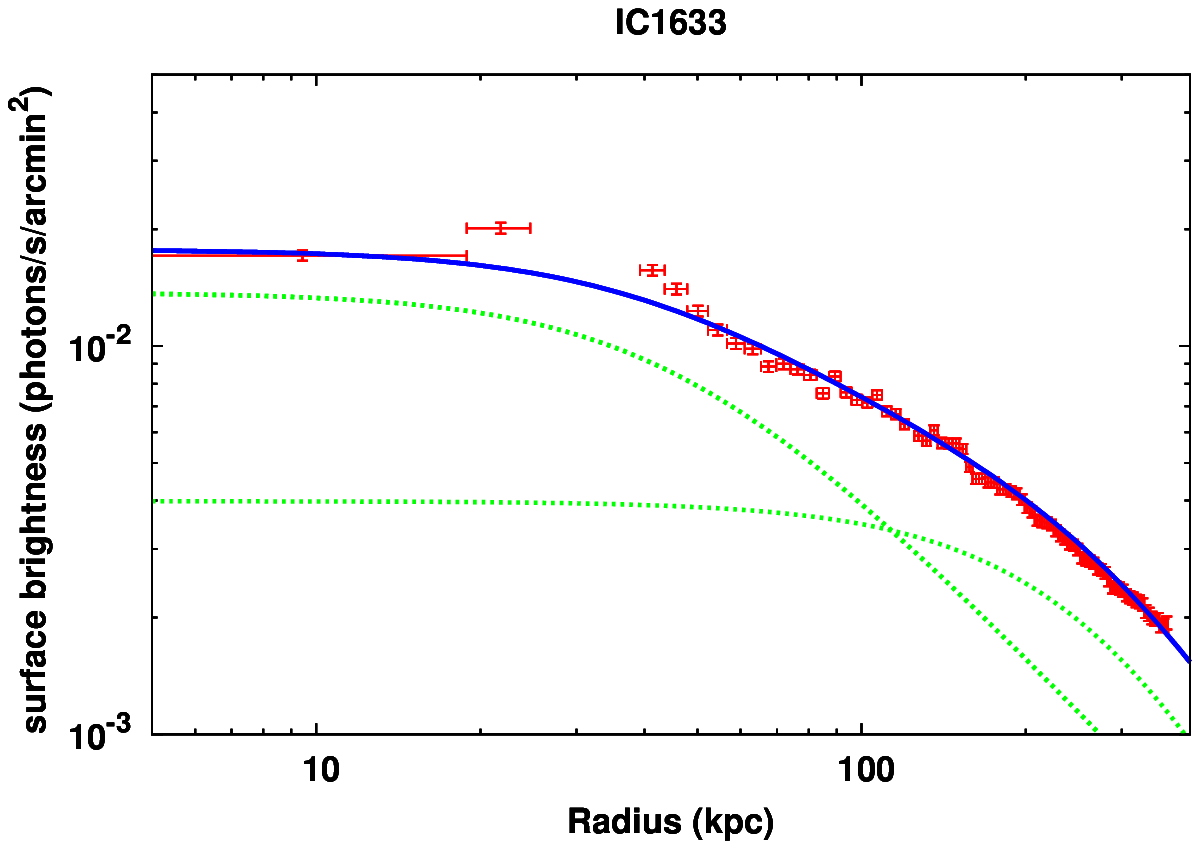,height=5cm,width=0.3\textwidth,angle=0}
\epsfig{figure=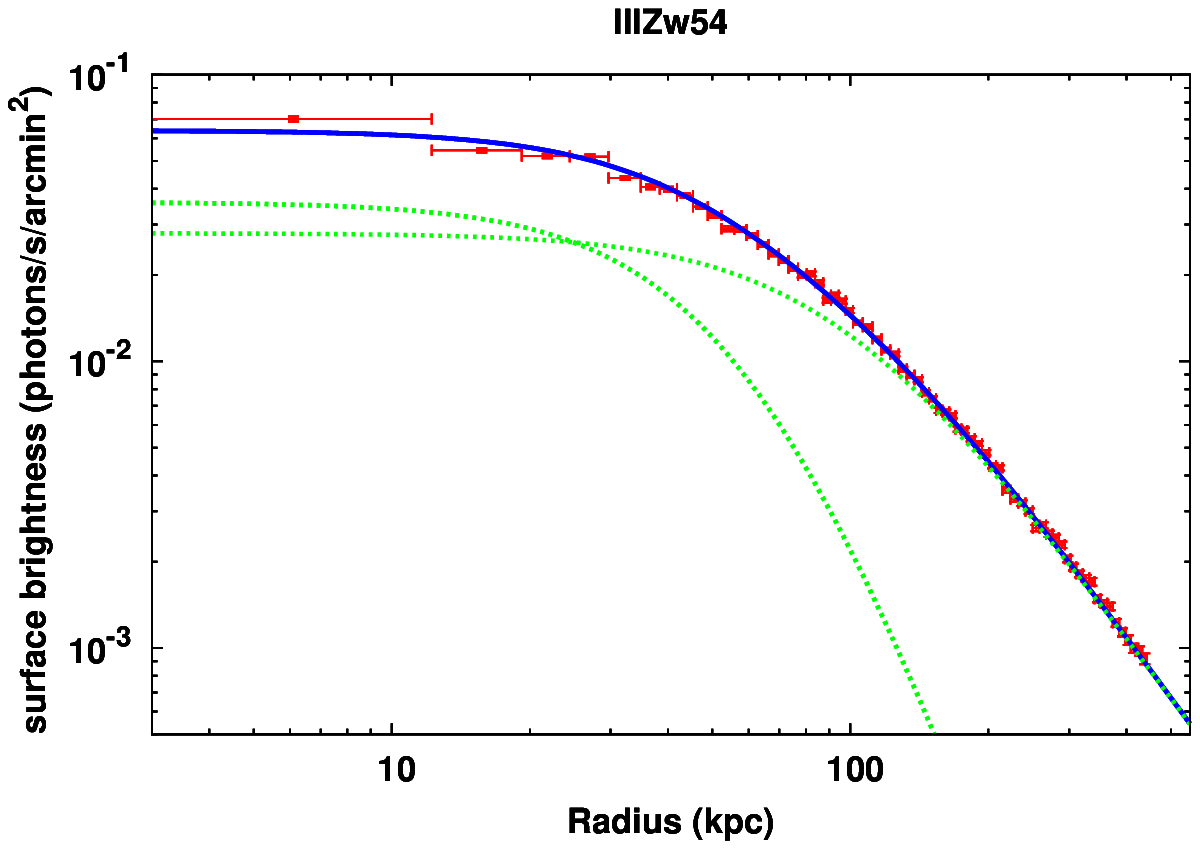,height=5cm,width=0.3\textwidth,angle=0}
\epsfig{figure=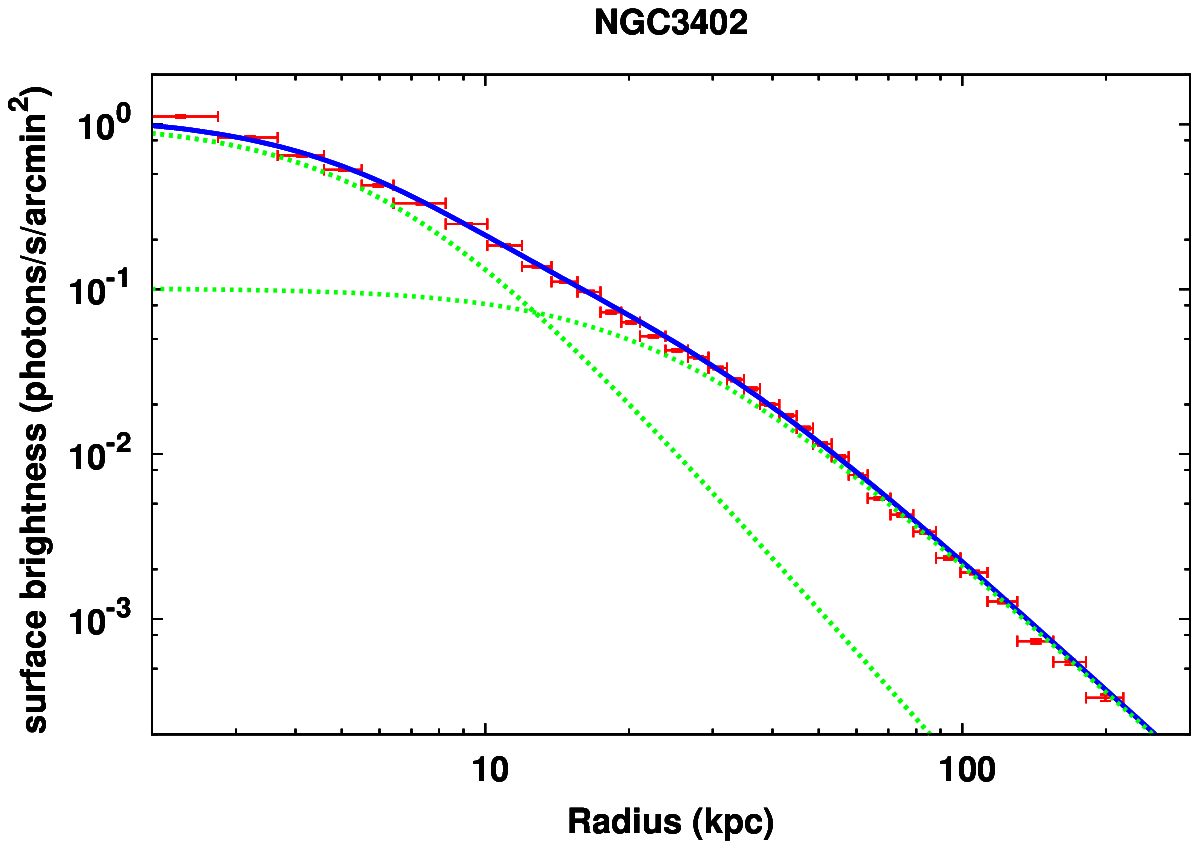,height=5cm,width=0.3\textwidth,angle=0}
}
\end{center}
\vspace{-20pt}
\caption{\footnotesize{\it Surface brightness profiles for IC1633, IIIZw054, and NGC3402.}}
\end{figure*}

%-----------------------------Figure End--------------------------------

%-----------------------------Figure Start------------------------------
\begin{figure*}[ht]
\begin{center}
\hbox{
\epsfig{figure=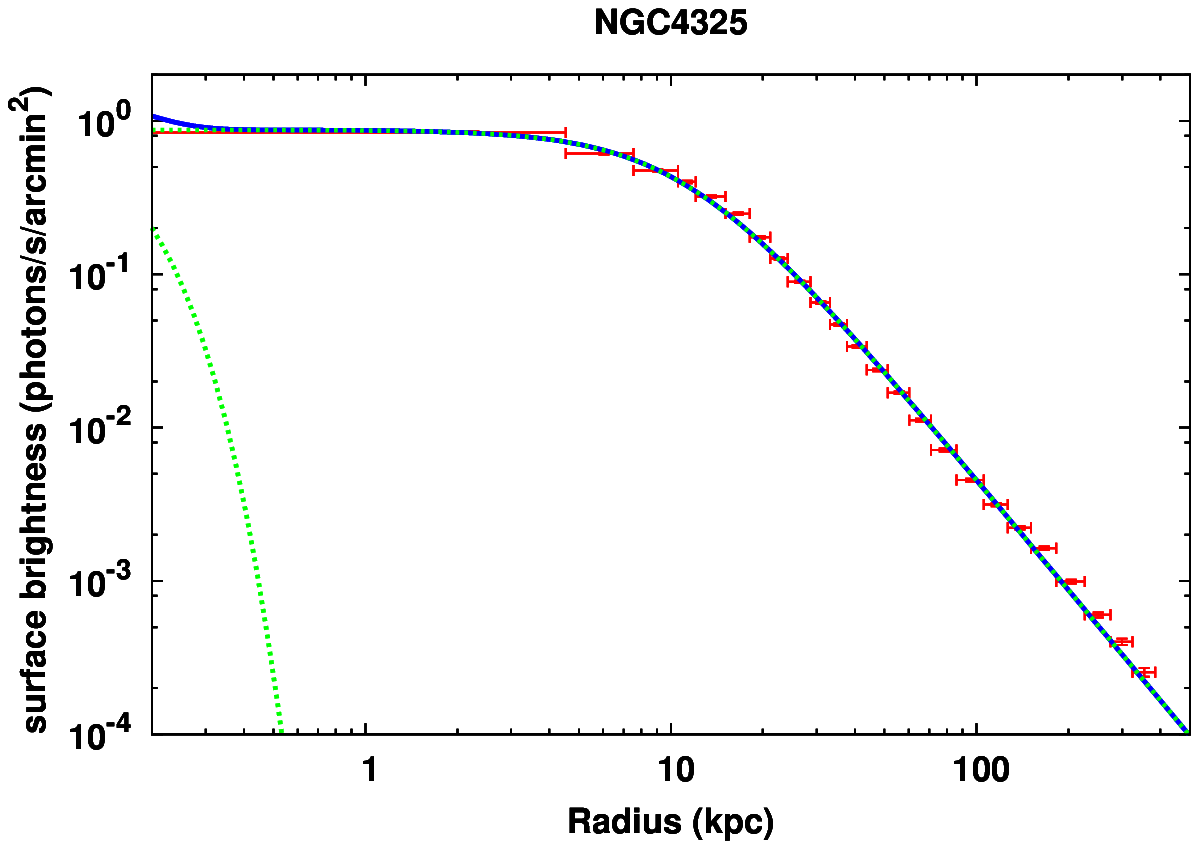,height=5cm,width=0.3\textwidth,angle=0}
\epsfig{figure=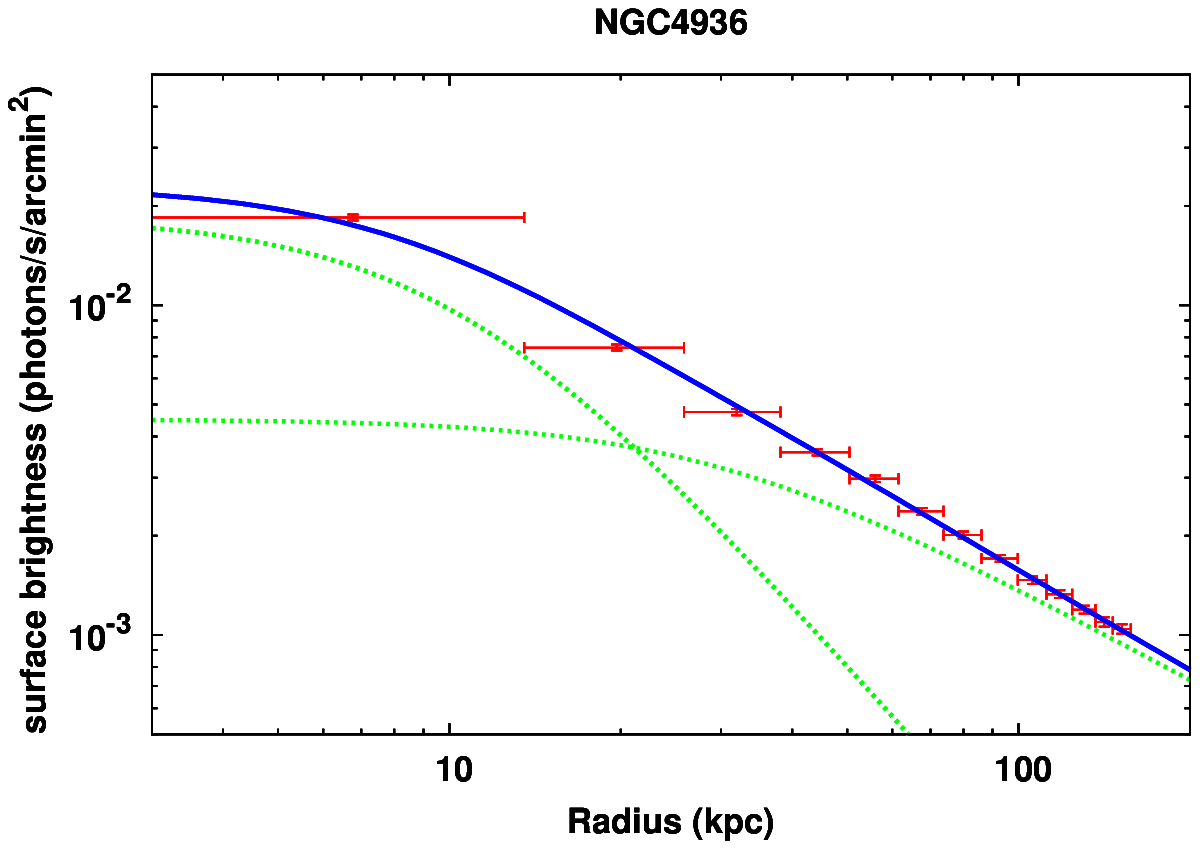,height=5cm,width=0.3\textwidth,angle=0}
\epsfig{figure=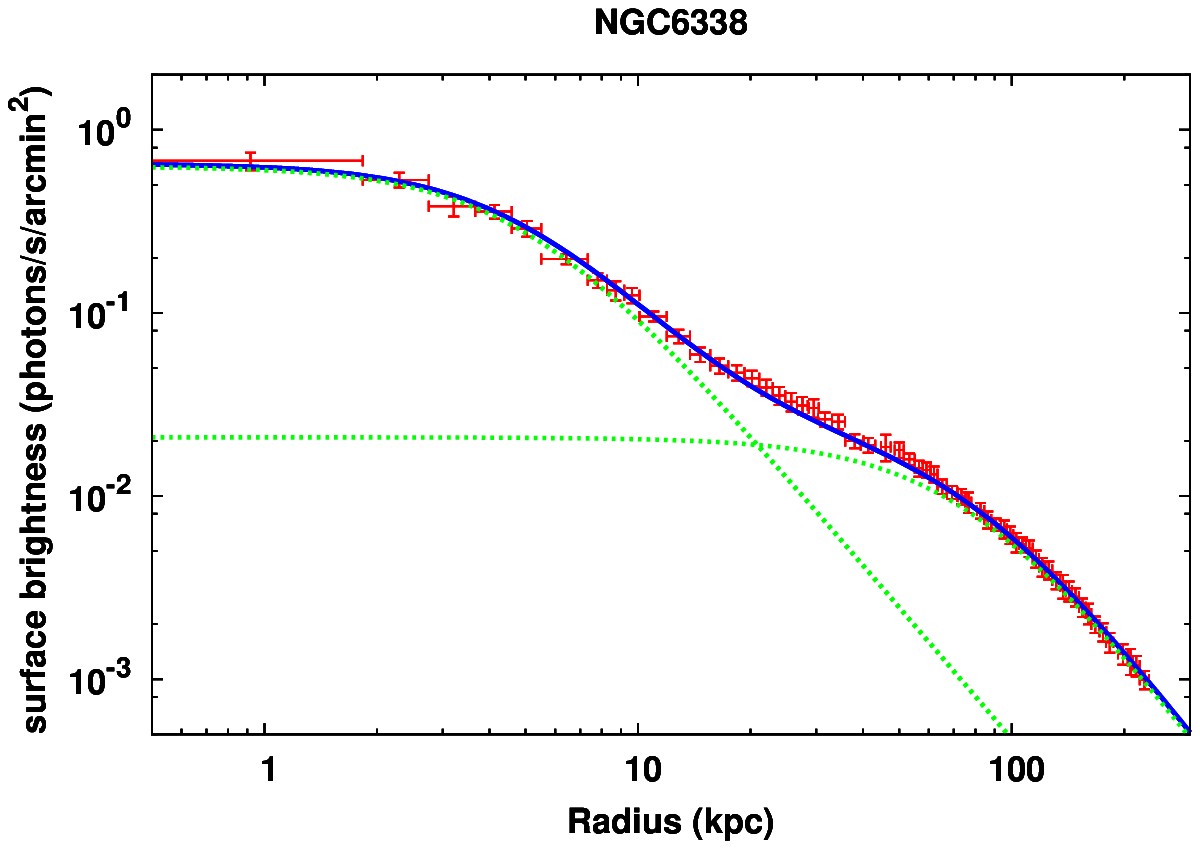,height=5cm,width=0.3\textwidth,angle=0}
}
\end{center}
\vspace{-20pt}
\caption{\footnotesize{\it Surface brightness profiles for NGC4325, NGC4936, and NGC6338.}}
\end{figure*}

%-----------------------------Figure End--------------------------------

%-----------------------------Figure Start------------------------------
\begin{figure*}[ht]
\begin{center}
\hbox{
\epsfig{figure=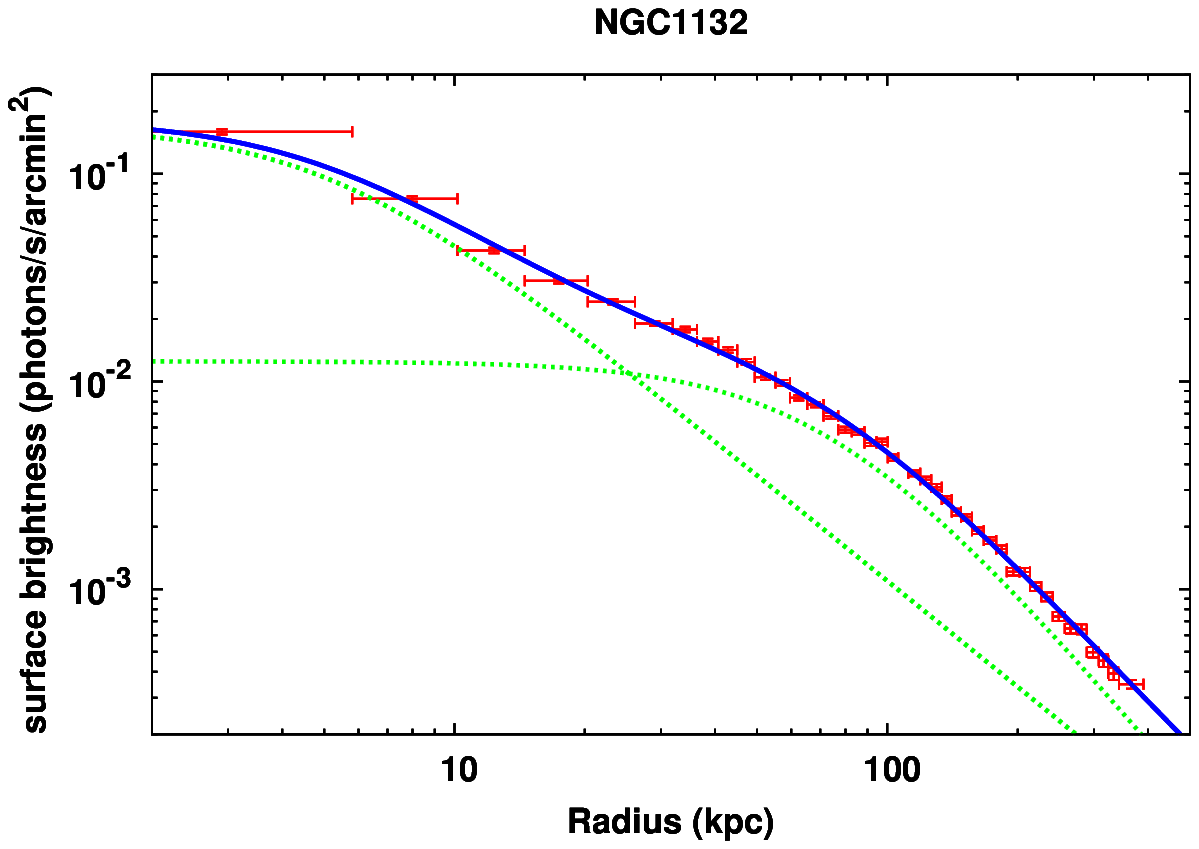,height=5cm,width=0.3\textwidth,angle=0}
\epsfig{figure=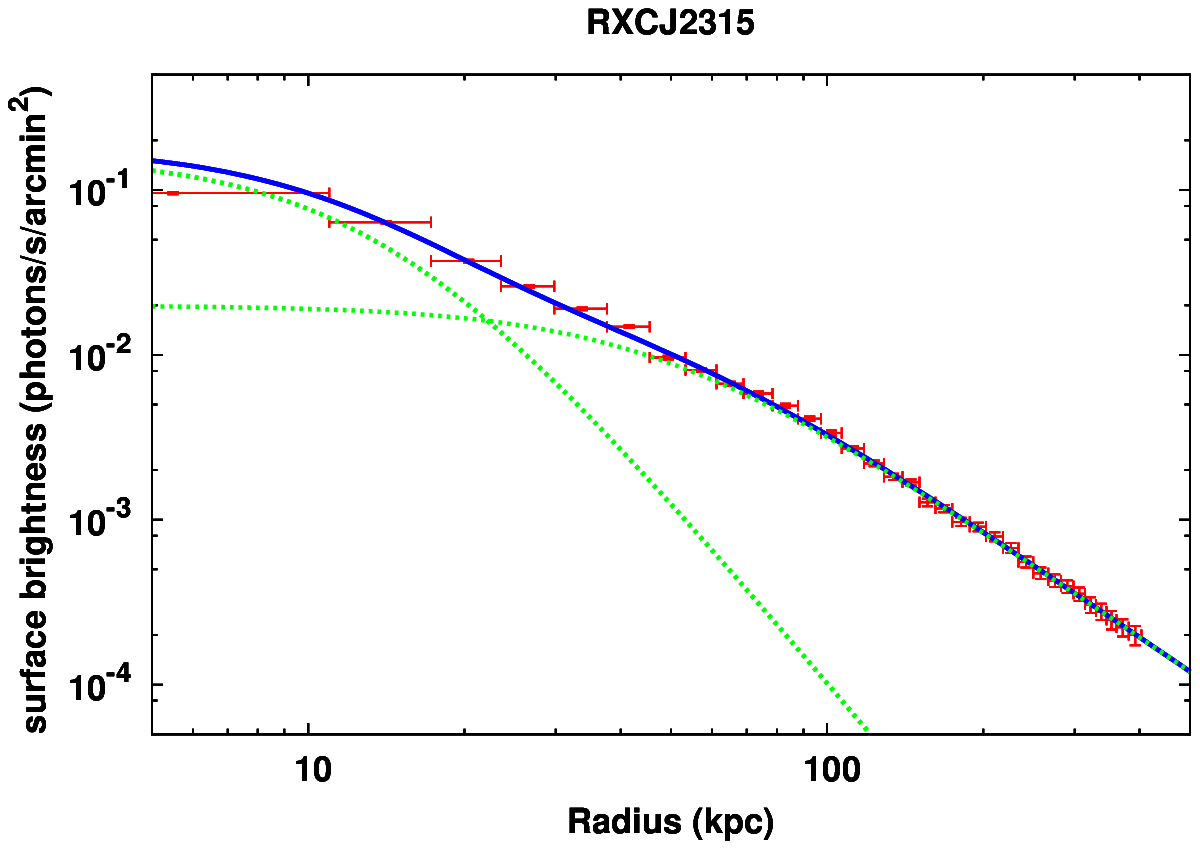,height=5cm,width=0.3\textwidth,angle=0}
\epsfig{figure=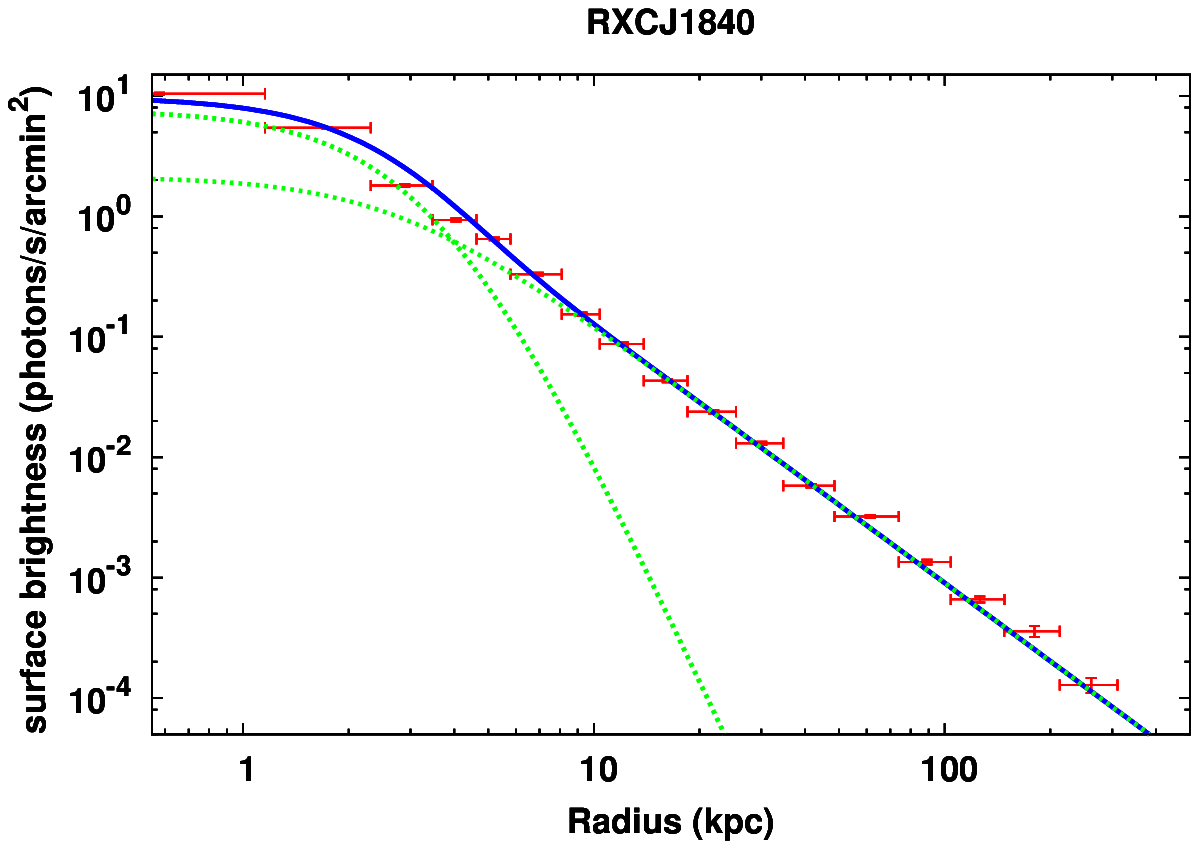,height=5cm,width=0.3\textwidth,angle=0}
}
\end{center}
\vspace{-20pt}
\caption{\footnotesize{\it Surface brightness profiles for NGC1132, RXCJ2315.7-0222, and RXCJ1840.6-7709.}}
\end{figure*}

%-----------------------------Figure End--------------------------------

%-----------------------------Figure Start------------------------------
\begin{figure*}[ht]
\begin{center}
\hbox{
\epsfig{figure=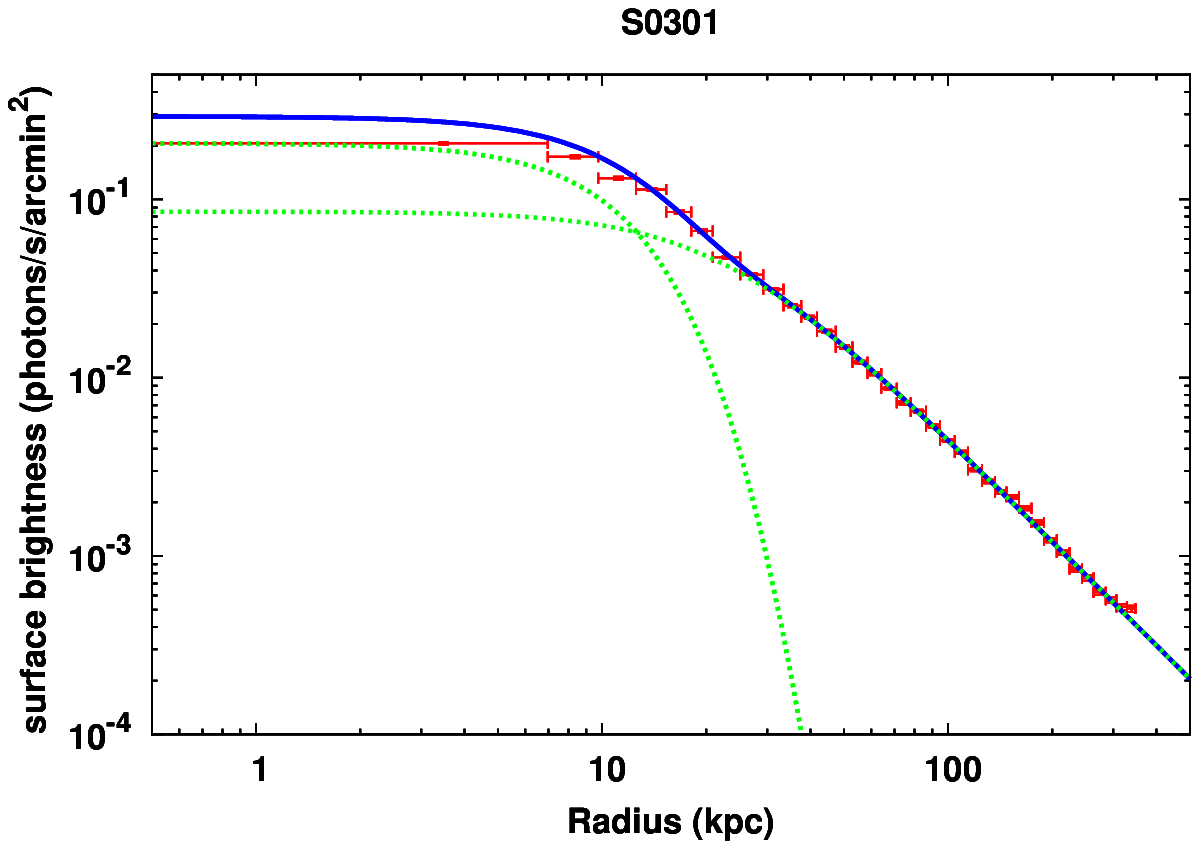,height=5cm,width=0.3\textwidth,angle=0}
\epsfig{figure=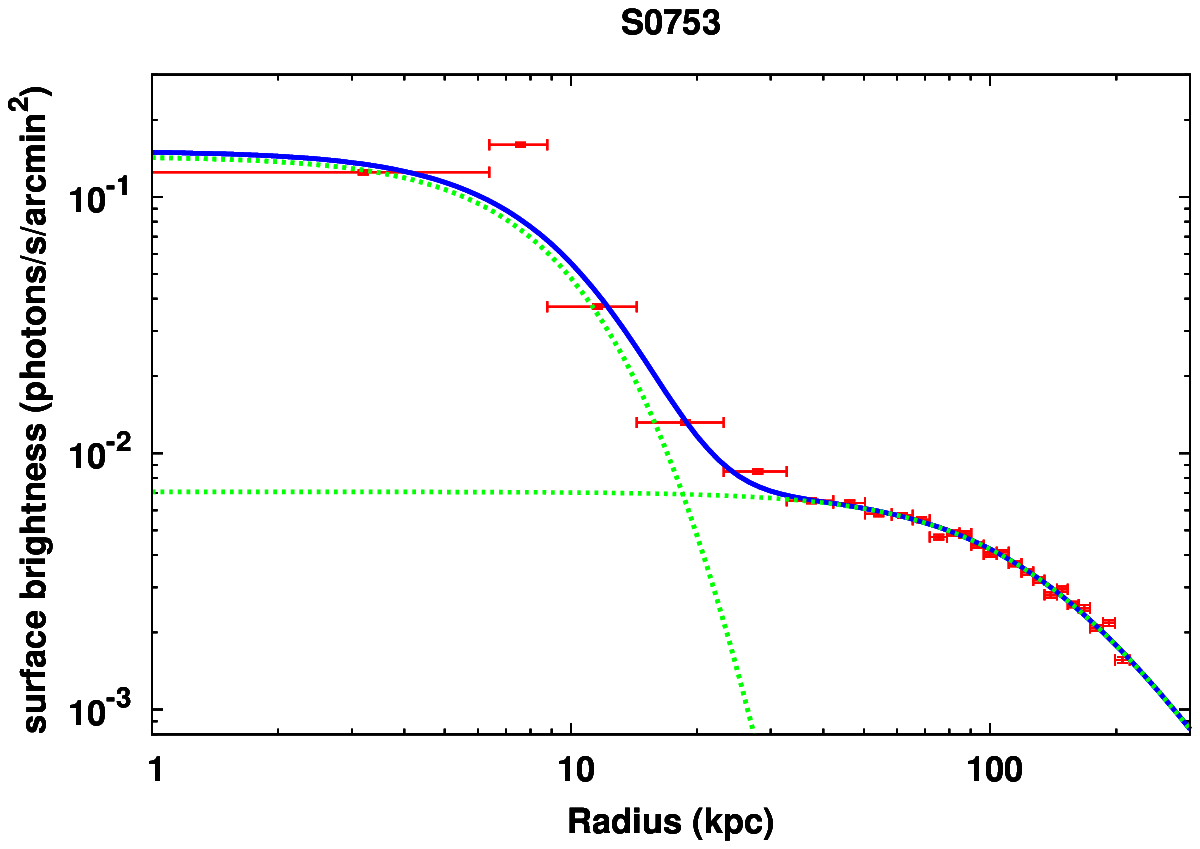,height=5cm,width=0.3\textwidth,angle=0}
\epsfig{figure=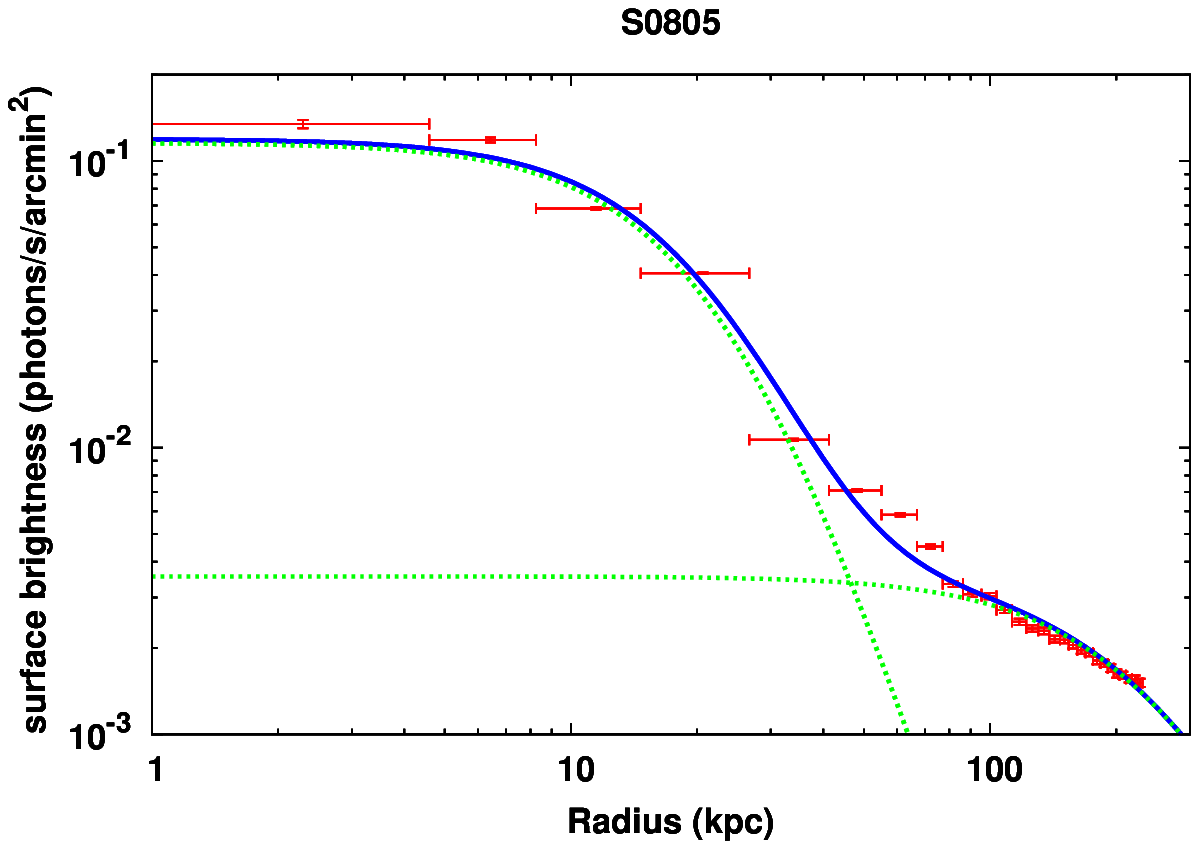,height=5cm,width=0.3\textwidth,angle=0}
}
\end{center}
\vspace{-20pt}
\caption{\footnotesize{\it Surface brightness profiles for S0301, S0753, and S0805.}}
\end{figure*}

%-----------------------------Figure End--------------------------------

%-----------------------------Figure Start------------------------------
\begin{figure*}[ht]
\begin{center}
\hbox{
\epsfig{figure=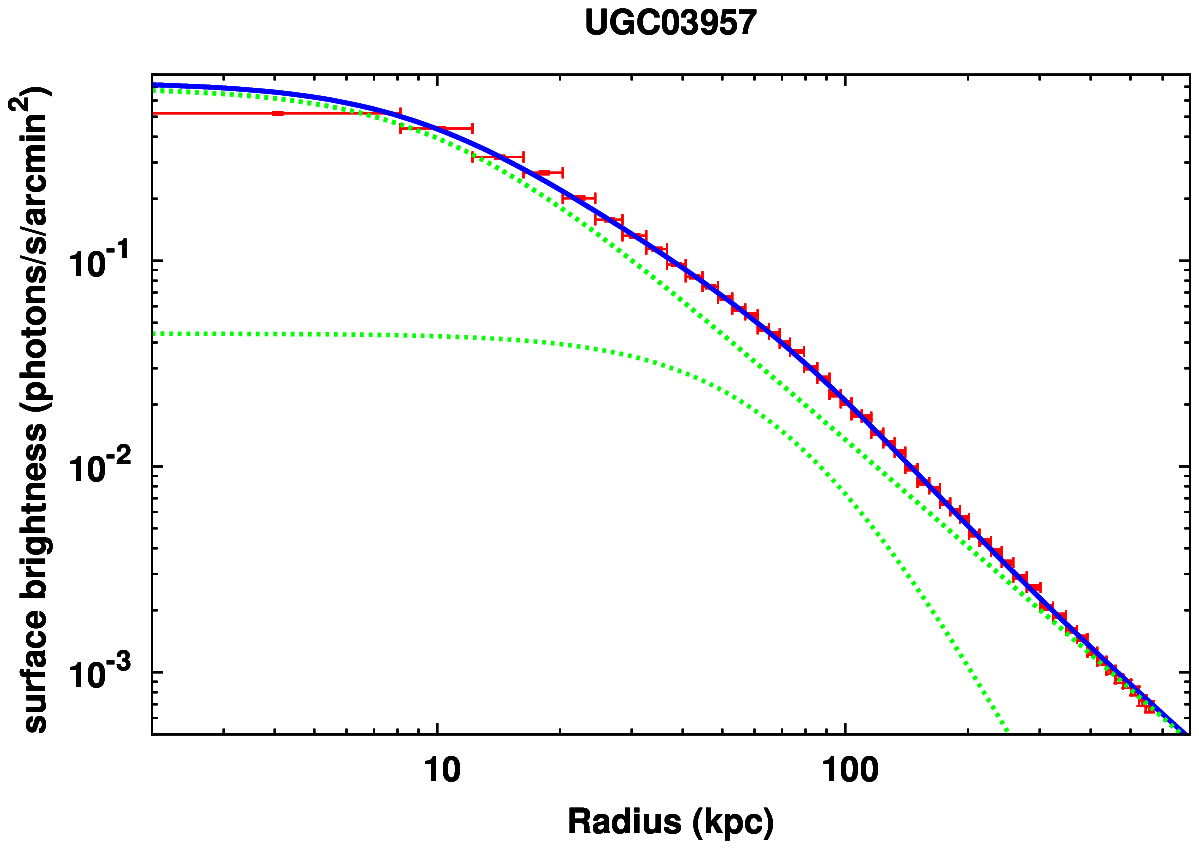,height=5cm,width=0.3\textwidth,angle=0}
\epsfig{figure=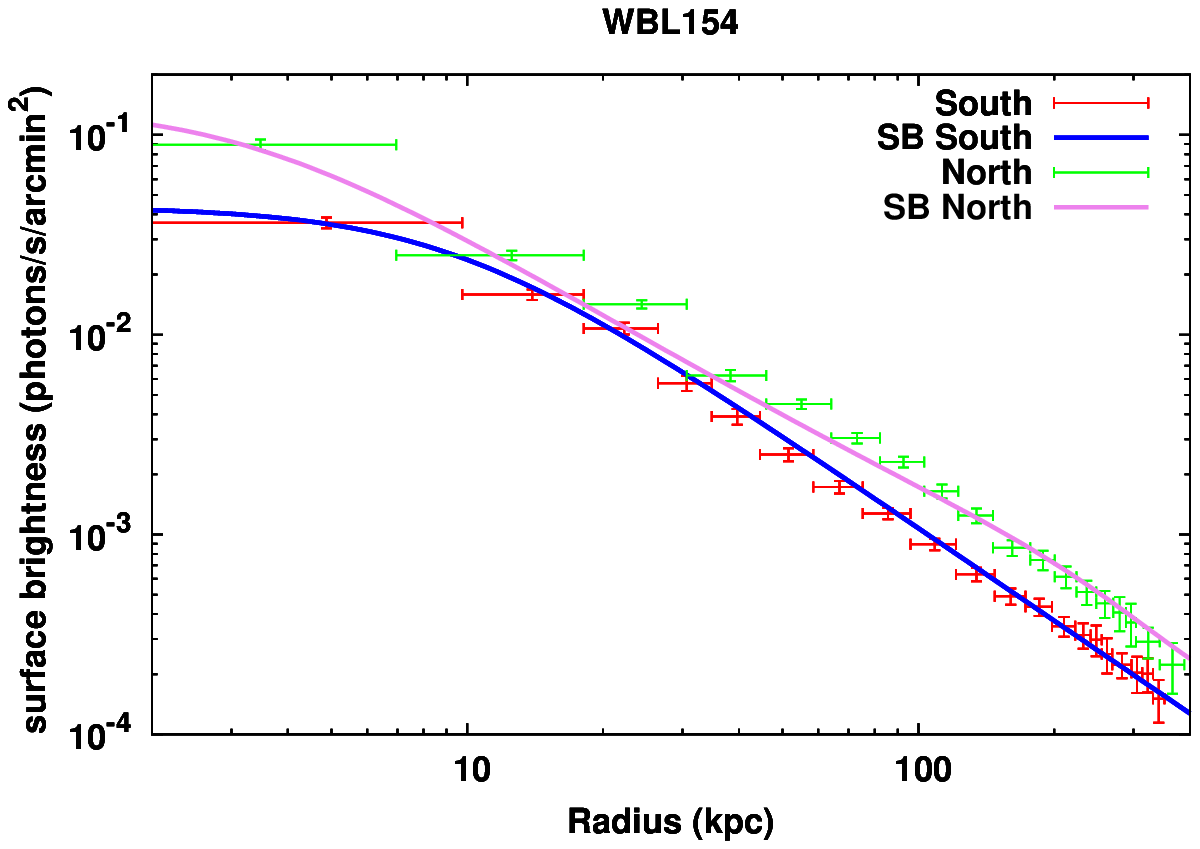,height=5cm,width=0.3\textwidth,angle=0}
}
\end{center}
\vspace{-20pt}
\caption{\footnotesize{\it Surface brightness profiles for UGC03957 and WBL154.}}
\end{figure*}

%-----------------------------Figure End--------------------------------

\begin{table}[t]
  \caption{Fit parameters with a double $\beta$ model.}
$$
\centering
\begin{array}{ccccccc}
\hline\hline
\noalign{\smallskip}
{\rm Name} &  {\rm S_{01}} & {\rm S_{02}} & {\rm \beta_1} & {\rm \beta_2} & {\rm r_{c,1}} & {\rm r_{c,2}} \\
     	   &  10^{-2} {\rm cts/s/armin^2} & 10^{-2} {\rm cts/s/armin^2} & & & {\rm kpc} & {\rm kpc} \\  
\hline
{\rm NGC4936} 	 	 & 1.9\pm0.1 & 0.5\pm0.1 & 0.494\pm0.051 & 0.320\pm0.013 & 11\pm1 & 28\pm2 \\ 
{\rm S0753}  		 & 195.0\pm34.0 & 0.7\pm0.0 & 1.948\pm0.342 & 0.583\pm0.085 & 21\pm3 & 140\pm24 \\ 
{\rm HCG62}  	  	 & 34.6\pm2.6 & 0.7\pm0.2 & 0.861\pm0.207 & 0.484\pm0.051 & 13\pm3 & 74\pm20 \\ 
{\rm S0805}  		 & 91.5\pm7.4 & 0.4\pm0.0 & 0.915\pm0.074 & 0.658\pm0.276 & 15\pm2 & 247\pm88 \\ 
{\rm NGC3402}		 & 100.2\pm3.2 & 9.7\pm1.6 & 0.720\pm0.007 & 0.607\pm0.020  & 7\pm1 & 24\pm6 \\ 
{\rm A194 } 		 & 0.2\pm0.0 & 0.1\pm0.0 & 0.405\pm0.097 & 0.737\pm0.103 & 93\pm23 & 422\pm81 \\ 
{\rm RXCJ1840.6-7709} & 995.0\pm47.5 & 18.3\pm10.0 & 0.813\pm0.071 & 0.518\pm0.009 & 3\pm1 & 8\pm6 \\ 
{\rm S0301  }		 & 8.0\pm2.0 & 22.5\pm6.4 & 0.490\pm0.005 & 2.544\pm1.565 & 23\pm4 & 30\pm25 \\ 
{\rm NGC1132 }	 	 & 16.2\pm0.8 & 1.2\pm0.1 & 0.475\pm0.007 & 0.612\pm0.020 & 6\pm1 & 84\pm6 \\ 
{\rm IC1633 }		 & 1.6\pm0.1 & 0.3\pm0.1 & 0.390\pm0.042 & 0.710\pm0.020 & 42\pm5 & 327\pm29 \\ 
{\rm NGC4325 }	 	 & 87.7\pm14.2 & 87.4\pm1.1 & 6.194\pm1.177 & 0.566\pm0.003 & 1\pm1 & 11\pm1 \\ 
{\rm RXCJ2315.7-0222 } & 8.8\pm0.3 & 2.0\pm0.0 & 0.959\pm0.073 & 0.526\pm0.063 & 23\pm3 & 47\pm8 \\ 
{\rm NGC6338  }		 & 65.0\pm4.8 & 2.9\pm1.0 & 0.640\pm0.154 & 0.573\pm0.061 & 6\pm2 & 59\pm18 \\ 
{\rm IIIZw054	}	 & 3.6\pm0.6 & 2.8\pm1.1 & 0.907\pm0.261 & 0.540\pm0.010 & 63\pm24 & 96\pm22 \\ 
{\rm IC1262    }     & 16.8\pm1.2 & 1.1\pm0.2 & 0.891\pm0.120 & 0.600\pm0.020 & 50\pm6 & 221\pm21 \\ 
{\rm AWM4  	 }	 	 & 10.9\pm3.0 & 0.7\pm0.1 & 0.600\pm0.020 & 0.990\pm0.090 & 36\pm5 & 247\pm33 \\ 
{\rm CID28 	}		 & 5.9\pm2.0 & 1.4\pm0.1 & 0.590\pm0.020 & 0.580\pm0.010 & 24\pm2 & 113\pm4 \\ 
{\rm UGC03957 	}	 & 67.7\pm3.3 & 4.4\pm0.4 & 0.457\pm0.007 & 0.776\pm0.025 & 10\pm2 & 78\pm6 \\ 
\hline
\end{array}
\label{tab:SB}
$$
\tablefoot{
In some cases the errors on the amplitudes ${\rm S_{01}}$ and ${\rm S_{02}}$ were smaller than $\unit[0.1] \times 10^{-2} {\rm cts/s/armin^2}$.
}
\end{table}

\clearpage
\section{Temperature profiles} \label{kTprofiles}

%-----------------------------Figure Start------------------------------
\begin{figure*}[ht]
\begin{center}
\hbox{
\epsfig{figure=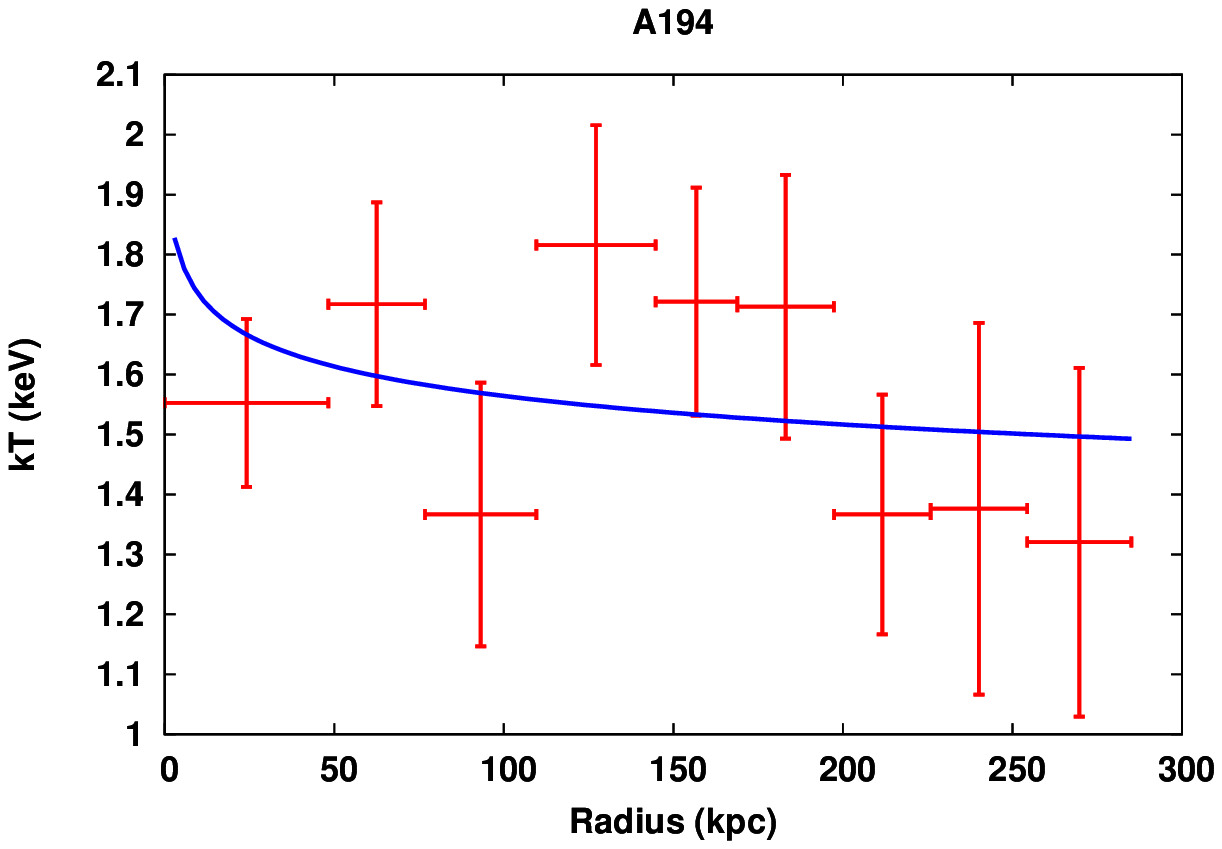,height=5cm,width=0.3\textwidth,angle=0}
\epsfig{figure=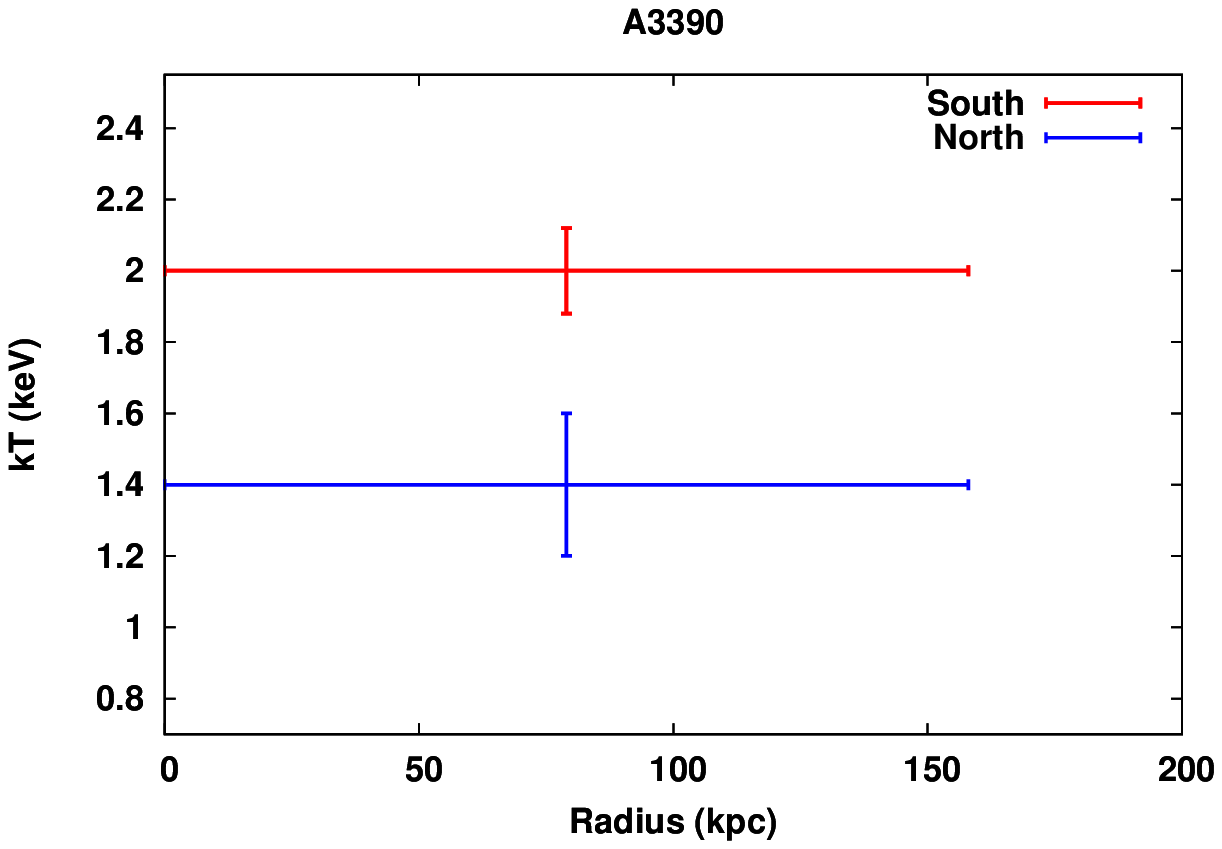,height=5cm,width=0.3\textwidth,angle=0}
\epsfig{figure=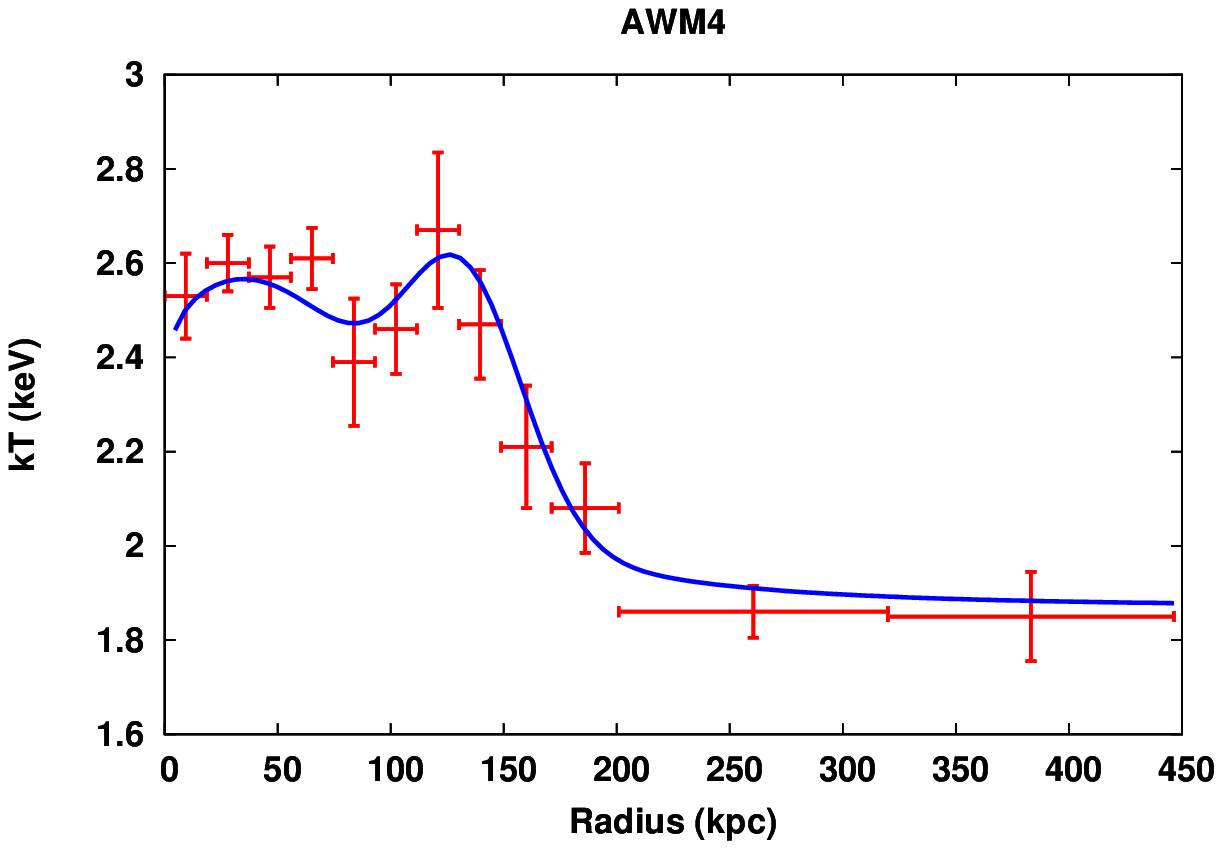,height=5cm,width=0.3\textwidth,angle=0}
}
\end{center}
\vspace{-20pt}
\caption{\footnotesize{\it Temperature profiles for A194, A3390, and AWM4.}}
\end{figure*}

%-----------------------------Figure End--------------------------------

%-----------------------------Figure Start------------------------------
\begin{figure*}[ht]
\begin{center}
\hbox{
\epsfig{figure=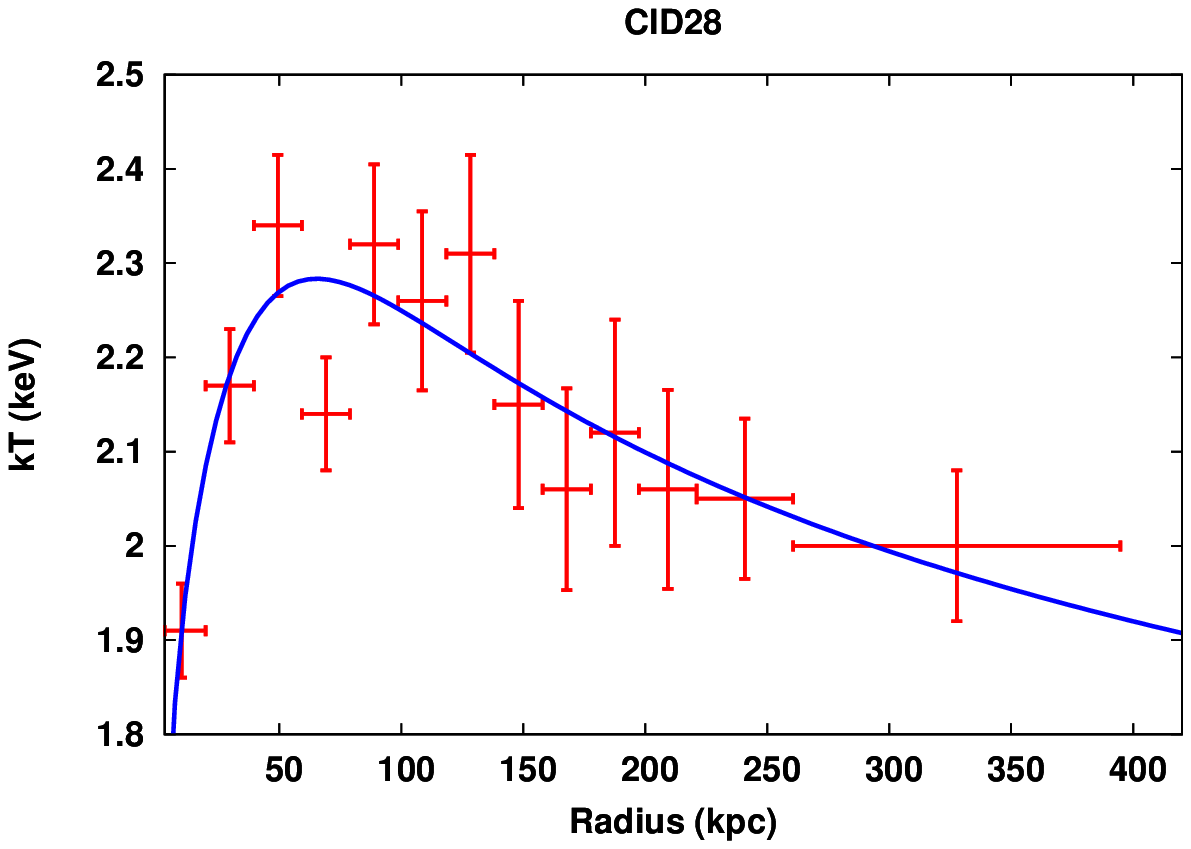,height=5cm,width=0.3\textwidth,angle=0}
\epsfig{figure=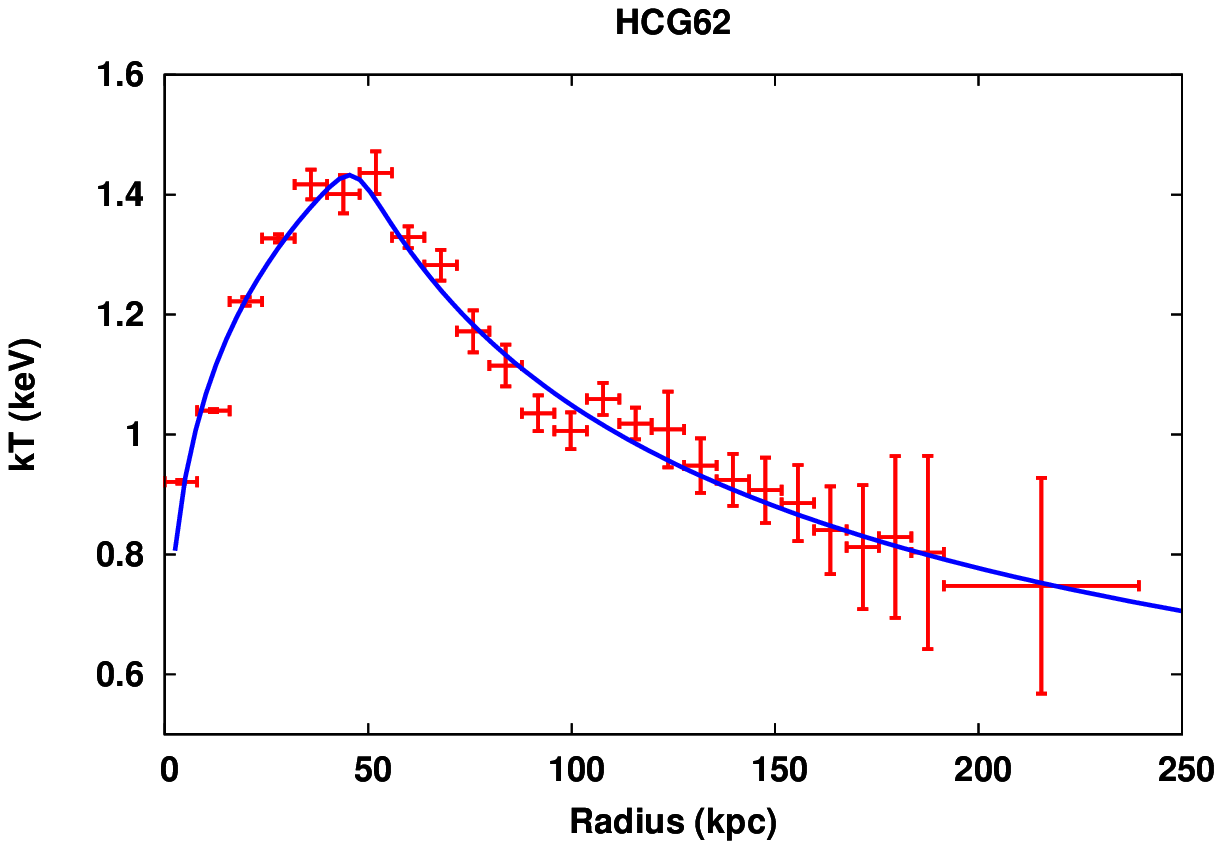,height=5cm,width=0.3\textwidth,angle=0}
\epsfig{figure=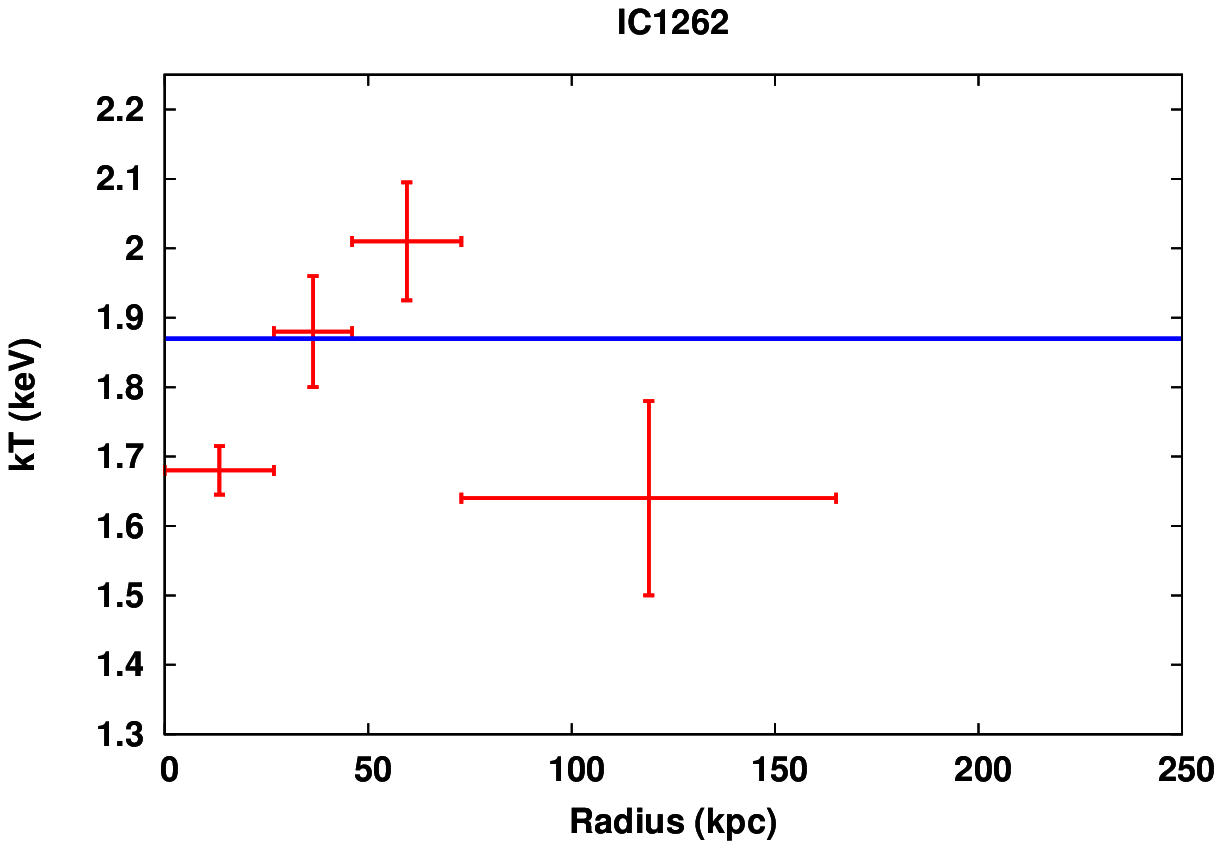,height=5cm,width=0.3\textwidth,angle=0}
}
\end{center}
\vspace{-20pt}
\caption{\footnotesize{\it Temperature profiles for CID28, HCG62 and IC1262.}}
\end{figure*}

%-----------------------------Figure End--------------------------------

%-----------------------------Figure Start------------------------------
\begin{figure*}[ht]
\begin{center}
\hbox{
\epsfig{figure=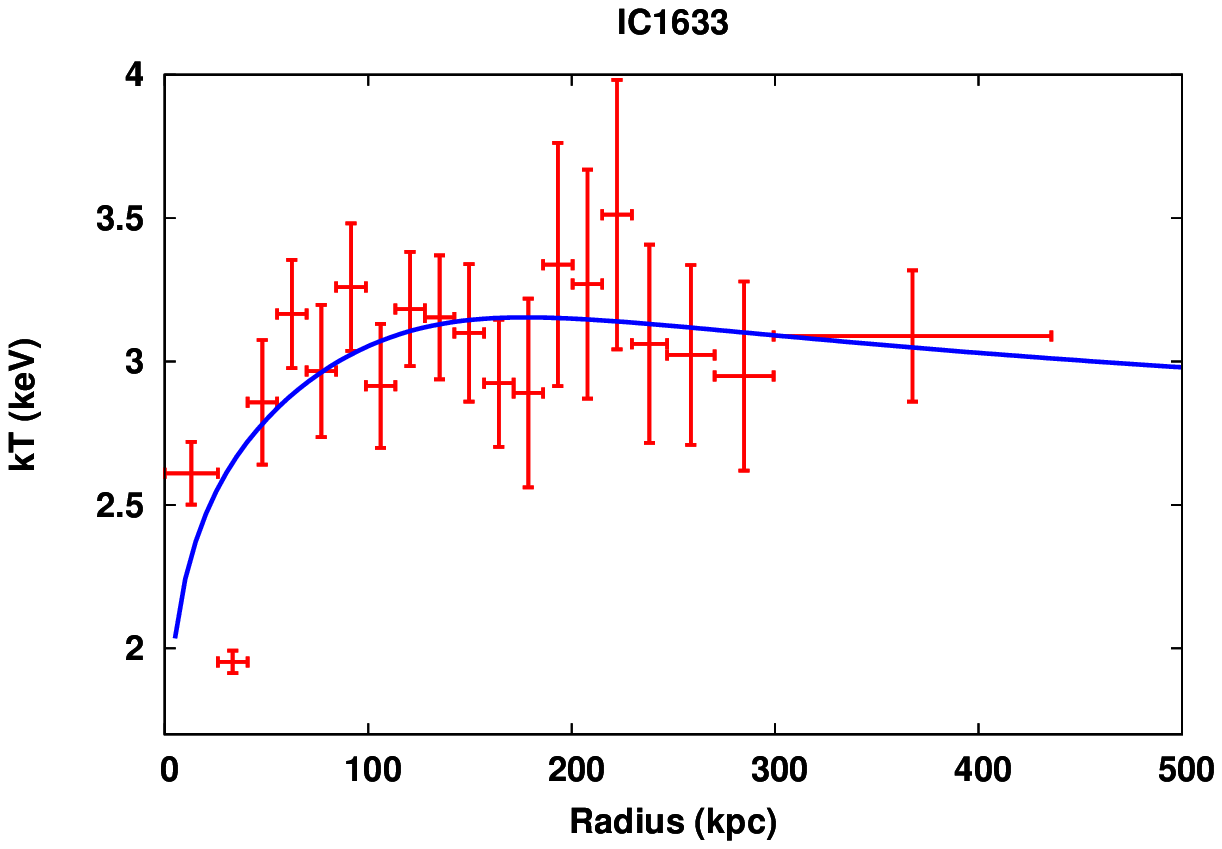,height=5cm,width=0.3\textwidth,angle=0}
\epsfig{figure=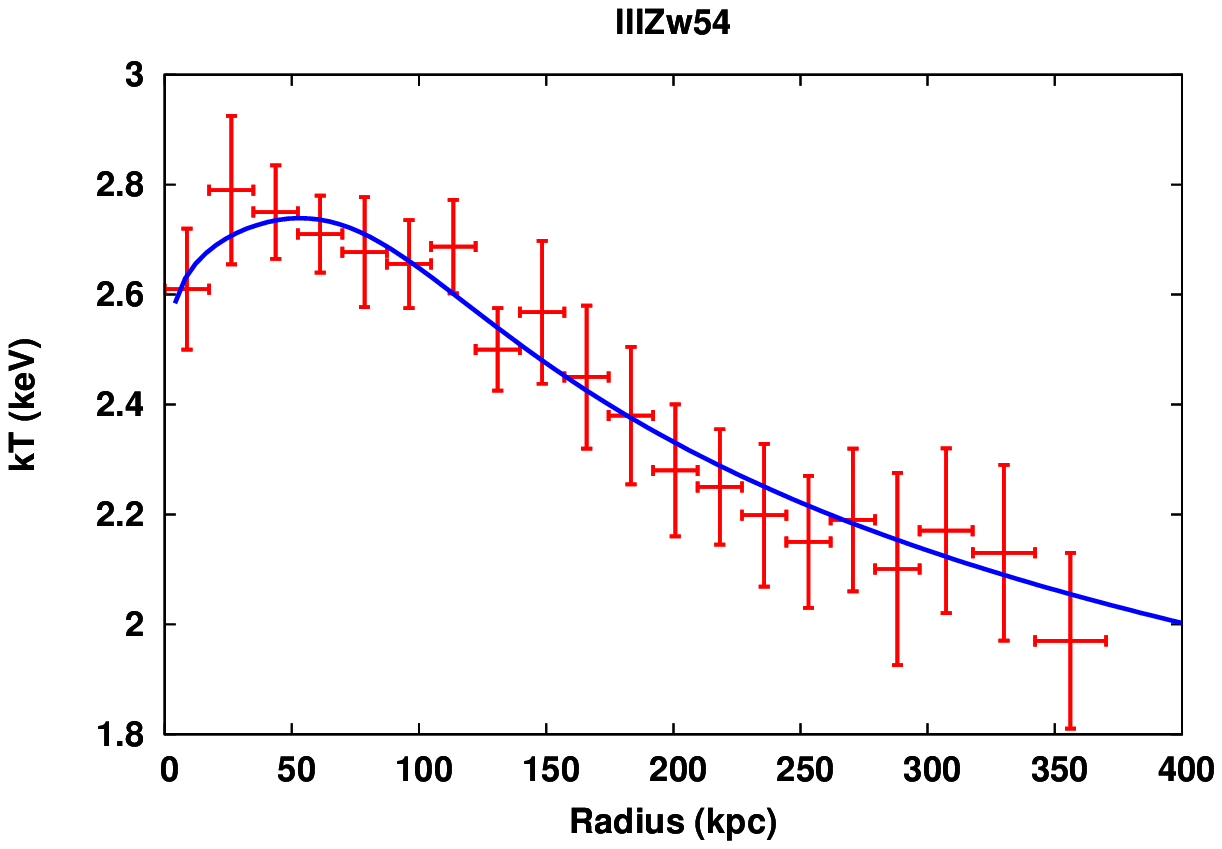,height=5cm,width=0.3\textwidth,angle=0}
\epsfig{figure=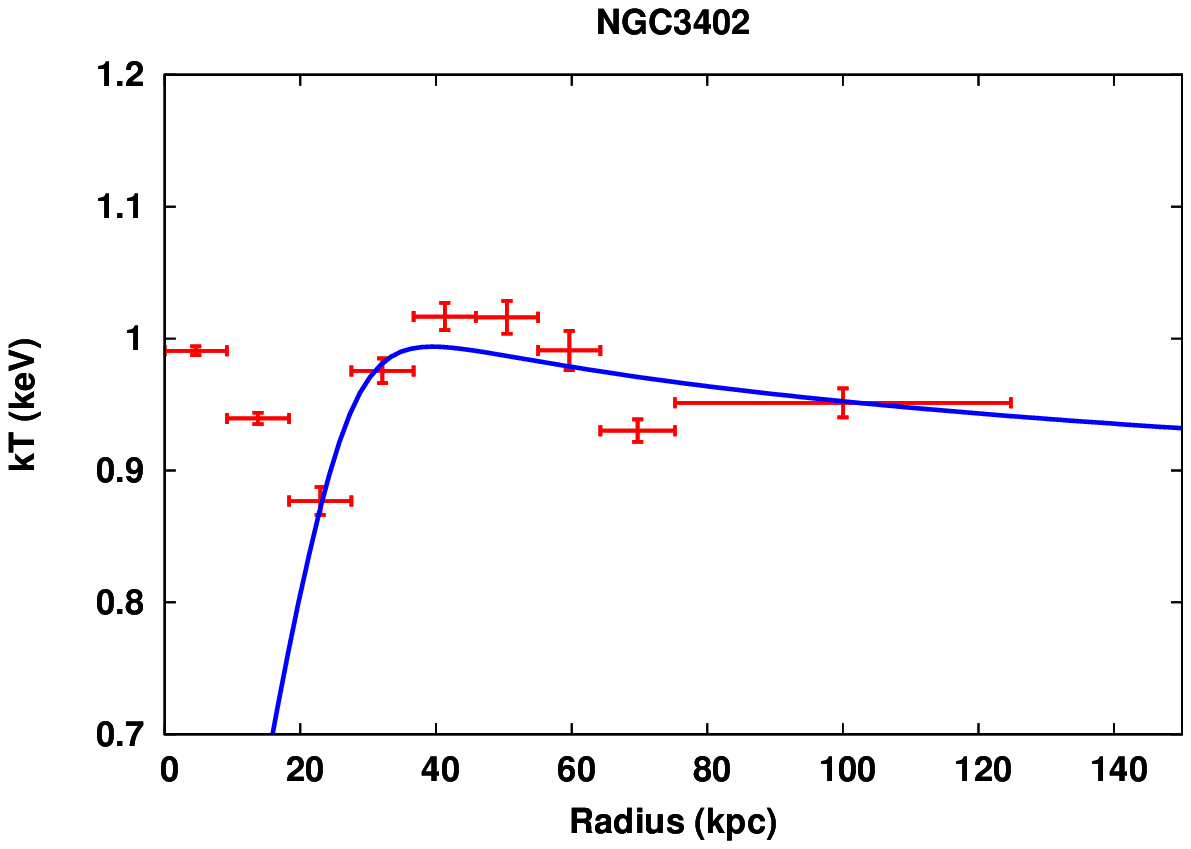,height=5cm,width=0.3\textwidth,angle=0}
}
\end{center}
\vspace{-20pt}
\caption{\footnotesize{\it Temperature profiles for IC1633, IIIZw54, and NGC3402.}}
\end{figure*}

%-----------------------------Figure End--------------------------------

%-----------------------------Figure Start------------------------------
\begin{figure*}[ht]
\begin{center}
\hbox{
\epsfig{figure=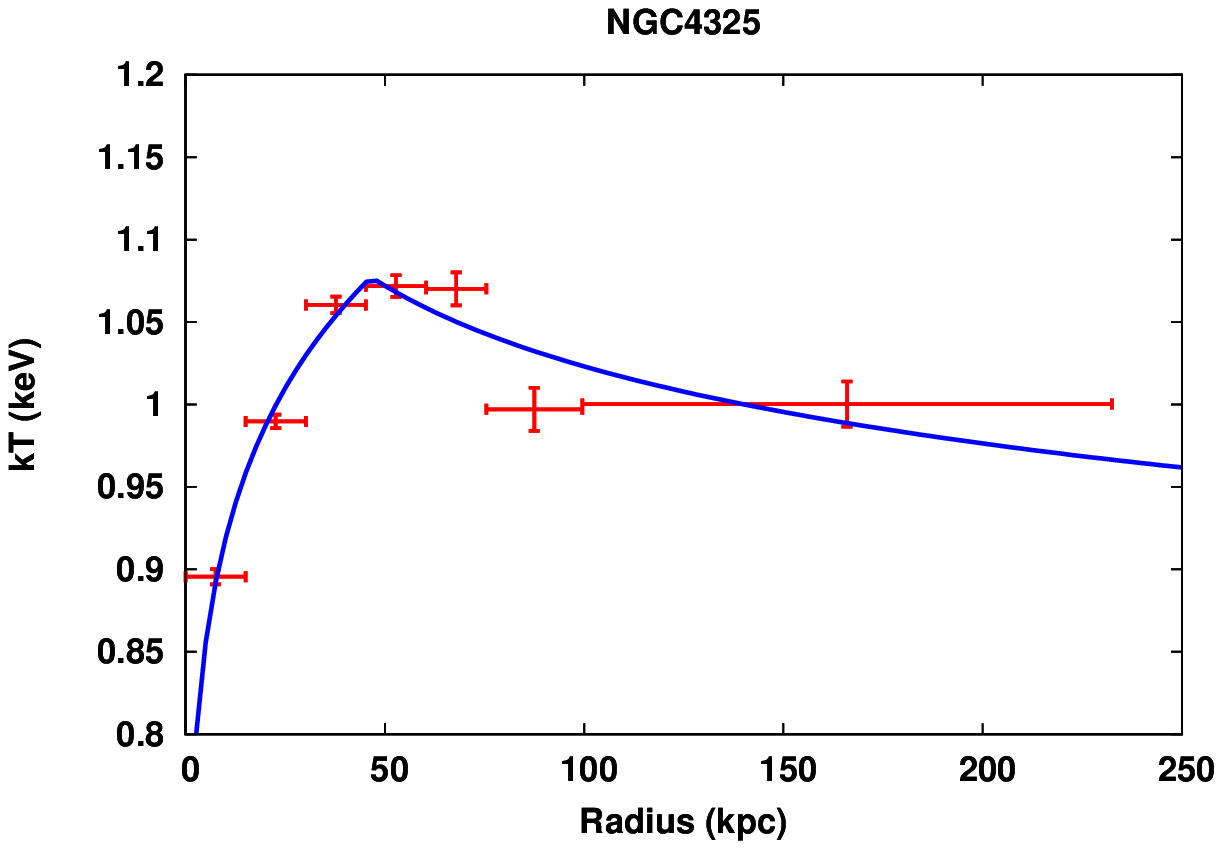,height=5cm,width=0.3\textwidth,angle=0}
\epsfig{figure=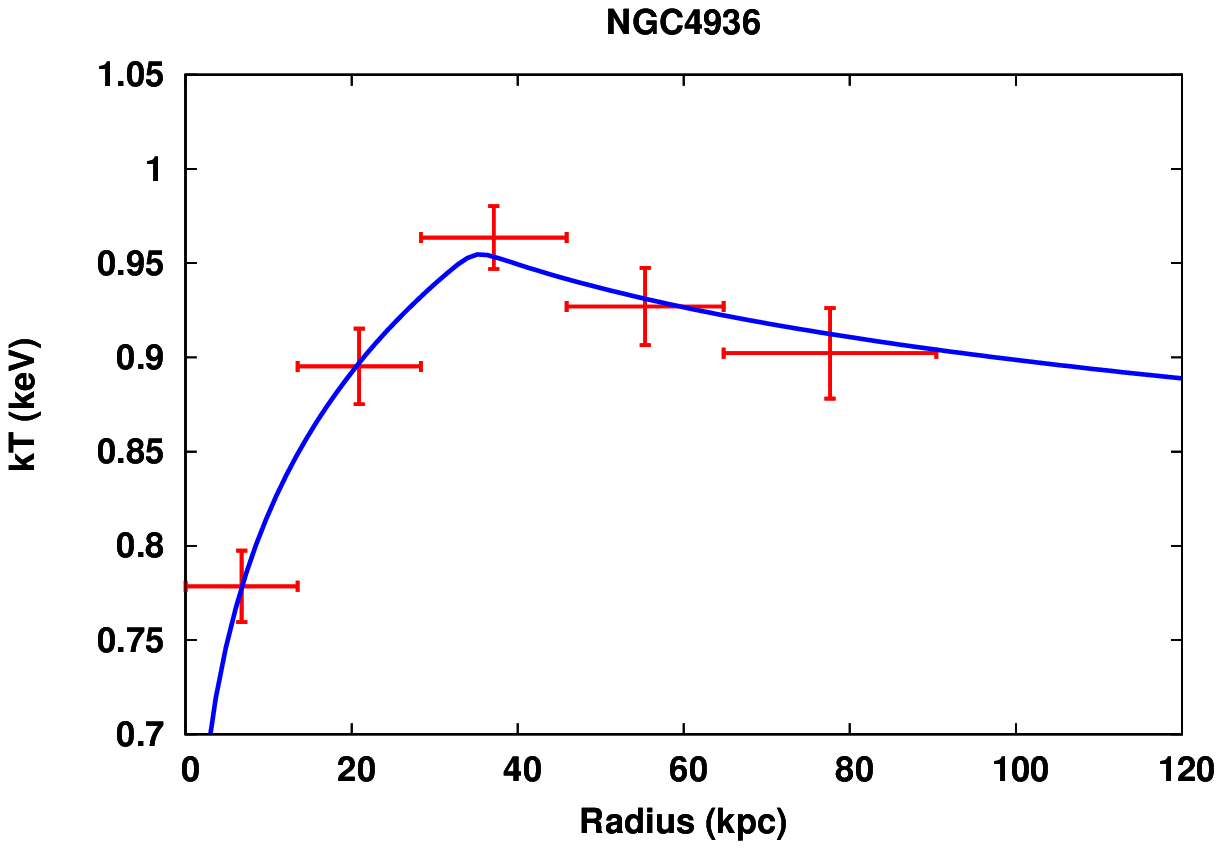,height=5cm,width=0.3\textwidth,angle=0}
\epsfig{figure=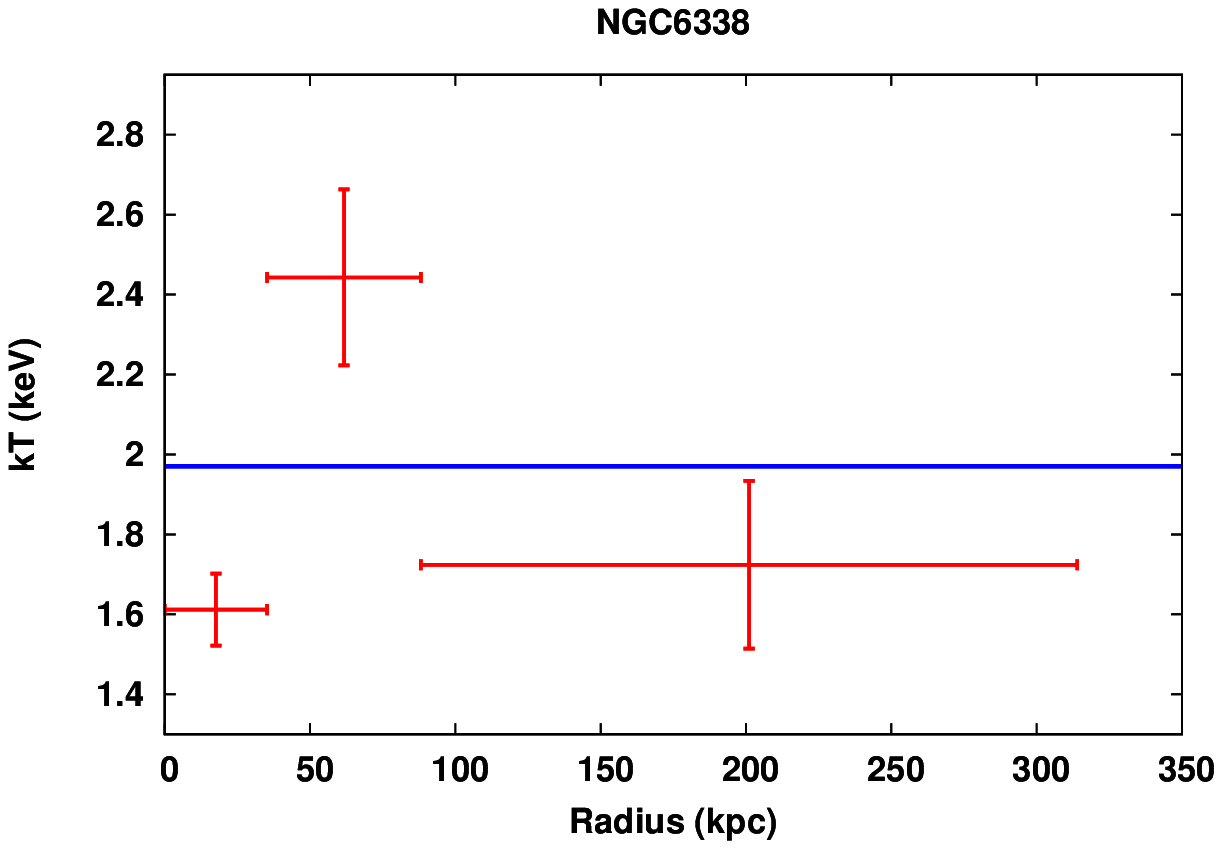,height=5cm,width=0.3\textwidth,angle=0}
}
\end{center}
\vspace{-20pt}
\caption{\footnotesize{\it Temperature profiles for NGC4325, NGC4936, and NGC6338.}}
\end{figure*}

%-----------------------------Figure End--------------------------------

%-----------------------------Figure Start------------------------------
\begin{figure*}[ht]
\begin{center}
\hbox{
\epsfig{figure=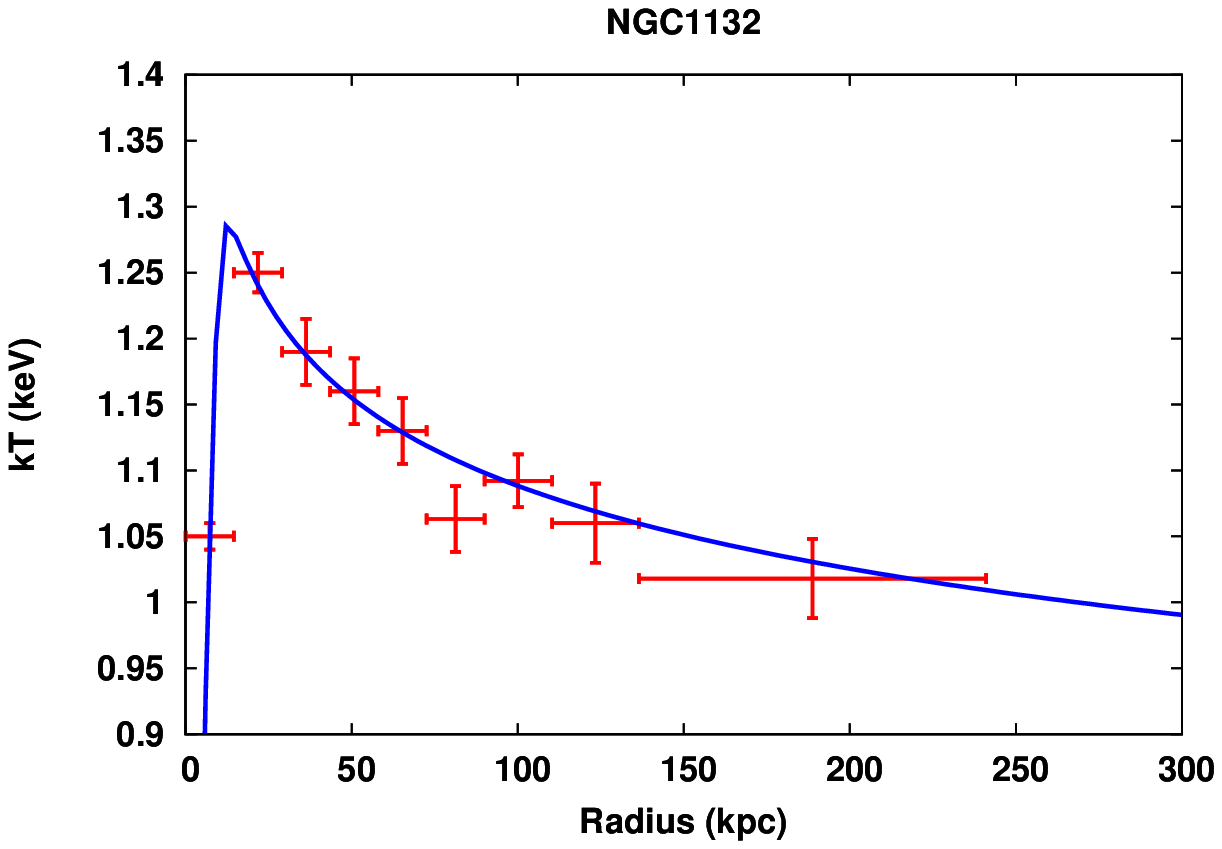,height=5cm,width=0.3\textwidth,angle=0}
\epsfig{figure=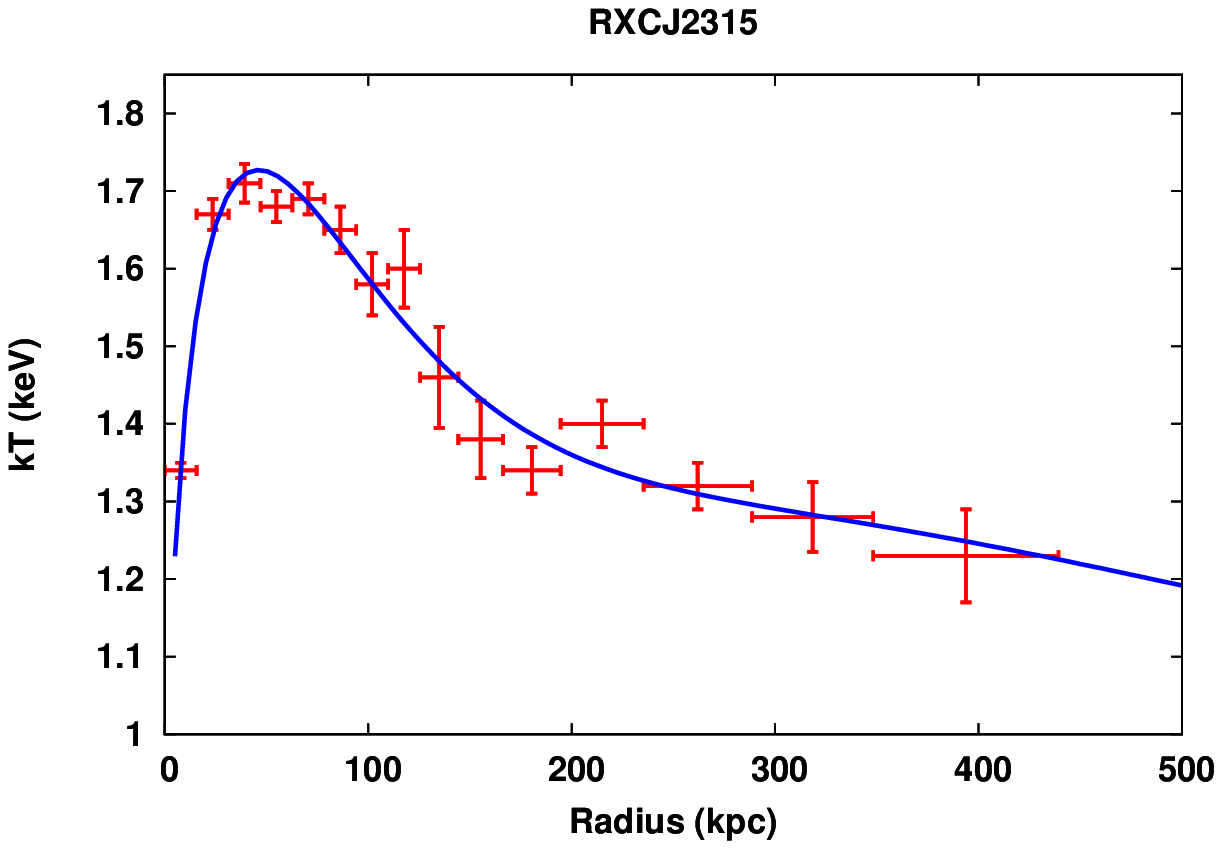,height=5cm,width=0.3\textwidth,angle=0}
\epsfig{figure=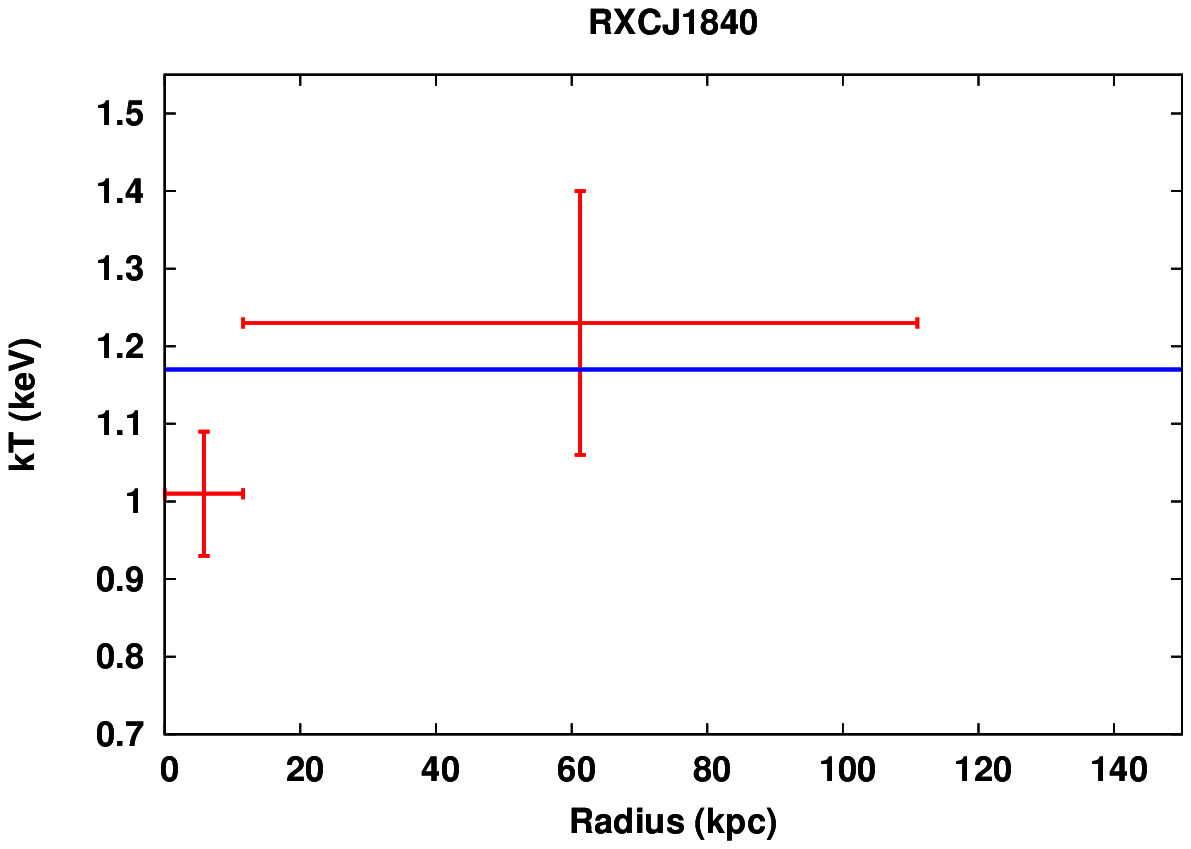,height=5cm,width=0.3\textwidth,angle=0}
}
\end{center}
\vspace{-20pt}
\caption{\footnotesize{\it Temperature profiles for NGC1132, RXCJ2315.7-0222, and RXCJ1840.6-7709.}}
\end{figure*}

%-----------------------------Figure End--------------------------------

%-----------------------------Figure Start------------------------------
\begin{figure*}[ht]
\begin{center}
\hbox{
\epsfig{figure=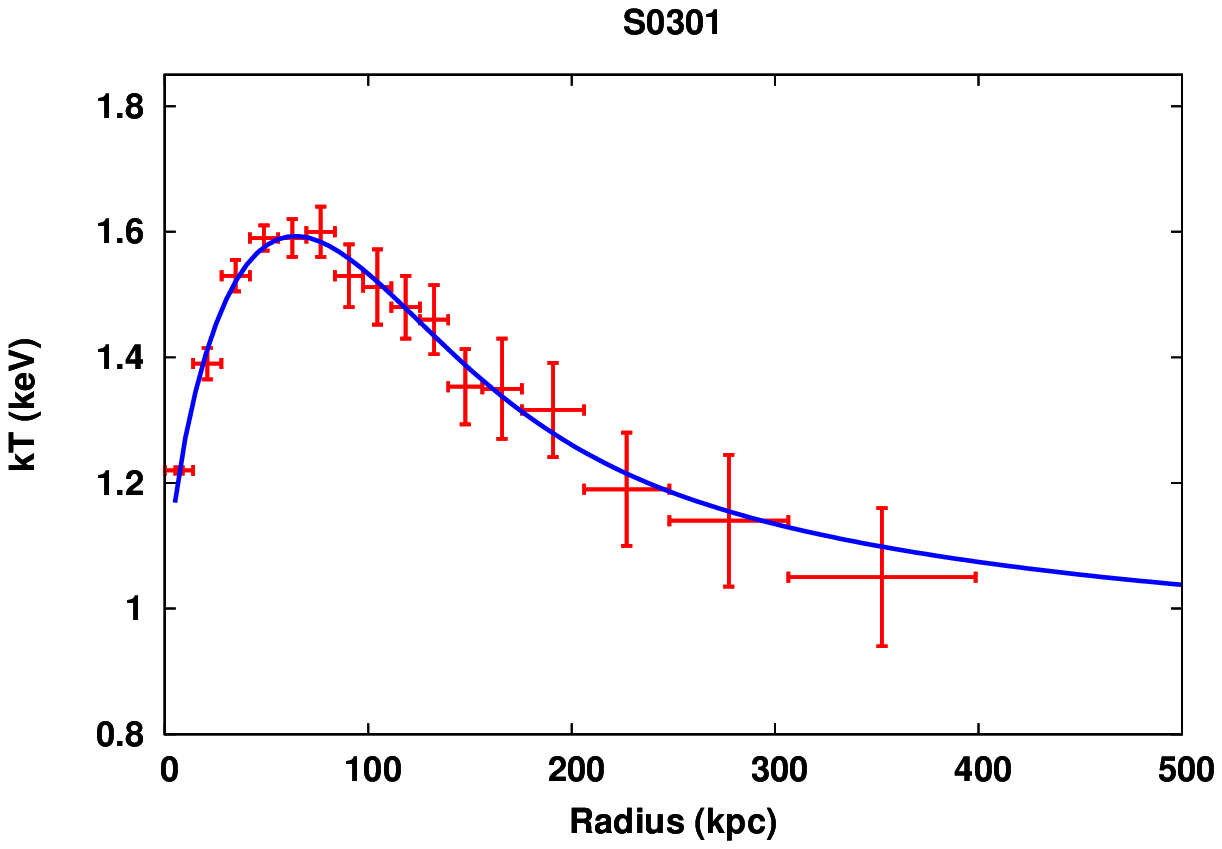,height=5cm,width=0.3\textwidth,angle=0}
\epsfig{figure=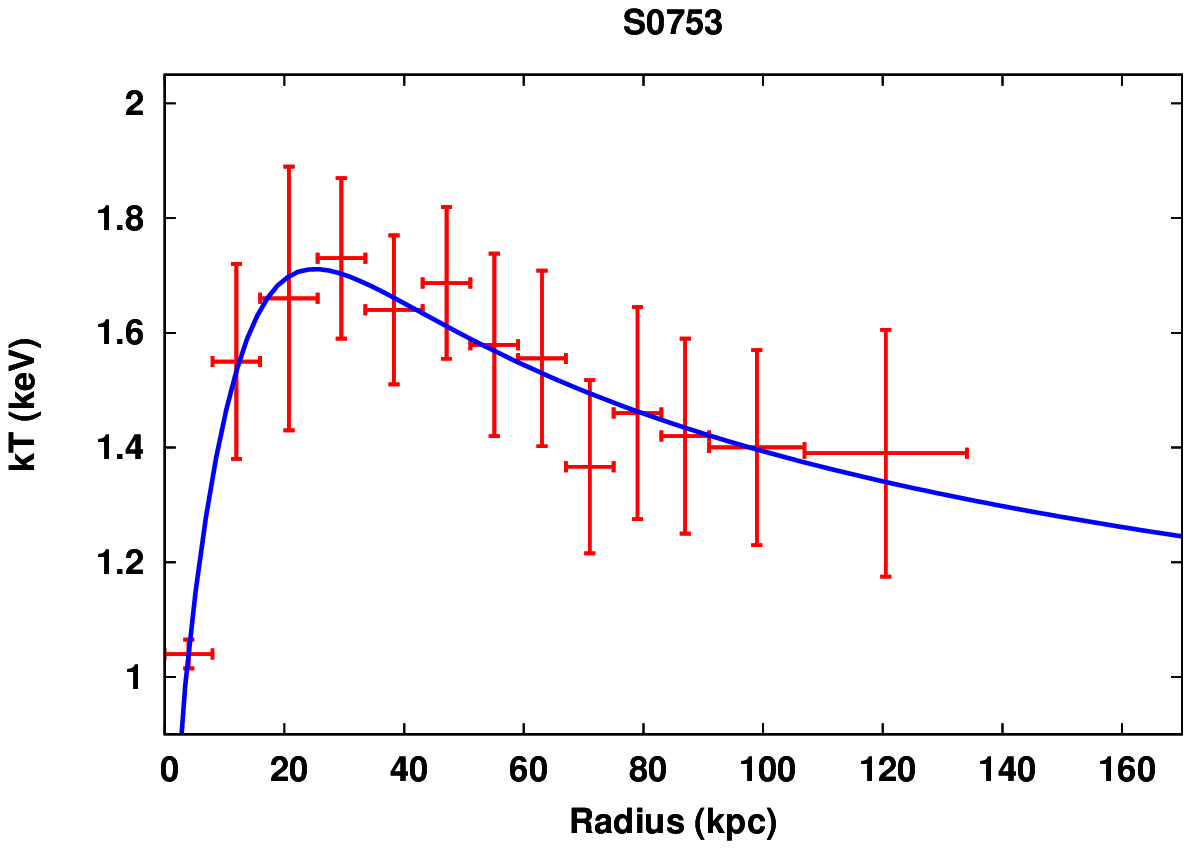,height=5cm,width=0.3\textwidth,angle=0}
\epsfig{figure=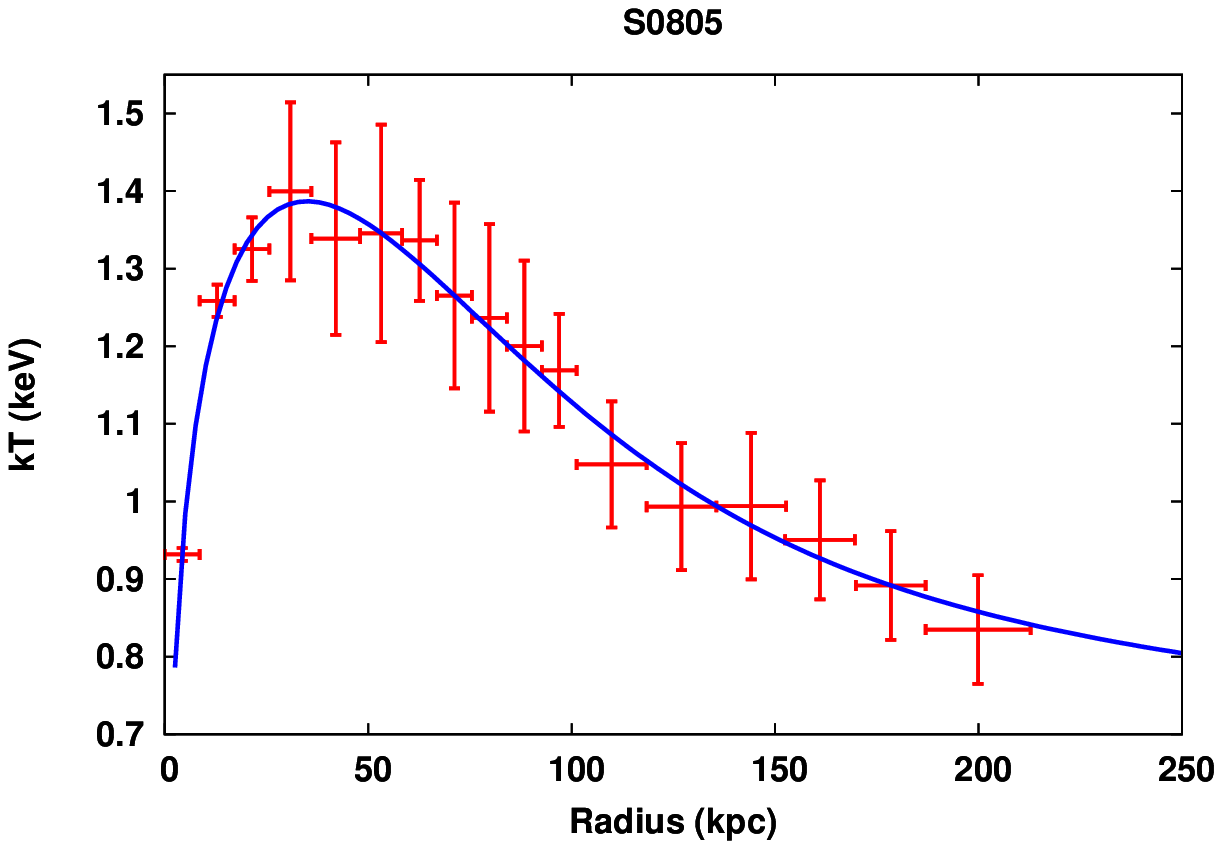,height=5cm,width=0.3\textwidth,angle=0}
}
\end{center}
\vspace{-20pt}
\caption{\footnotesize{\it Temperature profiles for  S0301, S0753, and S0805.}}
\end{figure*}

%-----------------------------Figure End--------------------------------

%-----------------------------Figure Start------------------------------
\begin{figure*}[ht]
\begin{center}
\hbox{
\epsfig{figure=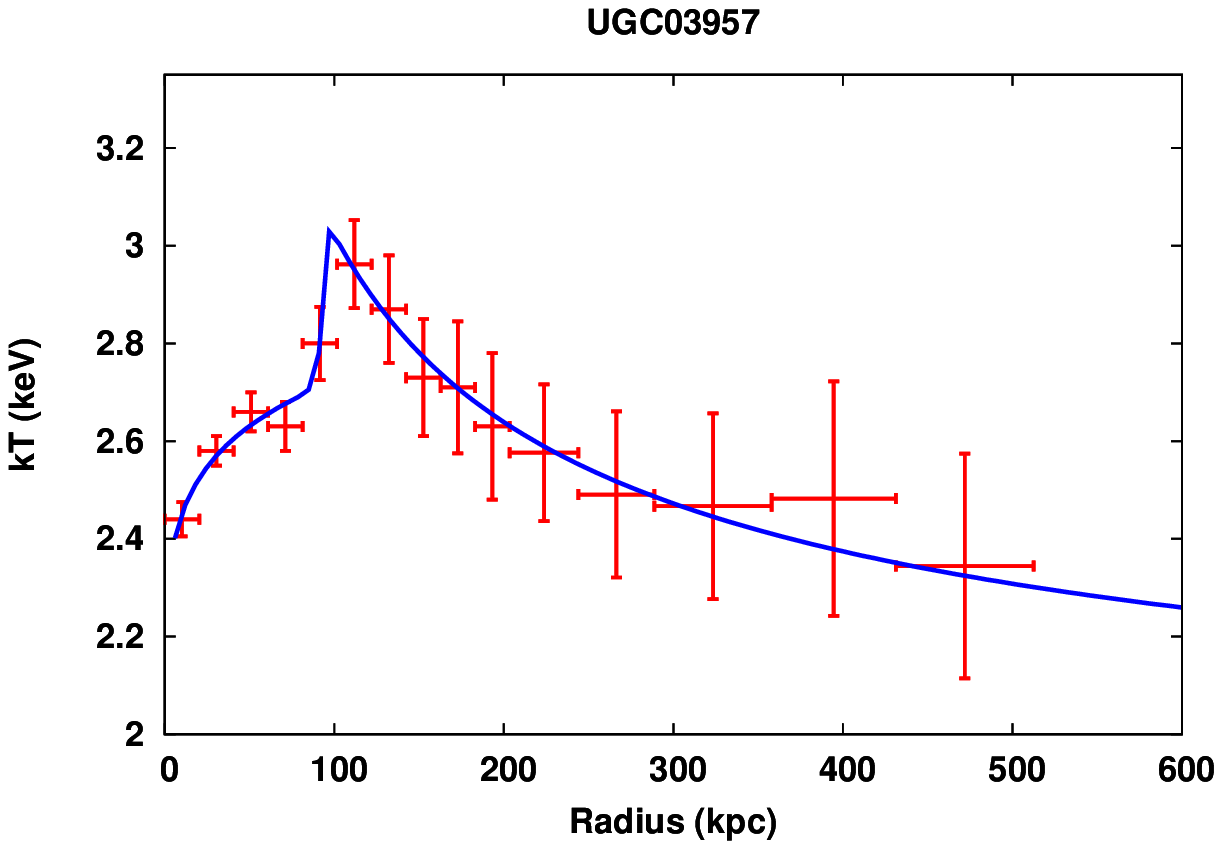,height=5cm,width=0.3\textwidth,angle=0}
\epsfig{figure=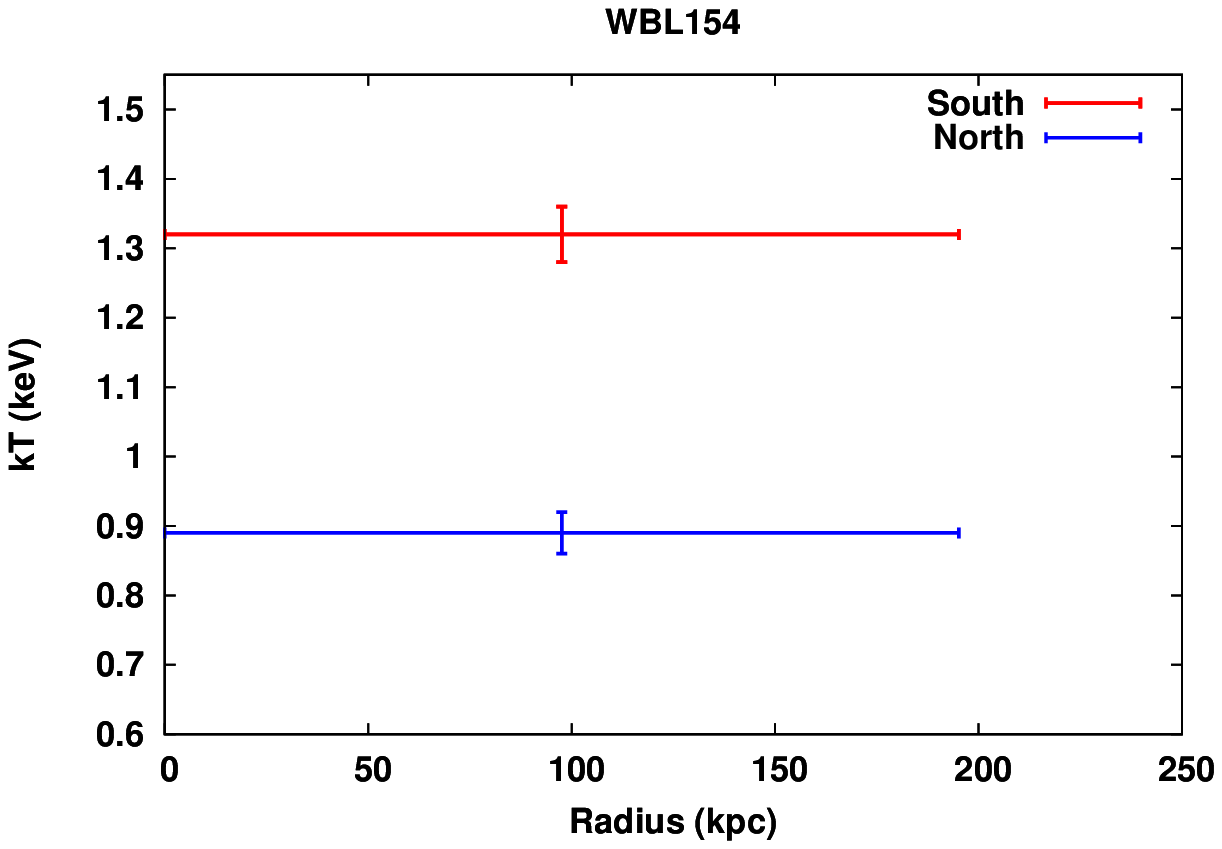,height=5cm,width=0.3\textwidth,angle=0}
}
\end{center}
\vspace{-20pt}
\caption{\footnotesize{\it Temperature profiles for UGC03957 and WBL154.}}
\end{figure*}

%-----------------------------Figure End--------------------------------

\end{appendix}

\end{document}